\documentclass[aps, prx, superscriptaddress, reprint]{revtex4-2}
\usepackage{graphicx}
\usepackage{dcolumn}
\usepackage{enumitem}
\usepackage{bm}
\usepackage{bbm}
\usepackage{amsmath}
\usepackage{hyperref}
\usepackage{braket}
\usepackage{xcolor}
\usepackage{xr}
\usepackage{mathrsfs}

\usepackage[ruled,vlined]{algorithm2e}

\DeclareMathOperator{\diag}{diag}
\DeclareMathOperator{\prob}{\mathbbm{P}}

\DeclareMathOperator{\R}{\mathbbm{R}}
\DeclareMathOperator{\N}{\mathbbm{N}}
\DeclareMathOperator{\one}{\mathbbm{1}}

\DeclareMathOperator*{\argmin}{arg\,min}
\def \({\left(}
\def \){\right)}

\makeatletter
\newbox\dtgme@widefloattext
\newcommand{\bottomnextwidefloat}{%
  \let\dtgme@combinedblfloats\@combinedblfloats
  \def\@combinedblfloats{%
    \ifx\@dbltoplist\@empty\else
      \setbox\dtgme@widefloattext\box\@outputbox
      \dtgme@combinedblfloats
      \setbox\@outputbox\vbox{%
        \unvbox\dtgme@widefloattext
        \vskip\dbltextfloatsep
        \unvbox\@outputbox\unskip
      }%
      \global\let\@combinedblfloats\dtgme@combinedblfloats
    \fi
  }%
}
\makeatother

\begin{document}

\title{Optimal parameterization of nonequilibrium generalized master equations from discrete-time experimental data}

\author{Chih-Wei Joshua Liu}
\affiliation{Biophysics Program, Stanford University, Stanford, California 94305, United States of America}

\author{J\'{e}r\'{e}mie Klinger}
\affiliation{Department of Chemistry, Stanford University, Stanford, California 94305, United States of America}

\author{Grant M. Rotskoff}
\email{rotskoff@stanford.edu}
\affiliation{Department of Chemistry, Stanford University, Stanford, California 94305, United States of America}
\affiliation{Institute for Computational and Mathematical Engineering, Stanford University, Stanford, California 94305, United States of America}

\date{November 7, 2026}

\begin{abstract}
    Kinetic analyses of experiments often require coarse-grained descriptions, but complex systems rarely conform to the widely used modeling assumptions of Markovianity and thermodynamic equilibrium.
    Memory is indeed a general and often inevitable consequence of coarse-graining.
    Markov state models (MSMs) are a popular choice of coarse-grained description, but require microstate assignments---which are rarely experimentally tunable---to macrostates that minimize memory.
    Generalized master equations (GMEs) circumvent this limitation of MSMs by explicitly capturing memory.
    However, GMEs are difficult to parameterize and usually formally approximate in the experimentally relevant discrete-time setting.
    Here we introduce a maximum-likelihood-based procedure to parameterize formally exact, physically feasible,
    discrete-time generalized master equations from experiments and simulations in and out of equilibrium.
    By adapting algorithms typically used in optimal transport, we construct physical-constraint-satisfying conditional-maximum-likelihood estimators of
    both exact Nakajima-Zwanzig memory kernels and time-convolutionless GME propagators in discrete time.
    Applying these estimators to three examples---
    experimental recordings of F\"{o}rster resonance energy transfer in an ion channel,
    experimental nanoparticle tracking of a processive molecular motor,
    and simulated folding of a benchmark protein domain---
    we recover kinetic parameters including relaxation rates, irreversibilities, dwell times, and first-passage times.
    These results establish discrete-time GMEs as a physically and statistically principled alternative to MSMs for kinetic analyses of experimental and simulated biomolecular systems.
\end{abstract}

\maketitle

\section{INTRODUCTION}
Coarse-grained models, essential for capturing kinetics from biomolecular experiments and simulations \cite{schilling2022coarse,schilling2024evolution}, often incorporate unphysical approximations.
In particular, the Markov assumption---future coarse-grained states depending only on the present and lacking memory of the past---holds only when microstate transitions within a macrostate are much faster than those between macrostates \cite{zwanzig2001nonequilibrium,elber2020molecular}.
By jointly optimizing microstate assignments and macrostate transition probabilities to satisfy this assumption \cite{husic2018minimum,bowman2012improved,ph2013identification,diez2022correlation},
Markov state models (MSMs) \cite{chodera2015markov,husic2018markov} can provide valuable insights into reversible processes in molecular dynamics simulations \cite{noe2020machine,noe2015kinetic,beauchamp2012simple,schwantes2013improvements}.
However, MSMs must use timesteps long enough to neglect remaining memory even with careful microstate assignment \cite{swope2004describing,prinz2014spectral},
obscuring fast transitions of physical interest.
MSMs face further challenges when describing nonequilibrium steady states \cite{knoch2019nonequilibrium}, which do not benefit from the regularizing assumption of detailed balance~\cite{prinz2011markov}.
Most importantly, microstates with indistinguishable experimental observables necessarily partition into the same macrostate \cite{dill1991denatured}: memoryless macrostates and thus valid MSMs are usually impossible for experimental data.

Generalized master equations (GMEs) avoid optimizing microstate assignments by explicitly incorporating memory \cite{gb2024tutorial,sartore2025markov}.
Though GMEs can evolve physically and experimentally meaningful macrostates at fine temporal resolution \cite{dominic2023building},
both time-local \cite{tokuyama1975statistical} and convolution-integral \cite{nakajima1958quantum,zwanzig1961memory} GMEs are difficult to parameterize in continuous time \cite{kidon2015exact,gu2024diagrammatic,qiang2003new,vroylandt2022likelihood}.
As experiments \cite{oppenheim2009discrete} and numerics \cite{larsson2003partial} usually discretize time,
discrete-time GMEs have seen increasing popularity and success as coarse-grained models of biomolecular kinetics \cite{goonetilleke2025targeting,yue2024tutorials,lorpaiboon2024accurate,unarta2024submillisecond}.
However, previous methods \cite{dominic2023memory} for parameterizing discrete-time GMEs are fraught.
Most methods use Taylor \cite{cao2023igme} or quadrature approximations \cite{makri2025discrete} that introduce systematic error;
these approximations are often prerequisites to complementary parameterizations of the continuous-time memory kernel \cite{meyer2020numerical,cao2020advantages}.
More recent methods are formally exact but invert transition matrices to solve linear least-squares systems \cite{sayer2023compact,dominic2023building}.
Estimates may be unphysically undefined when these matrices are noninvertible.
Even when defined, these estimates can include negative probabilities and fail to parameterize valid evolution equations.
Finally, no previous approach maximizes model likelihood or incorporates physical constraints such as stationarity or reversibility \cite{dominic2023building}.
Without satisfying statistical and physical criteria, estimated discrete-time GME parameters are not interpretable as optimal or even feasible inferences from data.

In this paper, we overcome all of these limitations through a principled, efficient, and broadly applicable approach to parameterizing discrete-time GMEs:
\begin{enumerate}
    \item We derive physical constraints in and out of equilibrium for the time-dependent propagators that generalize time-local GMEs to discrete time \cite{dominic2023building}.
    We then exactly generalize the convolutional GME \cite{nakajima1958quantum,zwanzig1961memory} to discrete time by Zwanzig-Mori projection \cite{mori1965transport,zwanzig1960ensemble} of Markov chains.
    Using a diagrammatic decomposition, we nonperturbatively reexpress discrete-time memory kernels in terms of readily estimated macrostate transition matrices.
    \item Combining mirror descent \cite{nemirovsky1979problem}---a generalization of gradient descent to non-Euclidean statistical distances \cite{bregman1967relax}---with the Sinkhorn-Knopp algorithm \cite{knopp1967concerning} widely used in entropic optimal transport \cite{cuturi2013sinkhorn},
    we introduce conditional-maximum-likelihood estimators \cite{meng1993mle} for both time-dependent propagators and discrete-time memory kernels.
    Our estimators use only coarse-grained trajectories, are always defined, 
    always validly parameterize evolution equations, and satisfy physical constraints to numerical tolerance in and out of equilibrium.
\end{enumerate}

Importantly, we bring our methods to bear on real experimental and molecular-dynamics examples with and without equilibrium constraints.
Parameterizing discrete-time GME models of experimentally recorded F\"{o}rster resonance energy transfer \cite{foerster1948zwischenmolekulare} in human cystic-fibrosis transmembrane-conductance regulator \cite{levring2023CFTR}, experimentally tracked rotary catalysis by bovine mitochondrial F$_1$-ATPase \cite{kobayashi2020BMF1}, and simulated folding of the benchmark chicken villin headpiece \cite{piana2012HP35}, we recover experimentally validated kinetic parameters including relaxation rates, irreversibilities, dwell times, and first-passage times.
These results demonstrate the utility of our approach relative to MSM baselines. 

\section{GENERALIZED MASTER EQUATIONS IN DISCRETE TIME}
\subsection{Preliminaries}
\begin{figure*}
\includegraphics[width=\textwidth]{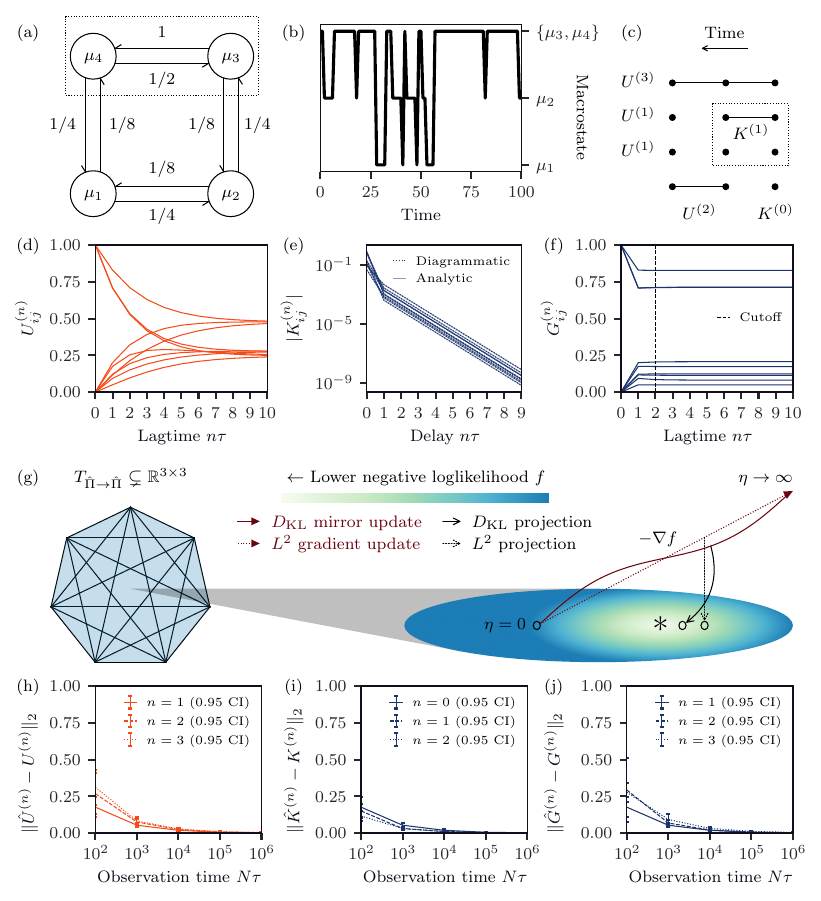}
\caption{
    Coarse-grained MJP and parameterization of its discrete-time GMEs.
    Toy ergodic MJP (a) generates example macrostate series (b).
    Microstates $\mu_3$ and $\mu_4$ are coarse-grained into a single macrostate and thus indistinguishable.
    (c) $2^2$ diagrams with $2+1$ vertices represent delay-$2\tau$ memory kernel $K^{(2)}$.
    A chain of $n$ vertices represents lag-$n\tau$ transition matrix $U^{(n)}$;
    empty spaces in rows of vertices represent factors of $-1$.
    Observe that $K^{(2)}=U^{(3)}-U^{(1)}K^{(1)}-U^{(2)}K^{(0)}$,
    giving the recursion in Eq. \ref{eq:recursion} for $n=2$.
    (d) $U^{(n)}$ for the coarse-grained MJP in (a).
    (e) Analytical memory kernels following Eq. \ref{eq:nzgmedt-kernel-def} coincide with those computed from $U^{(n)}$ using the diagrammatic recursion.
    (f) Lag-$n\tau$ TCL propagators for the coarse-grained MJP in (a) plateau at a cutoff lagtime.
    (g) $D_\mathrm{KL}$ projected mirror descent.
    Feasible fluxes have marginals equaling estimated stationary vector $\hat{\Pi}$ and belong to transportation polytope $T_{\hat{\Pi}\to\hat{\Pi}}$ embedded in $\mathbbm{R}^{3\times 3}$.
    In either $D_\mathrm{KL}$ or $L^2$ divergence, updates increase proximity to high-likelihood parameterizations while projections restore feasibility.
    Optimization is faster and stabler in $D_\mathrm{KL}$ than in $L^2$ for probability distributions such as flux matrices.
    Projected-mirror-descent estimates of (h) $\hat{U}^{(n)}$,
    (i) $\hat{K}^{(n)}$ (plugging $\hat{U}^{(n)}$ into Eq. \ref{eq:recursion}), and
    (j) $\hat{G}^{(n)}$ converge to ground truth in the large-sample limit.
    $\|\cdot\|_2$ is the spectral norm.
}
\label{fig:schematic}
\end{figure*}

Consider an ergodic Markov jump process (MJP) [Fig. \ref{fig:schematic}(a)] on microstates $\mu$ with generator $\mathcal{L}\in\R^{|\mu|\times |\mu|}$ and stationary probability vector $\pi\in\Delta^{|\mu|}$.
Let macrostates $M$ be a partition of $\mu$ with $\mathcal{M}: \mu\to M$ mapping microstates to macrostates.
We may define the Mori-Zwanzig projector \cite{mori1965transport,zwanzig1961memory} onto the relevant subspace as
\begin{equation}
    \mathcal{P} :=  A \Omega^{\mathsf{T}} \in \R^{|\mu|\times |\mu|},
\end{equation}
which is also known as the Hummer-Szabo projector \cite{hummer2014optimal} in classical stochastic systems.
Here $A\in\R^{|\mu |\times |M|}$ is such that $ A_{ij} := \one[\mathcal{M}(\mu_i)=M_j]$,
$ \Omega := \mathsf{D}_\pi A \mathsf{D}_{\Pi}^{-1}\in\R^{|\mu |\times |M|}$ where $\mathsf{D}_\cdot := \diag (\cdot)$,
and $\Pi :=  A^{\mathsf{T}}\pi\in\Delta^{|M|}$ is the probability vector representing the stationary density over macrostates $M$.
Note that $A^{\mathsf{T}} \Omega= \Omega^{\mathsf{T}} A=I\in\R^{|M|\times |M|}$;
we also consider only coarse-grainings $A$ that commute with time-reversal \cite{hartich2024comment,bisker2024reply}. 
The complement of $\mathcal{P}$ is the Mori-Zwanzig projector $\mathcal{Q} := I-\mathcal{P}$ onto the irrelevant subspace.
$\mathcal{P}$ is self-adjoint with respect to the $\pi$-weighted inner product $\braket{\cdot,\cdot}_\pi$ (Appendix \ref{app:identities});
we may define the complement of the adjoint projector $\mathcal{P}^\dagger := \mathcal{P}^{\mathsf{T}}$ as $\mathcal{Q}^\dagger  := I-\mathcal{P}^\dagger  = \mathcal{Q}^{\mathsf{T}}$.
As expected, projectors $\mathcal{P}$ and $\mathcal{Q}$ are orthogonal with respect to the $\pi$-weighted inner product, that is,
$\mathcal{P}^\dagger \mathsf{D}_\pi\mathcal{Q} = \mathcal{Q}^\dagger \mathsf{D}_\pi\mathcal{P} = 0$ (Appendix \ref{app:identities}).
Let $L := e^{\tau\mathcal{L}}$ be the single-step transition matrix of the Markov chain $\{x^{(n)}:x^{(n)}\in\mu\}_{n=0}^\infty$
with probability vectors $\{\rho^{(n)}:\rho^{(n)}\in\Delta^{|\mu|}\}_{n=0}^\infty$ representing its densities over $\mu$ at times $n\tau$.

We may now define the coarse-grained sequence $\{X^{(n)}:X^{(n)}=\mathcal{M}(x^{(n)})\in M\}_{n=0}^\infty$
with probability vectors $\{P^{(n)}:P^{(n)}=A^{\mathsf{T}}\rho^{(n)}\in\Delta^{|M|}\}_{n=0}^\infty$ representing its densities over $M$ at times $n\tau$.
Denote
\begin{equation}
    U^{(n)} := A^{\mathsf{T}} L^n \Omega
    \label{eq:U_explicit_def}
\end{equation}
and observe that
\begin{align}
    P^{(n)} = U^{(n)} P^{(0)} +  A^{\mathsf{T}} L^n (I-\Omega A^{\mathsf{T}})\rho^{(0)} \label{eq:transfer-transition}
\end{align}
for all $n\in\N$ where $\N$ denotes nonnegative integers.
Under the approximation
\begin{equation}
    \mathcal{Q}^\dagger \rho^{(0)} := (I-\Omega A^{\mathsf{T}})\rho^{(0)} = 0
    \label{eq:initialization_assumption}
\end{equation}
typical of generalized master equations \cite{schilling2022coarse},
Eq. \ref{eq:transfer-transition} simplifies to
\begin{equation}
    P^{(n)} = U^{(n)}P^{(0)}
\end{equation}
such that
\begin{equation}
    U_{ij}^{(n)} = \prob (X^{(n)} = M_i|X^{(0)} = M_j)
\end{equation}
and $U^{(n)}$ is the lag-$n\tau$ transition matrix.
Note that Eq. \ref{eq:initialization_assumption} holds for initial densities $\rho^{(0)}$ in the image of $\mathcal{P}^\dagger$, which include $\rho^{(0)}=\pi$ (Appendix \ref{app:identities}).
Thus we may estimate $U^{(n)}$ from the $n$-timestep transition counts of stationary coarse-grained trajectories [Fig. \ref{fig:schematic}(b)]
such as those of steady-state experiments or ``burned-in'' \cite{stella1998equilibration,vanvliet2006equilibration} MD simulations.
$U^{(n)}$ is also stationary with respect to $\Pi$ at all lags;
$U^{(n)}$ is detailed-balance with respect to $\Pi$ if $L$ is detailed balance with respect to $\pi$ (Appendix \ref{app:identities}).
Unlike $L$, however, $U^{(1)}$ generally fails to satisfy the Chapman-Kolmogorov equation \cite{chapman1928brownian,kolmogoroff1931ueber} and is not the propagator of a homogeneous Markov chain \cite{sartore2025markov}.

\subsection{Time-convolutionless generalized master equation in discrete time}
The time-convolutionless generalized master equation (TCL GME) \cite{tokuyama1975statistical,tokuyama1976statistical} in continuous time
\begin{equation}
    \frac{\partial}{\partial t}(A^{\mathsf{T}} e^{t\mathcal{L}}\rho^{(0)}) = \mathscr{G}(t)A^{\mathsf{T}}e^{t\mathcal{L}}\rho^{(0)}
    \label{eq:tcl-gme-ct}
\end{equation}
models the evolution of coarse-grained density $A^{\mathsf{T}} e^{t\mathcal{L}}\rho^{(0)}$ using a time-local but explicitly time-dependent generator $\mathscr{G}(t)$.
Assuming Eq. \ref{eq:initialization_assumption}, the continuous-time TCL-GME generator usually takes the form \cite{chaturvedi1979timeconvolutionless}
\begin{equation*}
    \mathscr{G}(t) := A^{\mathsf{T}}\mathcal{L}[I-\Sigma (t)]^{-1}\Omega
\end{equation*}
where
\begin{equation*}
    \Sigma (t) := \int_0^t e^{(t-s)\mathcal{Q}^\dagger\mathcal{L}}\mathcal{Q}^\dagger\mathcal{L}\mathcal{P}^\dagger e^{(s-t)\mathcal{L}}\mathrm{d}s
\end{equation*}
in the classical stochastic context.
Parameterization of $\mathscr{G}(t)$ is generally difficult and unstable \cite{sayer2023compact,dominic2023building}
except with detailed knowledge of microscopic dynamics \cite{timm20115tcl,kidon2015exact,gu2024diagrammatic}.
We assume existence of $\mathscr{G}(t)$ following recent chemical kinetics literature \cite{dominic2023building,sinclair2025influence,bement2025models,yue2024tutorials}, though we note that it is difficult or impossible to establish without recourse to microscopic generator $\mathcal{L}$ as
extensively discussed elsewhere \cite{strasberg2019nonmarkovianity,maldonado2012investigating,chruscinski2010nonmarkovian,hou2012singularity,sayer2023compact}.

As $A^{\mathsf{T}} e^{t\mathcal{L}}\Omega$ is the lagtime $t$ transition matrix for any $\rho^{(0)}$ satisfying Eq. \ref{eq:initialization_assumption},
each of its columns is a probability vector representing densities over macrostates at time $t$ given an initial macrostate.
Thus by Eq. \ref{eq:tcl-gme-ct},
\begin{equation*}
    A^{\mathsf{T}} e^{u\mathcal{L}}\Omega = \(\mathcal{T}_\leftarrow\exp\int_s^u\mathscr{G}(t)\mathrm{d}t\) A^{\mathsf{T}} e^{s\mathcal{L}}\Omega
\end{equation*}
for times $u\geq s$ where $\mathcal{T}_\leftarrow$ is the chronological time-ordering operator.
Then letting $u=n\tau$ and $s=(n-1)\tau$,
\begin{equation}
    U^{(n)} = G^{(n)}U^{(n-1)}
    \label{eq:tcl-gme-dt-one-step-propagator-product}
\end{equation}
where we define the TCL propagator
\begin{equation}
    G^{(n)} := \mathcal{T}_\leftarrow\exp\int_{(n-1)\tau}^{n\tau}\mathscr{G}(t)\mathrm{d}t.
    \label{eq:tcl-gme-dt-propagator}
\end{equation}
for all $n\in\N_+$ with $\N_+$ denoting the positive integers.
Like $U^{(n)}$, $G^{(n)}$ encodes correlations between macrostate indicators at discrete times;
$G^{(n)}$ is often easier to parameterize with data than $\mathscr{G}(t)$ \cite{sayer2023compact,dominic2023building}.
Clearly, 
\begin{equation}
    G^{(n)} = U^{(n)}(U^{(n-1)})^{-1}
    \label{eq:G-from-inversion}
\end{equation}
if $U^{(n-1)}$ is invertible.
However, propagator $G^{(n)}$ may be defined by Eq. \ref{eq:tcl-gme-dt-propagator} even for $n$ such that $U^{(n-1)}$ is singular,
such as in the large-$n$ limit when $U^{(n-1)}$ approaches $\Pi\mathbf{1}^{\mathsf{T}}_{|M|}$ for $\mathbf{1}_\cdot$ the column vector of ones with $\cdot$ elements (Appendix \ref{app:identities}).
Indeed, $\mathscr{G}(t)$ generally plateaus to long-time effective Markovian generator $\mathscr{G}(\infty)$ \cite{nestmann2021quantum}.
In the long-time limit, $G^{(n)}$ thus converges [Fig. \ref{fig:schematic}(f)] to effective Markovian propagator
\begin{equation}
    G^{(\infty)} := e^{\tau\mathscr{G}(\infty)}
\end{equation}
that is well-defined even when $U^{(n-1)}$ becomes singular.
We note that $G^{(n)}$ permits an alternative definition by direct projection of the microscopic (discrete-time) Markov chain,
but this definition is less general (Appendix \ref{app:alternative-tcl-gme-dt-def}) than the definition in Eq. \ref{eq:tcl-gme-dt-propagator} that we hereafter adopt.
Neither definition invokes a quadrature or finite-difference approximation.

$G^{(n)}$ with nonnegative entries for all $n\in\N_+$ permit a straightforward probabilistic interpretation.
Consider independent, identically distributed (i.i.d.) initial microstates $x^{(0)}\sim \rho^{(0)}$ where $\rho^{(0)}$ satisfies the assumption in Eq. \ref{eq:initialization_assumption}.
Evolving the macrostates of $x^{(0)}$ using $G^{(n)}$ as a time-dependent but time-local propagator,
\begin{align}
    \prob(&X^{(n)} =M_i|X^{(0)}=M_k) = U^{(n)}_{ik}\notag\\
    &= \sum_{j=1}^{|M|}G_{ij}^{(n)}U^{(n-1)}_{jk}\notag\\
    &= \sum_{j=1}^{|M|}G_{ij}^{(n)}\prob(X^{(n-1)}=M_j|X^{(0)}=M_k)\implies\notag\\
    G_{ij}^{(n)} &= \prob(X^{(n)}=M_i|X^{(n-1)}=M_j,X^{(0)}=M_k)\label{eq:G_interpretation_start}\\
    &= \prob(X^{(n)}=M_i|X^{(n-1)}=M_j)\label{eq:G_interpretation_end}
\end{align}
and the probability of a transition from $M_j$ to $M_i$ between times $(n-1)\tau$ and $n\tau$ is conditionally independent of $X^{(0)}$ taking values $M_k$ for all $i,j,k\in\{1,\cdots,|M|\}$ and $n>1$.
Equivalently, $\{X^{(n)}\}_{n=0}^\infty$ is a first-order inhomogeneous Markov chain:
the time-convolutionless generalized master equation in discrete time (TCL-GME-DT) expresses non-Markovian evolution in the time-local form
\begin{equation}
    P^{(n)} = G^{(n)}P^{(n-1)}
    \label{eq:tcl-gme-dt}
\end{equation}
for all $n\in\N_+$.
At positive lags, the TCL-GME-DT parameterizes path observables such as dwell and first-passage times under the assumption that transitions are conditionally independent of $X^{(0)}$.
Note that $\mathscr{G}_{ij}(t)$ may be negative for $i\neq j$ and represent scaled additive terms of positive $M_i\to M_j$ transition rates \cite{laine2012local,brandner2025dynamics}.
Thus $G^{(n)}$ may have negative entries, though its definition in terms of $\mathscr{G}(t)$ implies its columns always sum to unity.
To be useful for simulations and kinetic characterization, $G^{(n)}$ must have only nonnegative entries:
we seek nonnegative approximate $G^{(n)}_{ij}$ analogous to approximate $\mathscr{G}(t)$ consistent with completely positive trace-preserving evolution in the quantum-mechanical context \cite{shabani2005completely,matthew2023full,nestmann2021quantum}.

Let us now characterize the steady states of TCL-GME-DT propagation.
From Eq. \ref{eq:tcl-gme-dt-one-step-propagator-product} and the stationarity of $U^{(n)}$ and $U^{(n-1)}$ with respect to $\Pi$,
\begin{equation}
    G^{(n)}\Pi = G^{(n)}U^{(n-1)}\Pi = U^{(n)}\Pi = \Pi
    \label{eq:G-stationary-vec}
\end{equation}
so that $\Pi$ is also the stationary vector of TCL propagators $G^{(n)}$ for all $n\in\N_+$.
In the equilibrium case, the microscopic Markov chain is detailed-balance with respect to $\pi$.
This has several consequences.
Firstly, the microscopic trajectory $\{x^{(n)}\}_{n=0}^{N-1}$ must have all stationary statistics equaling those of its time reverse $\{x^{(n)}\}_{n=N-1}^{0}$ \cite{feller1991introduction}.
As proven \cite{pomeau1982symmetry} and reviewed \cite{dieball2025time} elsewhere, the coarse-grained time series $\{X^{(n)}\}_{n=0}^{N-1}$ inherits the statistical reversibility of the microscopic time series.
At a minimum, then, we require $G^{(n)}$ that are detailed-balance with respect to $\Pi$ so that local irreversibility vanishes \cite{ro2022modelfree,ge2006reversibility,lynn2022emergence,me}.
We also require $U^{(n)}$ and $U^{(n-1)}$ that are detailed-balance with respect to $\Pi$ (Appendix \ref{app:identities});
detailed-balance $G^{(n)}$, $U^{(n-1)}$, and $U^{(n)}$ respectively imply reversible statistics of the pairs $\{X^{(n)}, X^{(n-1)}\}$, $\{X^{(n-1)}, X^{(0)}\}$, and $\{X^{(n)}, X^{(0)}\}$.
Then by Eq. \ref{eq:tcl-gme-dt-one-step-propagator-product}
\begin{align}
    \mathsf{D}_\Pi U^{(n-1)\mathsf{T}}G^{(n)\mathsf{T}} &= G^{(n)}U^{(n-1)}\mathsf{D}_\Pi\notag\\
    &= G^{(n)}\mathsf{D}_\Pi U^{(n-1)\mathsf{T}}\notag\\
    &= \mathsf{D}_\Pi G^{(n)\mathsf{T}}U^{(n-1)\mathsf{T}}\notag\\
    \iff G^{(n)}U^{(n-1)} &= U^{(n-1)}G^{(n)}\label{eq:reversibility-commutation}
\end{align}
and $[G^{(n)}, U^{(n-1)}]=0$ for all $n\in\N_+$.

Having discussed consequences of reversible two-time statistics,
we will now see that GMEs constrain higher-order statistics and thermodynamic properties not accessible at the level of Markovian description.
Master equations---even those with memory---typically define statistics at only single and paired times \cite{modi2021quantum}.
However, the assumption that $X^{(n)}$ conditioned on $X^{(n-1)}$ is independent of $X^{(0)}$ is explicit in Eqs. \ref{eq:G_interpretation_start}-\ref{eq:G_interpretation_end}.
Unlike master equations in general, the TCL-GME-DT then also defines statistics of the triplet $\{X^{(n)},X^{(n-1)},X^{(0)}\}$.
Let us consider the thermodynamic consequences.
Reversibility implies $\{X^{(n)},X^{(n-1)},X^{(0)}\}$ has the same stationary statistics as its time reverse $\{X^{(0)},X^{(1)},X^{(n)}\}$.
Then for all $n\in\N_+$,
\begin{multline}
    \prob_\Pi (X^{(n)}=M_k|X^{(1)}=M_j,X^{(0)}=M_i)\\
    = \frac{G_{ij}^{(n)}U^{(n-1)}_{jk}\Pi_k}{U^{(1)}_{ji}\Pi_i}
    \label{eq:triplet-reversibility}
\end{multline}
where $\prob_\Pi(\cdot) = \prob(\cdot|X^{(0)}\sim\Pi)$.
However, as derived with Eq. \ref{eq:triplet-reversibility} in Appendix \ref{app:triplet-irreversibility},
the TCL-GME-DT specifies the preceding left-hand-side probability only for $n=2$.
Reversible $\{X^{(2)}, X^{(1)}, X^{(0)}\}$ statistics is equivalent to $G^{(2)} = G^{(1)}$.
We may estimate third-order irreversibility \cite{roldan2012entropy} $\sigma_\mathrm{GME}$ as
\begin{equation}
    \hat{\sigma}_\mathrm{GME} = \frac{1}{2\tau}\sum_{ijk}\hat{G}_{ij}^{(2)}\hat{G}_{jk}^{(1)}\hat{\Pi}_k\ln\frac{\hat{G}_{ij}^{(2)}\hat{G}_{jk}^{(1)}\hat{\Pi}_k}{\hat{G}_{kj}^{(2)}\hat{G}_{ji}^{(1)}\hat{\Pi}_i},
    \label{eq:epr_estimator}
\end{equation}
which lower-bounds \cite{ehrich2021tightest} the stationary entropy-production rate (EPR)---provided
memory is negligible past a delay of $2\tau$ \cite{schwarz2025consistent}---using
estimates $\{\hat{G}^{(n)}\}_{n=1}^2$ of the TCL-GME-DT propagators and $\hat{\Pi}$ of the macrostate stationary vector.
As expected, $\sigma_\mathrm{GME}$ vanishes for homogeneous Markov chains that have $G^{(2)}=G^{(1)}$.
It does not generally vanish for unequal detailed-balance $G^{(2)}$ and $G^{(1)}$:
we may use Eq. \ref{eq:epr_estimator} to lower-bound the stationary EPRs of, for example,
two-macrostate coarse-grained processes in which broken detailed balance is never observable in individual propagators.

\subsection{Nakajima-Zwanzig generalized master equation in discrete time}
Continuous-time Nakajima-Zwanzig generalized master equations of the form \cite{nakajima1958quantum,zwanzig1961memory}
\begin{equation}
    \frac{\partial}{\partial t}(A^{\mathsf{T}} e^{t\mathcal{L}}\rho^{(0)}) = \int_0^t\mathscr{K}(t-s)A^{\mathsf{T}} e^{s\mathcal{L}}\rho^{(0)}\mathrm{d}s
    \label{eq:nzgmect}
\end{equation}
capture memory effects by convolving a delay-dependent memory kernel $\mathscr{K}(t-s)$ with past coarse-grained densities $A^{\mathsf{T}} e^{s\mathcal{L}}\rho^{(0)}$ under the assumption of Eq. \ref{eq:initialization_assumption}.
In classical contexts, the memory kernel usually takes the form \cite{espanol2002coarse}
\begin{equation*}
    \mathscr{K}(t) := A^{\mathsf{T}}\mathcal{L}\mathcal{Q}^\dagger\mathcal{L}e^{t\mathcal{Q}^\dagger\mathcal{L}}\Omega + \delta (t)A^{\mathsf{T}}\mathcal{L}\Omega
\end{equation*}
where $\delta (\cdot)$ is the Dirac delta function; like that of $\mathscr{G}(t)$,
its parameterization is difficult and unstable \cite{qiang2003new,mulvihill2021roadmap,vroylandt2022likelihood} except with detailed knowledge of microscopic dynamics \cite{cohen2011memory,tian2022explicit,netz2024derivation}.
To lessen the burden of kernel estimation, Eq. \ref{eq:nzgmect} is often approximated as
\begin{equation}
    P^{(n)} = \sum_{m=0}^n \tau b_m\mathscr{K}[(n-m)\tau]P^{(m)}
    \label{eq:transfer-tensor-method}
\end{equation}
where the $b_m$ are quadrature weights \cite{cerrillo2014nonmarkovian,buser2017initial,makri2025discrete}.

Here we present an exact Nakajima-Zwanzig generalized master equation in discrete time (NZ-GME-DT)
that takes a form resembling Eq. \ref{eq:transfer-tensor-method} and is easily parameterized from stationary coarse-grained trajectories.
A detailed proof is presented in Appendix \ref{app:nzgmedt};
in brief, we first observe that the microscopic Markov chain
\begin{align}
    \rho^{(n+1)} = L\rho^{(n)}
\end{align}
has formal solution
\begin{equation}
    \rho^{(n+1)} = (L\mathcal{Q}^\dagger )^{n+1}\rho^{(0)} + \sum_{m=0}^n (L\mathcal{Q}^\dagger )^{n-m}L\mathcal{P}^\dagger \rho^{(m)}
    \label{eq:formal-solution}
\end{equation}
by Dyson's identity \cite{dyson1949S}.
Though generality of Dyson's identity remains an area of active research \cite{zhu2018estimation,widder2025generalized},
we note that Eq. \ref{eq:formal-solution} holds for the bounded propagators $L$ we consider \cite{gill1990survey}.
Substituting Eq. \ref{eq:formal-solution} into the second term of the projected microscopic Markov chain
\begin{equation}
    \mathcal{P}^\dagger\rho^{(n+1)} = \mathcal{P}^\dagger L\mathcal{P}^\dagger\rho^{(n)} + \mathcal{P}^\dagger L\mathcal{Q}^\dagger\rho^{(n)}
\end{equation}
we obtain
\begin{equation}
    \mathcal{P}^\dagger \rho^{(n+1)} = \mathcal{P}^\dagger  L\sum_{m=0}^{n}(\mathcal{Q}^\dagger L)^{n-m}\mathcal{P}^\dagger\rho^{(m)}
    \label{eq:P-markov-chain-main}
\end{equation}
upon assuming Eq. \ref{eq:initialization_assumption}.
Then as $A^{\mathsf{T}}\mathcal{P}^\dagger = A^{\mathsf{T}}$,
\begin{equation}
    A^{\mathsf{T}}\rho^{(n+1)} = \sum_{m=0}^{n}A^{\mathsf{T}} L(\mathcal{Q}^\dagger L)^{n-m}\mathcal{P}^\dagger\rho^{(m)}
    \label{eq:unsimplified-nzgmedt}
\end{equation}
by coarse-graining both sides of Eq. \ref{eq:P-markov-chain-main}.
Eq. \ref{eq:unsimplified-nzgmedt} at last simplifies to the desired form
\begin{equation}
    P^{(n+1)} = \sum_{m=0}^{n}K^{(n-m)} P^{(m)}
    \label{eq:nz-gme-dt}
\end{equation}
upon defining
\begin{equation}
    K^{(n)} :=  A^{\mathsf{T}} L(L-\mathcal{P}^\dagger L)^{n} \Omega
    \label{eq:nzgmedt-kernel-def}
\end{equation}
and identifying $P^{(n)} := A^{\mathsf{T}}\rho^{(n)}$ for all $n\in\N$.
Eq. \ref{eq:nz-gme-dt} evolves distributions over macrostates to stationarity,
thus characterizing population-level relaxation kinetics.

The term $L(L-\mathcal{P}^\dagger L)^n$ in Eq. \ref{eq:nzgmedt-kernel-def} appears difficult to compute but permits a diagrammatic decomposition, enabling efficient parameterization of Eq. \ref{eq:nz-gme-dt}.
Constructing the diagrams is straightforward: each term of $L(L-\mathcal{P}^\dagger L)^n=L(IL-\mathcal{P}^\dagger L)^n$ has $n+1$ factors of $L$ and a factor of either $-\mathcal{P}^\dagger$ or (implicit) $I$ between each factor of $L$.
To represent these terms, we may draw a vertex for each of the $n+1$ factors of $L$ in a row.
Time increases from right to left so that operator ordering coincides with that of Eq. \ref{eq:nzgmedt-kernel-def}.
Drawing an edge between successive factors of $L$ for each factor of $I$---such
that a chain of $m$ vertices represents $m$ applications of $(I)L$---we
may enumerate the $2^n$ unique diagrams for terms of $K^{(n)}$ [Fig. \ref{fig:schematic}(c)].
Note that we may recover terms by writing $L$ for each vertex and a factor of $-\mathcal{P}^\dagger $ for each ``cut'' between disconnected successive vertices,
multiplying all terms with $ A^{\mathsf{T}}$ from the left and $ \Omega$ from the right.
$K^{(n)}$ is then the sum of terms from the $2^n$ unique diagrams with $n+1$ vertices.

This decomposition reveals a recursion for $K^{(n)}$ in terms of the much simpler $U^{(m)}$.
Note that $-\mathcal{P}^\dagger  =  \Omega(-1) A^{\mathsf{T}}$ and recall the definition of $U^{(m)}$ given in Eq. \ref{eq:U_explicit_def}.
We may equivalently write from left to right $U^{(m)}$ for each chain of $m$ vertices and a factor of $-1$ for each cut.
The sum of terms for the $2^n$ unique diagrams with $n+1$ vertices is again $K^{(n)}$.
We may construct unique diagrams with $n+1$ vertices by drawing, from left to right, a chain of $m<n+1$ vertices, a cut, and a diagram with $n-m+1$ vertices.
A chain of $m$ vertices and a cut together represent a factor of $-U^{(m)}$;
the $2^{n-m}$ possible diagrams with $n-m+1$ vertices represent terms that sum to $K^{(n-m)}$.
The $2^{n-m}$ diagrams with a chain of $m$ vertices on the left, a cut, and a diagram with $n-m+1$ vertices on the right thus represent terms summing to $-U^{(m)}K^{(n-m)}$.
This procedure enumerates $\sum_{m=1}^n 2^{n-m} = 2^n-1$ unique diagrams that together represent the sum $-\sum_{m=1}^n U^{(m)}K^{(n-m)}$;
with the chain of $n+1$ vertices representing $U^{(n+1)}$, these are the $2^n$ possible diagrams with $n+1$ vertices representing terms that sum to $K^{(n)}$.
Thus for all $n\in\N_+$ we obtain a recursion for the memory kernels
\begin{equation}
    K^{(n)} = U^{(n+1)} - \sum_{m=1}^{n}U^{(m)} K^{(n-m)}
    \label{eq:recursion}
\end{equation}
with base case $K^{(0)} = U^{(1)}$ represented diagrammatically by the trivial graph.
As $U^{(n)}$ are merely stationary macrostate transition matrices [Fig. \ref{fig:schematic}(d)],
discrete-time memory kernels $K^{(n)}$ [Fig. \ref{fig:schematic}(e)] are readily estimated.

Eq. \ref{eq:recursion} closely resembles the Yule-Walker \cite{yule1927method,walker1931periodicity} form
\begin{equation}
    K^{(n)} = U^{(n+1)} - \sum_{m=1}^{n}K^{(n-m)}U^{(m)}
    \label{eq:flipped_recursion}
\end{equation}
previously derived for all $n\in\N_+$ in a quantum-mechanical context \cite{pollock2018tomograph},
suggesting a complementary nondiagrammatic derivation (Appendix \ref{app:nzgmedt}).
We immediately perceive that the sums in Eqs. \ref{eq:recursion} and \ref{eq:flipped_recursion} share diagrammatic representations and thus values.
Discrete-time memory kernels estimated from transition matrices using Eq. \ref{eq:recursion} thus also satisfy Eq. \ref{eq:flipped_recursion}.
Furthermore, discrete-time memory kernels satisfy a fluctuation-dissipation theorem (FDT) of the second kind \cite{callen1951irreversibility,kubo1956fdt} when $L$ is detailed-balance,
though this FDT is of limited utility in memory-kernel parameterization (Appendix \ref{app:fdt}).
In what follows, we describe conditional-maximum-likelihood strategies for computing both transition matrices $U^{(n)}$ and $G^{(n)}$.

\section{PARAMETERIZING NAKAJIMA-ZWANZIG GENERALIZED MASTER EQUATIONS IN DISCRETE TIME}
\subsection{Parameterizing discrete-time memory kernels}
Given the recursion in Eq. \ref{eq:recursion},
discrete-time memory kernels $K^{(n)}$ are readily formed from transition matrices $U^{(m)}$.
There exist many methods to parameterize transition matrices using lag-$m$ transition-count matrices
\begin{equation}
    C^{(m)}_{ij} := \sum_{n=0}^{N-m-1}\mathbbm{1}(X^{(n+m)}=M_i)\mathbbm{1}(X^{(n)}=M_j)
    \label{eq:transition_counts}
\end{equation}
constructed from macrostate trajectories $\{X^{(n)}\}_{n=0}^{N-1}$.
We refer readers to comprehensive reviews in Refs. \cite{chodera2015markov,husic2018markov}.
However, transition probability matrices from the same Markov jump process share properties crucial for physically feasible memory-kernel parameterization.
As shown in Appendix \ref{app:identities}, 
all $U^{(m)}$ of the same coarse-grained MJP share a stationary vector $\Pi$.
$\Pi$ is usually more easily estimated than $U^{(m)}$ at any lag.
We thus require a method for parameterizing maximum-likelihood transition matrices conditioned on a stationary-vector estimate $\hat{\Pi}$ shared by all lags.

When the further constraint of detailed balance is desired, a simple iterative scheme forms maximum-likelihood reversible estimates of $U^{(m)}$ from $C^{(m)}$ given $\hat{\Pi}$ \cite{prinz2011markov,trendelkampschroer2015estimation}.
However, there are to our knowledge no previously presented methods for conditional-maximum-likelihood estimation of general $U^{(m)}$ from $C^{(m)}$ given $\hat{\Pi}$.
Here we present such a method to form estimates $\hat{U}^{(m)}$ of $U^{(m)}$,
imposing stationary vector $\hat{\Pi}$ without the generally inapplicable assumption of detailed-balance $U^{(m)}$.
Let $\odot$ denote elementwise multiplication and $\oslash$ elementwise division.
Crucially, recognize that the flux-matrix estimate at lag $m$
\begin{equation}
    \hat{\Phi}^{(m)} := \hat{U}^{(m)}\mathsf{D}_{\hat{\Pi}} = \hat{U}^{(m)}\odot\mathbf{1}_{|M|}\hat{\Pi}^{\mathsf{T}}
\end{equation}
maps uniquely to a transition-matrix estimate $\hat{U}^{(m)} = \hat{\Phi}^{(m)}\mathsf{D}_{\hat{\Pi}}^{-1} = \hat{\Phi}^{(m)}\oslash\mathbf{1}_{|M|}\hat{\Pi}^{\mathsf{T}}$.
Moreover, the loglikelihood
\begin{multline}
    \sum_{ij}C_{ij}^{(m)}\ln \hat{U}_{ij}^{(m)} =\\
    \sum_{ij}C_{ij}^{(m)}\ln\hat{\Phi}_{ij}^{(m)}-\sum_{ij}C_{ij}^{(m)}\ln\hat{\Pi}_j
\end{multline}
of the counts $C^{(m)}$ is strictly concave with respect to $\hat{\Phi}^{(m)}$.
Maximum-likelihood estimation of $U^{(m)}$ conditioned on $\hat{\Pi}$ is thus the optimization problem
\begin{align}
    \min_{\Phi'}\quad&f_U({\Phi'}; C^{(m)}|\hat{\Pi})\label{eq:flux_obj}\\
    \text{s.t.}\quad&  \Phi_{ij}' > 0\quad\forall i,j\label{eq:flux_nn}\\
    & \Phi'\mathbf{1}_{|M|} = \hat{\Pi}\label{eq:flux_rows}\\
    & \Phi'^{\mathsf{T}}\mathbf{1}_{|M|} = \hat{\Pi}\label{eq:flux_cols}
\end{align}
where the loss
\begin{equation}
    f_U(\Phi'; C^{(m)}|\hat{\Pi}) := -\sum_{ij} C_{ij}^{(m)}\ln \Phi_{ij}'
    \label{eq:mirrordescentu_loss}
\end{equation}
equals the negative loglikelihood of the counts to an additive constant.
Eq. \ref{eq:flux_rows} imposes the constraint that $\hat{U}^{(m)}\hat{\Pi}=\hat{\Pi}$
while Eq. \ref{eq:flux_cols} imposes the constraint that $\hat{U}^{(m)}=\hat{\Phi}^{(m)} \mathsf{D}_{\hat{\Pi}}^{-1}$ is column-stochastic given $\hat{U}^{(m)}\hat{\Pi}=\hat{\Pi}$.

Projected mirror descent \cite{nemirovsky1979problem} generalizes projected gradient descent by measuring step sizes in a Bregman divergence \cite{bregman1967relax} other than Euclidean distance [Fig. \ref{fig:schematic}(g)].
With an appropriate choice of Bregman divergence,
projected mirror descent may converge to the optimum of a constrained convex problem more quickly and stably than Euclidean projected gradient descent \cite{beck2017first,vishnoi2021algorithms}.
As $\Phi'$ are probability distributions,
the generalized Kullback-Leibler divergence ($D_\mathrm{KL}$) \cite{kld}
\begin{equation}
    D_\mathrm{KL}(\Phi'\|\cdot) := \sum_{ij}\Phi_{ij}'\ln\frac{\Phi_{ij}'}{\cdot_{ij}}-\Phi_{ij}'+\cdot_{ij}
    \label{eq:KLD}
\end{equation}
is the natural Bregman divergence with which to minimize the strictly convex objective Eq. \ref{eq:flux_obj}
subject to linear constraints Eqs. \ref{eq:flux_nn}-\ref{eq:flux_cols} using projected mirror descent.
Algorithm \ref{alg:MDU} illustrates this:

\begin{algorithm}[H] \label{alg:MDU}
\caption{\protect\mbox{\texttt{MirrorDescentU}($C,\hat{\Pi},\epsilon,\eta$)}}
\textbf{input} count matrix $C$, stationary-vector estimate $\hat{\Pi}$, tolerance $\epsilon$, learning rate $\eta$\;
$\hat{\Phi}\gets\infty$\;
$\hat{f}_U\gets \infty$\;
$\hat{\Phi}^+\gets C\mathsf{D}_{\mathbf{1}^{\mathsf{T}}_{|M|} C}^{-1}\mathsf{D}_{\hat{\Pi}}$\;
$\hat{\Phi}^+\gets \texttt{SinkhornKnopp}(\hat{\Phi}^+,\hat{\Pi},\epsilon)$\;
$\nabla_{\Phi'} f_U\gets -C\oslash \hat{\Phi}^+$\;
\While{$\max\{|\hat{f}_U-f_U(\hat{\Phi}^+;C|\hat{\Pi})|,\max_{ij}\{|\hat{\Phi}^+-\hat{\Phi}|_{ij}\}\}>\epsilon$}{
    $\hat{\Phi}\gets\hat{\Phi}^+$\;
    $\hat{f}_U\gets f_U(\hat{\Phi}; C|\hat{\Pi})$\;
    $\hat{\Phi}^+\gets \argmin_{\Phi'}\{\eta\braket{\nabla_{\Phi'} f_U,\Phi'}_F+D_\mathrm{KL}(\Phi'\|\hat{\Phi})\}$\;
    $\hat{\Phi}^+\gets \texttt{SinkhornKnopp}(\hat{\Phi}^+,\hat{\Pi},\epsilon)$\;
    $\nabla_{\Phi'} f_U\gets -C\oslash \hat{\Phi}^+$\;
}
\textbf{return} conditional-maximum-likelihood feasible transition-matrix estimate $\hat{U}\gets\hat{\Phi}\mathsf{D}_{\hat{\Pi}}^{-1}$
\end{algorithm}

Here we use the exponentiated-gradient update \cite{beck2017first, vishnoi2021algorithms}
\begin{multline}
    \argmin_{\Phi'}\{\eta\braket{\nabla_{\Phi'} f_U,\Phi'}_F+D_\mathrm{KL}(\Phi'\|\hat{\Phi})\} =\\
    \hat{\Phi}\odot\exp_\circ(-\eta\nabla_{\Phi'} f_U)
    \label{eq:exponentiated_gradient}
\end{multline}
where $\braket{\cdot ,\cdot}_F$ denotes the Frobenius inner product \cite{Frobenius1878} and $\exp_\circ(\cdot)$ elementwise exponentiation.
Positivity of all $\hat{\Phi}$ entries endures as initial $\hat{\Phi}$ entries are all positive.
With projection of the iterate onto the feasible set after each exponentiated-gradient update,
Alg. \ref{alg:MDU} converges \cite{li2025convergence} to the solution of Eqs. \ref{eq:flux_obj}-\ref{eq:flux_cols} [Fig. \ref{fig:schematic}(h)-\ref{fig:schematic}(i)].

As we consider probability distributions and use mirror descent with the Kullback-Leibler divergence \cite{kld},
projection onto the feasible set is tantamount to minimizing $D_\mathrm{KL}(\Phi'\|\hat{\Phi})$ over $\Phi'$ subject to Eqs. \ref{eq:flux_nn}-\ref{eq:flux_cols}.
The Sinkhorn-Knopp algorithm \cite{knopp1967concerning} performs exactly such an ``information projection'' \cite{iproject} onto subsets of matrices defined by positive entries and fixed row and column marginals;
details of the algorithm are presented in Appendix \ref{app:projectors}.
We note that the Sinkhorn-Knopp algorithm has found wide application in computational optimal transport \cite{peyre2019computational}.
Indeed, feasible stationary flux matrices belong to ``transportation polytope'' $T_{\hat{\Pi}\to\hat{\Pi}}$ \cite{cuturi2013sinkhorn} embedded in $\R^{|M|\times|M|}$:
they are transport plans from $\hat{\Pi}$ to itself.

\subsection{Experimental CFTR relaxation}

\begin{figure*}
\includegraphics[width=\textwidth]{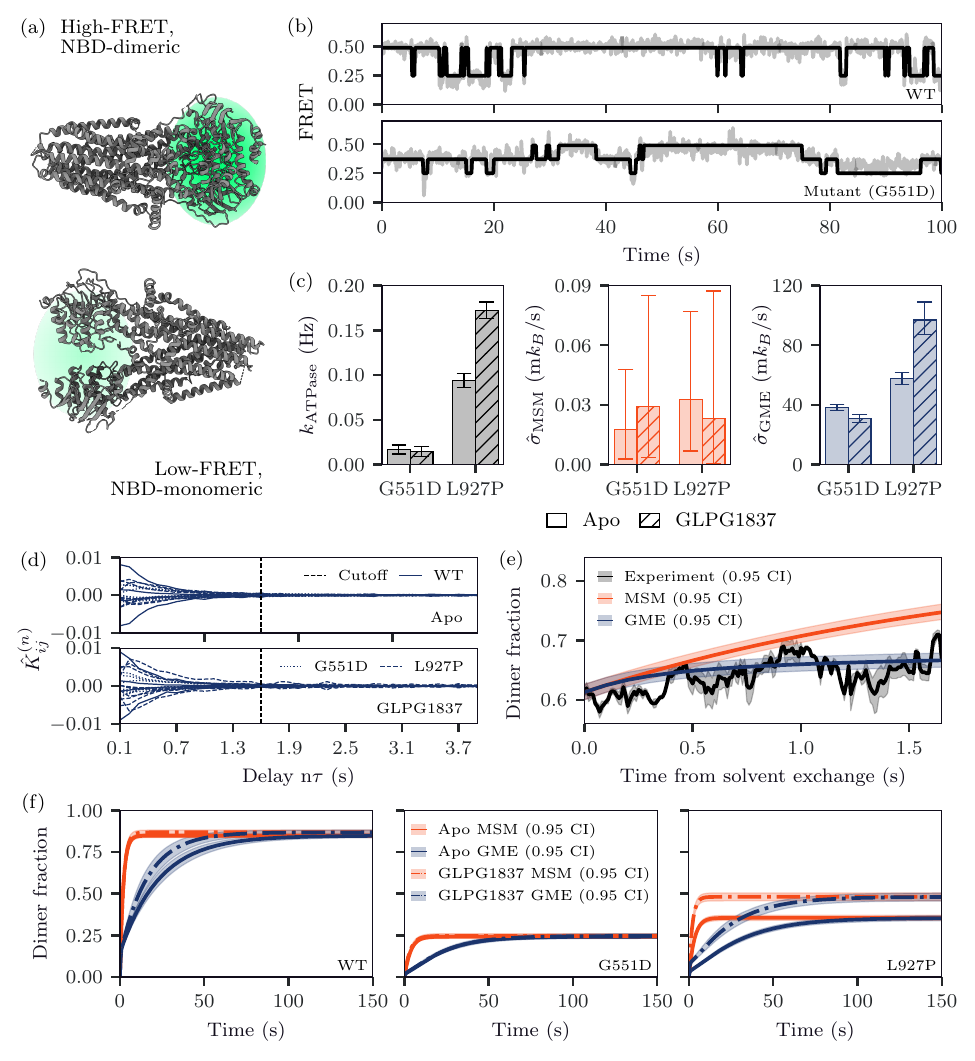}
\caption{CFTR.
    (a) Cystic-fibrosis transmembrane-conductance regulator CFTR [PDB: 6MSM (upper), 8FZQ (lower)] is a chloride channel with ATP-hydrolysis-dependent gating.
    Dimerization of its nucleotide-binding domains (NBDs) increases open probability as well as F\"{o}rster resonance energy transfer (FRET) between NBD labels.
    (b) As visualized experimentally, WT CFTR exhibits two idealized FRET efficiencies at 3 mM ATP;
    cystic-fibrosis-causing mutants (G551D and L927P) exhibit intermediate third efficiencies.
    (c) Three-time irreversibility estimates $\hat{\sigma}_{\mathrm{GME}}$ from TCL-GME-DT models trained on experimental three-state FRET traces reflect dependence of CFTR ATP-hydrolysis rates on mutations and 10 $\mu$M GLPG1837 drug.
    Two-time irreversibility estimates $\hat{\sigma}_\textrm{MSM}$ from MSMs trained on the same traces are much looser EPR bounds (note different axis scales) and do not reflect ATP hydrolysis rates.
    (d) Memory kernels of NZ-GME-DT models trained on experimental two-state WT and three-state mutant FRET traces are negligible by a cutoff delay of 1.6 s with and without GLPG1837.
    (e) An NZ-GME-DT trained on WT apo FRET traces recovers the dimerization rates upon ATP addition observed in an independent solvent-exchange experiment.
    An MSM trained on the same traces does not.
    (f) NZ-GME-DT models infer dimerization rates of initially fully monomeric CFTR populations, which are experimentally inaccessible.
}
\label{fig:CFTR}
\end{figure*}

We demonstrate Alg. \ref{alg:MDU}'s utility in extracting kinetic parameters from protein experiments using NZ-GME-DT models of domain dimerization in human cystic-fibrosis transmembrane-conductance regulator (CFTR).
CFTR is an anion channel regulating salt and water balance across epithelia \cite{csanady2019structure};
its congenital malfunction results in cystic fibrosis, a fatal and incurable disease \cite{cutting2015cystic}.
Upon phosphorylation by protein kinase A, CFTR exhibits nonequilibrium gating mediated by its two nucleotide-binding domains (NBDs).
The NBDs dimerize with ATP binding \cite{zhang2018molecular}, increasing conductance, and separate by approximately 20 \AA~with several cycles of ATP hydrolysis \cite{levring2023CFTR}.
F\"{o}rster-resonance-energy-transfer (FRET) \cite{foerster1948zwischenmolekulare} probes conjugated to the NBDs enable experimental imaging of dimerization in single molecules [Fig. \ref{fig:CFTR}(a)] \cite{levring2023CFTR}.
While wild-type (WT) phosphorylated CFTR exhibits only NBD-monomeric and NBD-dimeric states with FRET efficiencies around 0.25 and 0.49 respectively,
disease-causing mutants (G551D and L927P) rarely dimerize fully and exhibit third states with intermediate efficiencies [Fig. \ref{fig:CFTR}(b)] \cite{levring2023CFTR}.
ATP hydrolysis rates likewise differ between variants and in response to the drug GLPG1837 [Fig. \ref{fig:CFTR}(c)] \cite{vanderplas2018drug}.
Fully monomeric CFTR populations are experimentally inaccessible at physiological ATP \cite{levring2023CFTR},
obscuring CFTR's relaxation to nonequilibrium steady state.

The macrostates available for coarse-grained CFTR modeling are those distinguishable in single-molecule NBD FRET traces,
which were imaged in microfluidic chambers at 3 mM ATP with and without 10 $\mu$M GLPG1837 for both WT and mutant variants \cite{levring2023CFTR}.
FRET efficiencies recorded before photobleaching were previously idealized in SPARTAN \cite{juette2016single}.
Estimating stationary vectors and assembling transition-count matrices from idealized traces of each condition,
we parameterize conditional-maximum-likelihood NZ-GME-DT models with two (WT) or three (mutant) states and a timestep of 0.1 s in both drug-free (apo) and GLPG1837 conditions using Alg. \ref{alg:MDU} and the recursion in Eq. \ref{eq:recursion}.
Metadata, hyperparameters, and preprocessing details for all examples are given in Appendix \ref{app:estimation}.
We emphasize that, as CFTR conformational dynamics are out of equilibrium, Alg. \ref{alg:MDU} is essential for physically feasible conditional-maximum-likelihood modeling of the three-state mutants.
Two-state transition matrices are always detailed balance,
so we could in principle use the reversible estimator of Refs. \cite{prinz2011markov,trendelkampschroer2015estimation} for the WT.
Conditional-maximum-likelihood memory kernels estimated using Alg. \ref{alg:MDU} and Eq. \ref{eq:recursion} approximately vanish by a delay of 1.6 s in all six models, so we set all memory kernels to zero past this cutoff [Fig. \ref{fig:CFTR}(d)].

Remarkably, the NZ-GME-DT model trained on WT apo traces recovers the relaxation rate observed in an independent experiment performed under different imaging conditions [Fig. \ref{fig:CFTR}(e)].
Phosphorylated WT apo CFTR molecules in initially ATP-free solvent relax to a new stationary distribution with a higher NBD-dimerized fraction upon rapid titration of ATP to 3 mM \cite{levring2023CFTR}.
We define solvent exchange as occurring when 95 percent of ATP-free solvent has been replaced by 3 mM ATP solvent;
while the NZ-GME-DT captures the true slow relaxation to stationarity after this time,
an MSM with the same two macrostates and 0.1 s timestep infers much faster dynamics.
As fully monomeric CFTR populations are experimentally inaccessible,
we simulate relaxations of such populations to stationarity [Appendix \ref{app:estimation}, Fig. \ref{fig:CFTR}(f)].
Both MSM and GME simulations are inexpensive, but MSMs infer faster relaxation rates that are apparently overestimates in all six conditions.
Experimental confidence intervals are from standard errors over three experiments; model confidence intervals are from bootstrapping (Appendix \ref{app:estimation}).
All MSMs shown in Fig. \ref{fig:CFTR} use conditional-maximum-likelihood transition matrices parameterized to have observed stationary vectors with Alg. \ref{alg:MDU}.
We note that these MSMs enjoy a large timestep and regularizing constraints that were impossible to impose for nonreversible models before the introduction of Alg. \ref{alg:MDU},
but even then fail to extract slow dynamics from nonequilibrium experiments.

\section{PARAMETERIZING TIME-CONVOLUTIONLESS GENERALIZED MASTER EQUATIONS IN DISCRETE TIME}
\subsection{Parameterizing nonreversible TCL propagators}
With a conditional-maximum-likelihood scheme for estimating $U^{(n)}$ and thus $K^{(n)}$ in hand,
let us construct a similar scheme for general $G^{(n)}$.
Letting $\tilde{U}^{(n)}$ denote the na\"ive column-normalization of $C^{(n)}$,
Eq. \ref{eq:G-from-inversion} suggests the estimator \cite{sayer2023compact,dominic2023building}
\begin{equation}
    \tilde{G}^{(n)} = \tilde{U}^{(n)}(\tilde{U}^{(n-1)})^{-1}
    \label{eq:G-hat-from-inversion}
\end{equation}
that indeed converges to $G^{(n)}$ provided $U^{(n-1)}$ is invertible.
However, $G^{(n)}$ may remain defined when $U^{(n-1)}$ is singular and, indeed,
subasymptotic estimates $\tilde{U}^{(n-1)}$ of $U^{(n-1)}$ might not be invertible for full-rank ground-truth $U^{(n-1)}$.
Even when $\tilde{U}^{(n-1)}$ is invertible,
$\tilde{G}^{(n)}$ might have negative elements and fail to model the process of interest as a valid inhomogeneous Markov chain.
Finally, unless $\tilde{U}^{(n)}$ and $\tilde{U}^{(n-1)}$ happen to share a stationary vector,
subasymptotic $\tilde{G}^{(n)}$ will not share a stationary vector with $\tilde{U}^{(n)}$ and $\tilde{U}^{(n-1)}$ as required for physical feasibility by Eq. \ref{eq:G-stationary-vec}.

To extract interpretable kinetic parameters from data through a TCL-GME-DT model,
we seek an estimator of $G^{(n)}$ that satisfies physical constraints,
models the generating process as a valid inhomogeneous Markov chain,
and conditionally maximizes likelihood.
Eq. \ref{eq:tcl-gme-dt-one-step-propagator-product} implies
\begin{equation}
    U^{(n)} = \prod_{m=0}^{n-1}G^{(n-m)}
\end{equation}
whereby we may form the global negative loglikelihood
\begin{multline}
    f_\mathrm{NC}(\{G'^{(n)}\}_{n=1}^p;\{C^{(n)}\}_{n=1}^p)\\
    :=-\sum_{n=1}^p\sum_{ij}C_{ij}^{(n)}\ln\(\prod_{m=0}^{n-1} G'^{(n-m)}\)_{ij}\\
    =-\sum_{n=1}^p\sum_{ij}C_{ij}^{(n)}\ln U'^{(n)}_{ij}
    \label{eq:nonconvex_obj}
\end{multline}
parameterized by count matrices $\{C^{(n)}\}_{n=1}^p$ at lagtimes up to $p\tau$.
Unfortunately, Eq. \ref{eq:nonconvex_obj} is not amenable to convex minimization over $\{G'^{(n)}\}_{n=1}^p$ (Appendix \ref{app:nonconvex}).

Maximum-likelihood estimation of $\{G^{(n)}\}_{n=1}^p$ becomes tractable upon further conditioning.
To see this, observe that
\begin{multline}
    f_\mathrm{NC}(\{G'^{(n)}\}_{n=1}^p;\{C^{(n)}\}_{n=1}^p)\\
    :=-\sum_{n=1}^p\sum_{ij}C_{ij}^{(n)}\ln\(\prod_{m=0}^{n-1} G'^{(n-m)}\)_{ij}\\
    =-\sum_{n=1}^p\sum_{ij}C_{ij}^{(n)}\ln (G'^{(n)}U'^{(n-1)})_{ij}
\end{multline}
so if we fix propagator estimates $\hat{G}^{(m)}$ at lags $m<n$ and define the fixed previous-lag transition-matrix estimate
\begin{equation}
    \hat{U}^{(n-1)} := \(\prod_{m=1}^{n-1}\hat{G}^{(n-m)}\)
\end{equation}
we may form the simpler loss
\begin{multline}
    f_G(\Theta';C^{(n)}|\hat{\Pi},\hat{U}^{(n-1)})\\
    := -\sum_{ij}C^{(n)}_{ij}\ln(\Theta' \mathsf{D}_{\hat{\Pi}}^{-1}\hat{U}^{(n-1)}\mathsf{D}_{\hat{\Pi}})_{ij}\\
    = -\sum_{ij}C_{ij}^{(n)}\ln(G'\hat{U}^{(n-1)})_{ij} - \sum_{ij}C^{(n)}_{ij}\ln\hat{\Pi}_j
    \label{eq:convex_obj}
\end{multline}
that is to a constant the negative loglikelihood of the counts conditioned on $\hat{\Pi}$ and $\hat{U}^{(n-1)}$.
As a composition of a strictly convex function and linear transformations,
Eq. \ref{eq:convex_obj} is convex with respect to $\Theta' := G'\mathsf{D}_{\hat{\Pi}}$ and,
indeed, strictly convex whenever $\hat{U}^{(n-1)}$ is invertible.

With this simpler loss, we exploit the unique mapping of local flux $\Theta'$ to TCL propagator $G'$ and solve
\begin{align}
    \min_{\Theta'}\quad& f_G(\Theta';C^{(n)}|\hat{\Pi},\hat{U}^{(n-1)})\label{eq:convex_min}\\
    \text{s.t.}\quad& \Theta'_{ij} > 0\quad\forall i,j\label{eq:convex_orthant}\\
    & \Theta'\mathbf{1}_{|M|} = \hat{\Pi}\label{eq:convex_stationary}\\
    & \Theta'^{\mathsf{T}}\mathbf{1}_{|M|} = \hat{\Pi}\label{eq:convex_stochastic}
\end{align}
at increasing $n$, defining $\hat{U}^{(0)}:=I$ and fixing $\hat{G}^{(m)}$ at lags $m<n$.
We need not directly invert $\hat{U}^{(n-1)}$.
As Algorithm~\ref{alg:MDG} describes, Eqs. \ref{eq:convex_min}-\ref{eq:convex_stochastic} are also soluble by exponentiated-gradient mirror descent (Eq. \ref{eq:exponentiated_gradient}) with information projections onto the feasible set:

\begin{algorithm}[H] \label{alg:MDG}
\caption{\protect\mbox{\texttt{MirrorDescentG}($C,\hat{G},\hat{\Phi},\epsilon,\eta$)}}
\textbf{input} count matrix $C$, initial estimate $\hat{G}$, feasible previous-lag flux estimate $\hat{\Phi}$, tolerance $\epsilon$, learning rate $\eta$\;
$\hat{\Pi}\gets\hat{\Phi}\mathbf{1}_{|M|}$\;
$\hat{U}\gets\hat{\Phi}\mathsf{D}_{\hat{\Pi}}^{-1}$\;
$\hat{\Theta}\gets\infty$\;
$\hat{f}_G\gets\infty$\;
$\hat{\Theta}^+\gets \texttt{InfoProjectFeasible}(\hat{G}\mathsf{D}_{\hat{\Pi}},\epsilon,*)$\;
$\nabla_{\Theta'} f_G\gets -[C\oslash (\hat{\Theta}^+\mathsf{D}_{\hat{\Pi}}^{-1}\hat{U}\mathsf{D}_{\hat{\Pi}})]\hat{\Phi}^{\mathsf{T}} \mathsf{D}_{\hat{\Pi}}^{-1}$\;
\While{$\max\{|\hat{f}_G-f_G(\hat{\Theta}^+;C|\hat{\Pi},\hat{U})|$,
$\mathbbm{1}(\exists\hat{U}^{-1})\max_{ij}\{|\hat{\Theta}^+_{ij}-\hat{\Theta}_{ij}|\}\}>\epsilon$}{
    $\hat{\Theta}\gets\hat{\Theta}^+$\;    
    $\hat{f}_G\gets f_G(\hat{\Theta};C|\hat{\Pi},\hat{U})$\;
    $\hat{\Theta}^+\gets\argmin_{\Theta'}\{\eta\braket{\nabla_{\Theta'} f_G,\Theta'}_F+D_\mathrm{KL}(\Theta'\|\hat{\Theta})\}$\;
    $\hat{\Theta}^+\gets\texttt{InfoProjectFeasible}(\hat{\Theta}^+,\epsilon,*)$\;
    $\nabla_{\Theta'} f_G\gets -[C\oslash(\hat{\Theta}^+\mathsf{D}_{\hat{\Pi}}^{-1}\hat{U}\mathsf{D}_{\hat{\Pi}})]\hat{\Phi}^{\mathsf{T}} \mathsf{D}_{\hat{\Pi}}^{-1}$\;
}
\textbf{return} conditional-maximum-likelihood feasible TCL-propagator estimate $\hat{G}\gets\hat{\Theta}\mathsf{D}_{\hat{\Pi}}^{-1}$
\end{algorithm}

Here \texttt{InfoProjectFeasible} information-projects iterates $\hat{\Theta}^+$ onto the feasible set.
In nonreversible cases constrained by only Eqs. \ref{eq:convex_orthant}-\ref{eq:convex_stochastic}, the Sinkhorn-Knopp algorithm \cite{knopp1967concerning} again performs the necessary projection.
We initialize $\hat{G}^{(1)}$ as $C^{(1)}\mathsf{D}_{\mathbf{1}^{\mathsf{T}}_{|M|} C^{(1)}}^{-1}$ and $\hat{G}^{(n)}$ as $\hat{G}^{(n-1)}$ at subsequent lags,
exploiting large-$n$ convergence of $G^{(n)}$ to fixed propagator $G^{(\infty)}$.
As a further consequence of convergence to $G^{(\infty)}$,
we identify cutoff lags at which elements $\hat{G}^{(n)}_{ij}$ plateau as in previous discrete-time TCL GME approaches \cite{dominic2023building,sayer2023compact}.
Past these cutoffs, we may use the cutoff $\hat{G}^{(n)}$ as an approximation to $G^{(\infty)}$ [Fig. \ref{fig:schematic}(j)].
Modifying the toy model of Fig. \ref{fig:schematic},
the adversarial example in Appendix \ref{app:adversarial} slowly but eventually converges to ground truth even when $U^{(n-1)}$ is nearly singular and the solution to Eqs. \ref{eq:convex_min}-\ref{eq:convex_stochastic} is nearly nonunique.

\subsection{Experimental CFTR and BMF1 ATP hydrolysis}

\begin{figure*}
\includegraphics[width=\textwidth]{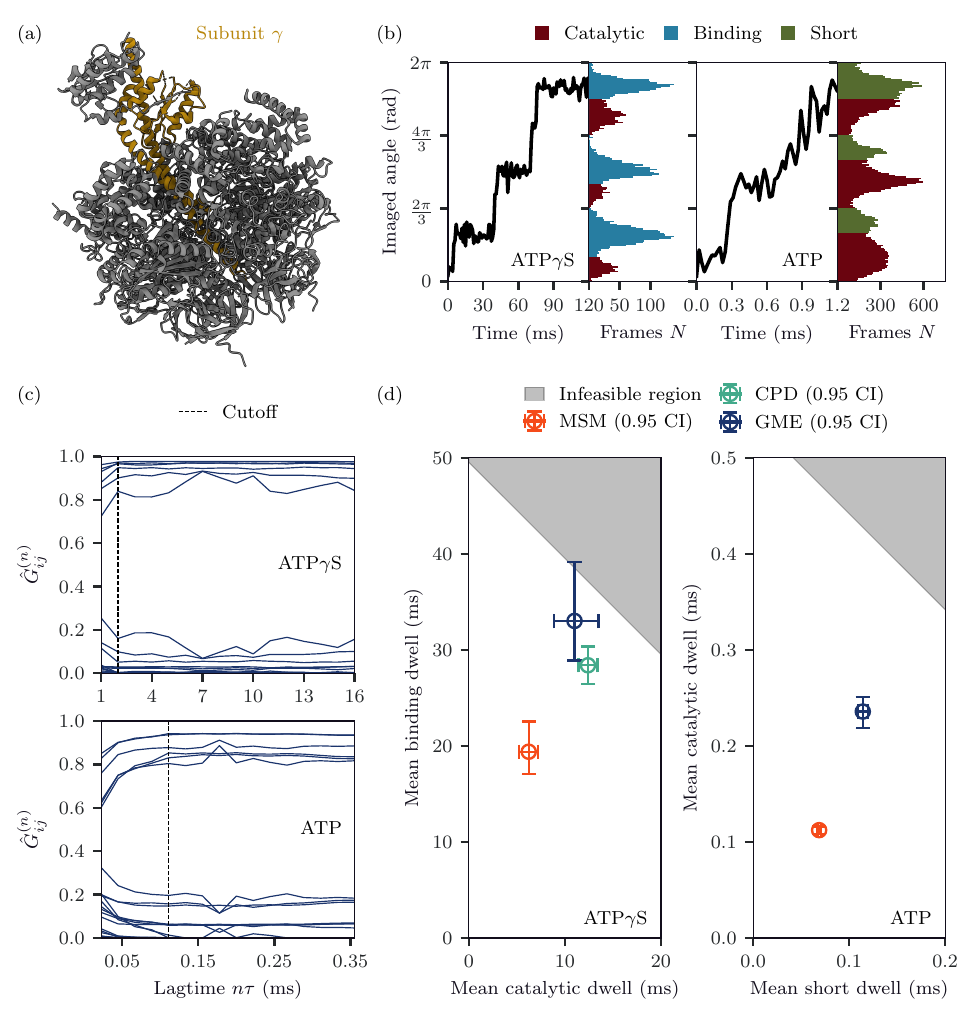}
\caption{BMF1.
    (a) Bovine mitochondrial F$_1$-ATPase BMF1 (PDB: 4YXW) consists of an $\alpha_3\beta_3$ stator surrounding a $\gamma\delta\epsilon$ rotor.
    The stator is immobilized on a coverslip;
    subunit $\gamma$ is labeled with a 40 nm gold probe to visualize rotary hydrolysis of ATP or ATP$\gamma$S.
    (b) Binding dwells represent nucleotide binding events; catalytic dwells occur $4\pi/9$ from binding dwells and represent hydrolysis events. 
    Short dwells occur $\pi/12$ from binding dwells and possibly represent phosphate release.
    Binding and catalytic dwells are easily observed in experimental angle traces and histograms at 1 $\mu$M ATP$\gamma$S.
    Catalytic and short dwells are observable in angle histograms and to a lesser extent traces at physiological 3 mM ATP.
    (c) Nonreversible TCL-GME-DT models are trained on experimental BMF1 counterrotation traces discretized to histogram peaks in 1 $\mu$M ATP$\gamma$S and 3 mM ATP conditions.
    TCL propagators for the ATP$\gamma$S condition have timestep 1 ms and plateau at 2 ms;
    TCL propagators for the ATP condition have timestep 1/45 ms and plateau at 1/9 ms.
    (d) In the ATP$\gamma$S condition, the TCL-GME-DT (GME) and change-point detection (CPD) identify similar mean dwells; a conditional-maximum-likelihood MSM underestimates both catalytic and binding dwells.
    In the ATP condition, the TCL-GME-DT identifies dwells comparable to those visible in the example trace in (b);
    an MSM underestimates both dwells and CPD cannot segment the trace.
    The infeasible region represents dwell estimates too long for the duration and total counterrotations of the trace.
}
\label{fig:BMF1}
\end{figure*}

We may now turn to Alg. \ref{alg:MDG} for further insight into CFTR.
CFTR exhibits mutation- and drug-dependent variation in ATP hydrolysis rates,
with L927P CFTR more responsive than G551D CFTR to potentiation by the drug GLPG1837 \cite{vanderplas2018drug}.
Though ATP hydrolysis is coupled to NBD dimerization \cite{jih2013vx},
NBD dimerization persists through multiple cycles of ATP hydrolysis \cite{levring2023CFTR}:
NBD state is not itself obviously cyclic.
It is thus unclear whether irreversibility is detectable in stationary FRET traces that track NBD state.
To address this question, we parameterize maximum-likelihood MSMs using Alg. \ref{alg:MDU} with timestep $\tau = 1.6$ s and TCL-GME-DT models using Alg. \ref{alg:MDG} with timestep $\tau = 0.8$ s.
All models are trained on the steady-state FRET traces used to parameterize NZ-GME-DT memory kernels in Fig. \ref{fig:CFTR}(d); timesteps are chosen so that MSM-based second-order and GME-based third-order irreversibility estimators \cite{roldan2012entropy} both saturate the 1.6 s of memory inferred from NZ-GME-DT memory kernels to validly lower-bound EPRs \cite{schwarz2025consistent}.
Second-order estimator
\begin{equation}
    \hat{\sigma}_\mathrm{MSM} := \frac{1}{\tau}\sum_{ij}\hat{U}_{ij}\hat{\Pi}_j\ln\frac{\hat{U}_{ij}\hat{\Pi}_j}{\hat{U}_{ji}\hat{\Pi}_{i}},
\end{equation}
indeed detects little to no irreversibility in either mutant with or without GLPG1837 [central panel of Fig. \ref{fig:CFTR}(c)].
However, third-order irreversibility estimator $\hat{\sigma}_\mathrm{GME}$ [Eq. \ref{eq:epr_estimator} and right panel of Fig. \ref{fig:CFTR}(c)] supplies an EPR lower bound orders of magnitude tighter than $\hat{\sigma}_\mathrm{MSM}$.
Interestingly, $\hat{\sigma}_\mathrm{GME}$ even recapitulates the irreversibility rank order across mutant-drug combinations implied by experimentally measured ATP hydrolysis rates [left panel of Fig. \ref{fig:CFTR}(c)].
We note that $\hat{\sigma}_\mathrm{GME}$ is $30\pm 3$ m$k_B$/s for WT apo and $20\pm 3$ m$k_B$/s for WT with 10 $\mu$M GLPG1837;
these values are not readily compared to those of three-state mutants as they are computed without access to leading-order currents.
Nevertheless, nonzero irreversibility estimates using only MSMs are impossible for the two-state CFTR WT as all two-state transition matrices have detailed balance.
These results demonstrate the advantages of TCL-GME-DT models in extracting irreversibilities from experimental data.

Alg. \ref{alg:MDG} is especially useful when data are scarce and noisy,
as the further experimental example of bovine mitochondrial F$_1$-ATPase (BMF1) illustrates \cite{kobayashi2020BMF1}.
Like other F$_1$-ATPases, BMF1 consists of a $\gamma\delta\epsilon$ rotor in an $\alpha_3\beta_3$ stator [Fig. \ref{fig:BMF1}(a)];
it functions primarily to phosphorylate ADP through chemiosmosis-driven rotation but also counterrotates upon hydrolyzing ATP \cite{kuehlbrandt2019structure}.
BMF1 is a model system for other molecular motors \cite{karplus2004biomolecular}.
As a representative mammalian F$_1$-ATPase \cite{suzuki2014chemomechanical}, BMF1 exhibits binding, catalytic, and short dwells that are interpretable as pauses in macrostates \cite{brandner2025dynamics}. 
Binding dwells occur at ATP binding angles and correspond to nucleotide-binding events,
catalytic dwells occur about $4\pi/9$ counterclockwise of binding dwells viewed from the rotor side and correspond to hydrolysis events,
and short dwells occur about $\pi/12$ counterclockwise of binding dwells and may correspond to phosphate-release events \cite{bason2015release,kobayashi2020BMF1}.
Each dwell occurs three times per counterrotation.
During experimental recordings of a single BMF1 complex immobilized on glass with subunit $\gamma$ labeled by a 40 nm gold probe, dwells are clearly visible in angle histograms [Fig. \ref{fig:BMF1}(b)].
Catalytic and binding dwells are also visible in angle traces of BMF1 driven by 1 $\mu$M slowly hydrolyzed ATP-analogue ATP$\gamma$S;
these dwells are estimated using change-point detection in Ref. \cite{kobayashi2020BMF1}.
However, though catalytic and short dwells are apparent in angle histograms of BMF1 driven by physiological 3 mM ATP (binding dwells are too short to be observed in these conditions),
these events are only occasionally visible in noisy angle traces, which the change-point-detection algorithm of Ref. \cite{kobayashi2020BMF1} fails to segment.
Estimating physiological BMF1 dwell times thus requires another approach.

We use TCL-GME-DT models to extract these dwell times.
For both 1 $\mu$M ATP$\gamma$S and 3 mM ATP models, we crudely coarse-grain by assigning imaged angles to six histogram peaks.
Three such macrostates correspond to catalytic dwell in each condition, three to binding dwell in 1 $\mu$M ATP$\gamma$S, and three to short dwell in 3 mM ATP.
The 1 $\mu$M ATP$\gamma$S trace and model both use a 1 ms timestep while the 3 mM ATP trace and model both use a $1 / 45$ ms timestep;
propagators estimated for these conditions using Alg. \ref{alg:MDG} respectively plateau at 2 ms and $1/9$ ms [Fig. \ref{fig:BMF1}(c)].
We estimate mean dwell times directly from propagators [Appendix \ref{app:estimation}, Fig. \ref{fig:BMF1}(d)].
Infeasible regions shaded in Fig. \ref{fig:BMF1}(d) represent mean-dwell-time estimates too long for consistency with the total observed counterrotations and trace durations;
dwell-time estimates of step-like processes should be close to the infeasible region.
Validating this approach, we find mean 1 $\mu$M ATP$\gamma$S catalytic and binding dwells of 11.0 ms (0.95 CI: [8.8 ms, 13.5 ms]) and 33.0 ms (0.95 CI: [28.9 ms, 39.2 ms]) respectively.
These values are close to the $12.4\pm 1.0$ ms and $28.4\pm 2.0$ ms estimated by Ref. \cite{kobayashi2020BMF1} using change-point detection.
A conditional-maximum-likelihood MSM trained using Alg. \ref{alg:MDU} on the same 1 $\mu$M ATP$\gamma$S trace with the same coarse-graining, 1 ms timestep, and stationary-vector constraint estimates much shorter catalytic and binding dwells of 6.2 ms (0.95 CI: [5.2 ms, 7.2 ms]) and 19.4 ms (0.95 CI: [17.0 ms, 22.6 ms]).
We emphasize that the estimates of Ref. \cite{kobayashi2020BMF1} represent averages over 45 independent traces where our estimates use only a single publicly available 6000 ms trace.
TCL-GME-DT models are evidently highly data-efficient in extracting dwell times.
Finally, the TCL-GME-DT model of BMF1 in 3 mM ATP finds mean short and catalytic dwells of 114 $\mu$s (0.95 CI: [109 $\mu$s, 120 $\mu$s]) and 236 $\mu$s (0.95 CI: [218 $\mu$s, 251 $\mu$s]);
a conditional-maximum-likelihood MSM trained using Alg. \ref{alg:MDU} on the same trace with the same coarse-graining, timestep, and stationary-vector constraint finds mean short and catalytic dwells of 69 $\mu$s (0.95 CI: [67 $\mu$s, 70 $\mu$s]) and 112 $\mu$s (0.95 CI: [108 $\mu$s, 116 $\mu$s]).
We lack reference change-point-detection estimates for these dwell times \cite{kobayashi2020BMF1},
but both TCL-GME-DT and MSM estimates are far from the infeasible region in line with the less step-like dynamics of BMF1 at physiological ATP.
However, while TCL-GME-DT estimates reflect the dwells visible in the 3 mM ATP BMF1 angle trace [Fig. \ref{fig:BMF1}(b)], MSM estimates are shorter.
These findings demonstrate the utility of TCL-GME-DT models in extracting kinetic parameters from nonequilibrium experiments using only scarce and noisy data.

\subsection{Parameterizing reversible TCL propagators}
Finally, extending the TCL-GME-DT approach of Alg. \ref{alg:MDG} to reversible dynamics, we estimate propagators $G^{(n)}$ at increasing lags $n$ under the further constraints of equilibrium.
In addition to Eqs. \ref{eq:convex_orthant}-\ref{eq:convex_stochastic}, propagator estimates must in the reversible case satisfy
\begin{align}
    \Theta' &= \Theta'^{\mathsf{T}}\label{eq:convex_db}\\
    G'^{(1)} = G'^{(2)}\iff \Theta'^{(1)} &= \Theta'^{(2)}\label{eq:vanishing_3pt}\\
    \hat{U}^{(n-1)}G' &= G'\hat{U}^{(n-1)}\label{eq:convex_commutation}
\end{align}
where Eq. \ref{eq:convex_db} imposes detailed-balance $G'^{(n)}$,
Eq. \ref{eq:vanishing_3pt} imposes vanishing third-order EPR (Eq. \ref{eq:epr_estimator}),
and Eq. \ref{eq:convex_commutation} imposes detailed-balance $U'^{(n)}$ (Eq. \ref{eq:reversibility-commutation}).
Fortunately, Eqs. \ref{eq:convex_orthant}-\ref{eq:convex_stochastic} and \ref{eq:convex_db} are imposed by the symmetry-inducing Knight-Ruiz-U\c{c}ar variant \cite{knight2014symmetry} of the Sinkhorn-Knopp algorithm (Appendix \ref{app:projectors}).
Eq. \ref{eq:vanishing_3pt} is imposed by letting $\hat{G}^{(2)}:=\hat{G}^{(1)}$.
The constraints in Eqs. \ref{eq:convex_orthant}-\ref{eq:convex_stochastic}, \ref{eq:convex_db}, and \ref{eq:vanishing_3pt} are all affine with respect to $\Theta'$;
as cyclic projections onto affine subsets converge to the projection onto the intersection of those subsets in any Bregman divergence \cite{bregman1967relax,benamou2015iterative},
to project onto the set of local fluxes corresponding to feasible equilibrium TCL-GME-DT propagators,
we need only express Eq. \ref{eq:convex_commutation} as an affine constraint on $\Theta'$ and construct a projector onto the subset of $\Theta'$ satisfying that constraint.
This is straightforward. As $\hat{U}^{(n-1)}$ is assumed detailed-balance with respect to $\hat{\Pi}$,
\begin{align}
    \hat{U}^{(n-1)}G' &= G'\hat{U}^{(n-1)}\notag\\
    \iff \hat{U}^{(n-1)}G'\mathsf{D}_{\hat{\Pi}} &= G'\hat{U}^{(n-1)}\mathsf{D}_{\hat{\Pi}}\notag\\
    &= G'\mathsf{D}_{\hat{\Pi}}\hat{U}^{(n-1)\mathsf{T}}\notag\\
    \iff \hat{U}^{(n-1)}\Theta' &= \Theta'\hat{U}^{(n-1)\mathsf{T}}
    \label{eq:affine_commutation}
\end{align}
whereby we have expressed Eq. \ref{eq:convex_commutation} as an affine constraint on local flux $\Theta'$.

To find a projector onto the set satisfying Eq. \ref{eq:convex_commutation},
we now express Eq. \ref{eq:affine_commutation} as
\begin{equation}
    \mathscr{C}\text{vec}(\Theta') :=(I\otimes \hat{U}^{(n-1)}-\hat{U}^{(n-1)}\otimes I)\text{vec}(\Theta')=0
\end{equation}
where $\otimes$ denotes tensor multiplication and $\text{vec}(\cdot)$ column-major flattening.
Information projection onto the set satisfying Eq. \ref{eq:convex_commutation} is then equivalent to solving the problem
\begin{align}
    \min_{\theta'}\quad &D_\mathrm{KL}\bigl(\theta'\| \text{vec}(\hat{\Theta})\bigr)\\
    \text{s.t.}\quad &\mathscr{C}\theta'=0
\end{align}
for which we may construct the Lagrangian
\begin{equation}
    \mathscr{L}(\theta',\lambda) := \sum_i\theta'_i\ln\frac{\theta'_i}{\text{vec}(\hat{\Theta})_i} - \theta'_i + \text{vec}(\hat{\Theta})_i + \lambda^{\mathsf{T}}\mathscr{C}\theta'
    \label{eq:lagrangian}
\end{equation}
minimized at
\begin{equation}
    \theta^*(\lambda) = \text{vec}(\hat{\Theta})\odot\exp_\circ(-\mathscr{C}^{\mathsf{T}}\lambda)
    \label{eq:theta_optimum}
\end{equation}
with respect to $\theta'$.
Substituting $\theta^*(\lambda)$ into Eq. \ref{eq:lagrangian}, we obtain the objective
\begin{align}
    \mathscr{L}(\lambda) &:= 1-\sum_i\text{vec}(\hat{\Theta})_i\exp[-(\mathscr{C}^{\mathsf{T}}\lambda)_i]\notag\\
    &= 1 - \mathbf{1}^{\mathsf{T}}_{|M|^2}\theta^*(\lambda)
    \label{eq:dual_objective}
\end{align}
with respect to dual variable $\lambda$.
As $\mathscr{L}(\lambda)$ has simple closed-form gradient $\mathscr{C}\theta^*(\lambda)$ and Hessian $-\mathscr{C}\mathsf{D}_{\theta^*(\lambda)}\mathscr{C}^{\mathsf{T}}$ with respect to $\lambda$,
we maximize it by iterating $\lambda$ with the Levenberg-Marquardt algorithm (Appendix \ref{app:projectors}) \cite{levenberg1944method,marquardt1963algorithm}.
Note that as all elements of $\theta^*(\lambda)$ are nonnegative $-\mathscr{C}\mathsf{D}_{\theta^*(\lambda)}\mathscr{C}^{\mathsf{T}}$ is negative semidefinite:
$\mathscr{L}(\lambda)$ is concave and its only optima are global.
The cyclic projector onto the full reversible constraints Eqs. \ref{eq:convex_orthant}-\ref{eq:convex_stochastic} and Eqs. \ref{eq:convex_db}-\ref{eq:convex_commutation} is then implemented by Algorithm~\ref{alg:IPRG}:

\begin{algorithm}[H] \label{alg:IPRG}
\caption{\protect\mbox{\texttt{InfoProjectReversibleG}($\hat{\Theta},\hat{\Phi},\epsilon$)}}
\textbf{input} infeasible local-flux estimate $\hat{\Theta}$, feasible previous-lag flux estimate $\hat{\Phi}$, tolerance $\epsilon$\;
$\hat{\Pi}\gets\hat{\Phi}\mathbf{1}_{|M|}$\;
$\hat{U}\gets\hat{\Phi}\mathsf{D}_{\hat{\Pi}}^{-1}$\;
$\hat{\Theta}\gets\texttt{KnightRuizU\c{c}ar}(\hat{\Theta},\hat{\Pi},\epsilon)$\;
\While{$\max\{|(\hat{U}\hat{\Theta}-\hat{\Theta}\hat{U}^{\mathsf{T}})_{ij}|\}>\epsilon$}{
    $\hat{\theta}\gets\argmin_{\theta'}\{D_\mathrm{KL}\bigl(\theta'\|\text{vec}(\hat{\Theta})\bigr):[\hat{U},\text{mat}(\theta')\mathsf{D}_{\hat{\Pi}}^{-1}]=0\}$\;
    $\hat{\Theta}\gets\text{mat}(\hat{\theta})$\;
    $\hat{\Theta}\gets\texttt{KnightRuizU\c{c}ar}(\hat{\Theta},\hat{\Pi},\epsilon)$\;
}
\textbf{return} feasible local-flux estimate $\hat{\Theta}$
\end{algorithm}
Here we compute $\hat{\theta}$ using the Levenberg-Marquardt algorithm for $\text{mat}(\cdot)$ the inverse of column-major vectorization $\text{vec}(\cdot)$.
Given counts $\{C^{(n)}\}_{n=1}^p$, we may now use Alg. \ref{alg:MDG} to estimate fully feasible conditional-maximum-likelihood estimates of TCL propagators $\{G^{(n)}\}_{n=1}^p$ in equilibrium cases by using \texttt{InfoProjectReversibleG} as \texttt{InfoProjectFeasible} instead of the Sinkhorn-Knopp algorithm \cite{knopp1967concerning} (Alg. \ref{alg:IPRG}).
Note that this procedure should, and indeed does, return the same maximum-likelihood estimate as the iterative scheme in Refs. \cite{prinz2011markov,trendelkampschroer2015estimation} for reversible $\hat{G}^{(1)} = \hat{U}^{(1)}$.

\subsection{HP35 folding benchmark}

\begin{figure*}
\includegraphics[width=\textwidth]{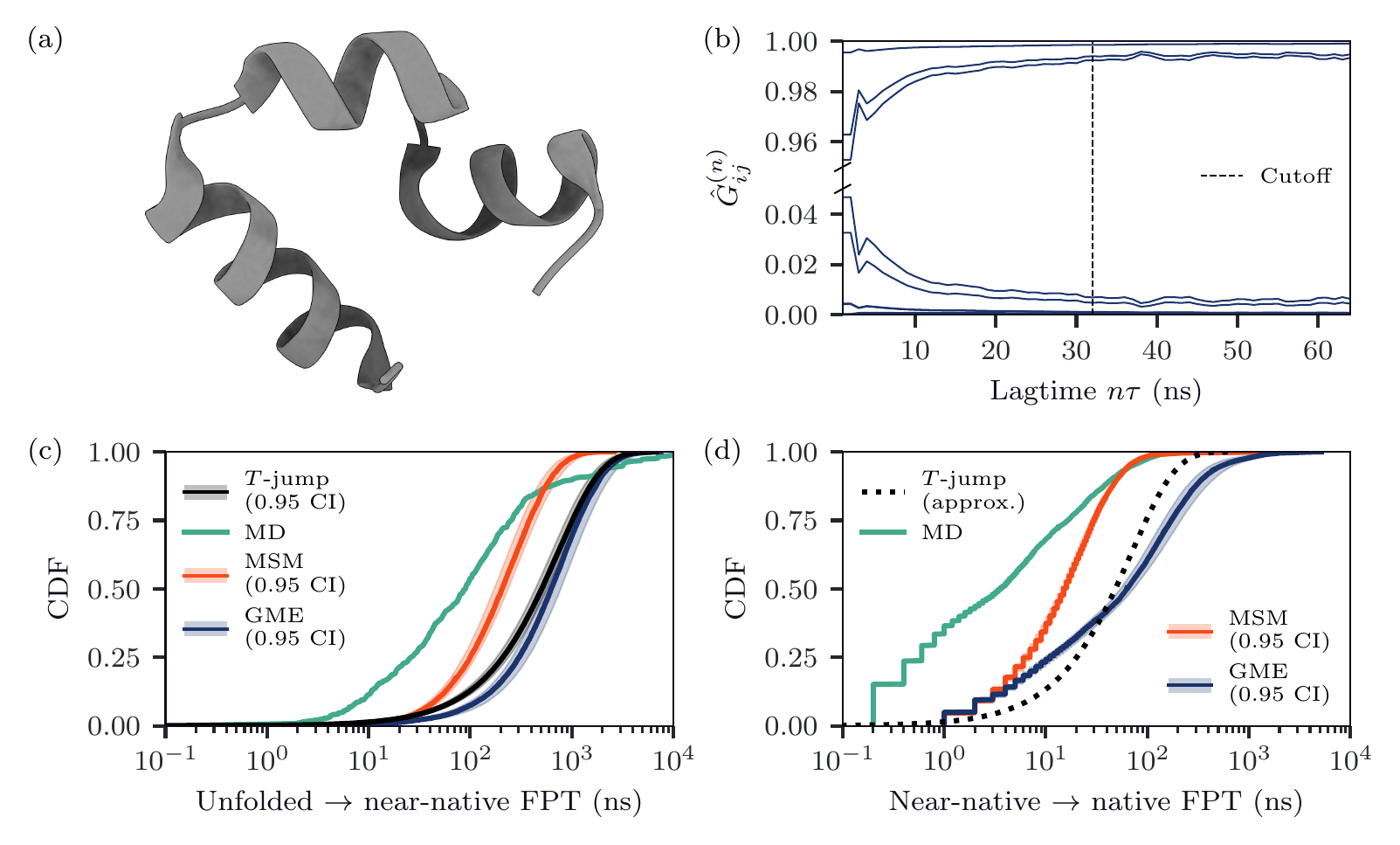}
\caption{HP35.
    (a) The Lys24Nle/Asn27His/Lys29Nle variant of villin headpiece HP35 (PDB: 2F4K) exhibits $\mu$s-scale folding.
    (b) An all-atom simulation of folding and unfolding at 360 K is coarse-grained into fully unfolded, near-native, and native macrostates.
    Reversible TCL-GME-DT propagators trained on the coarse-grained trajectory with timestep 1 ns plateau by a cutoff of 32 ns.
    (c) Temperature-jump spectroscopy ($T$-jump) of variant HP35 at 360 K resolves a $730\pm 98$ ns timescale for transitions from the unfolded basin to the folded basin.
    Though first-passage times (FPTs) in the simulated trajectory (MD) have not converged to a smooth distribution,
    the TCL-GME-DT (GME) approximately recovers the mean unfolded to near-native FPT.
    A conditional-maximum-likelihood MSM at the same temporal resolution underestimates mean FPT.
    (d) $T$-jump resolves an approximately 70 ns timescale for relaxations in the folded basin.
    Though near-native to native FPTs are again unconverged in MD,
    the GME approximately recovers the mean near-native to native FPT while the MSM underestimates it.
}
\label{fig:HP35}
\end{figure*}

At a fixed timestep and experimentally motivated coarse-graining, reversible TCL-GME-DT models accurately capture timescales that MSMs cannot.
We illustrate this through the benchmark \cite{eaton2021modern,nagel2023toward} folding of chicken villin headpiece.
Villin promotes actin bundling in intestinal epithelia \cite{bretscher1979villin}; its 35-residue headpiece (HP35) retains actin binding and is among the fastest-folding naturally occurring protein subdomains \cite{mcknight1997nmr}.
The Lys24Nle/Asn27His/Lys29Nle variant [Fig. \ref{fig:HP35}(a)], in which two lysine residues are substituted with norleucine, is hereafter referred to as HP35 and has a folding pathway characterized in temperature-jump ($T$-jump) experiments.
Upon 10 ns of laser heating, an HP35 population relaxes to a new equilibrium through folding and unfolding transitions that are inferred spectroscopically from conformation-dependent quenching of tryptophan fluorescence by the His27 substitution \cite{kubelka2006sub}.
At the folding midpoint of 360 K, such $T$-jump experiments identify a slow $730\pm 98$ ns timescale corresponding to transitions from the unfolded basin to the folded basin and a fast approximately $70$ ns timescale corresponding to relaxation of near-native molecules in the folded basin.
These transitions were simulated in a 305.208 $\mu$s all-atom MD trajectory with explicit solvent \cite{piana2012HP35};
this and other long Anton \cite{anton} simulations of HP35 variants \cite{lindorfflarsen2011how} accurately approximate experimental kinetics \cite{beauchamp2012simple,banushkina2013hires,henry2013comparing}.
An MSM previously constructed \cite{nagel2023selecting} from this trajectory finds timescales near those estimated from $T$-jump experiments,
but lacks reversibility or stationary-vector constraints and uses 10 ns timesteps and twelve macrostates instead of the experimentally modeled three.

However, a physically feasible TCL-GME-DT trained on the same MD trajectory recovers both experimentally estimated timescales while using a much smaller timestep and more interpretable macrostates.
We first coarse-grain the MD trajectory from \cite{piana2012HP35}, assigning all states described as native by Ref. \cite{nagel2023selecting} to a single native macrostate, all states described as near-native by Ref. \cite{nagel2023selecting} to a single near-native macrostate, and all states described as partly or fully unfolded by Ref. \cite{nagel2023selecting} to a single unfolded macrostate.
These macrostates reflect the three implied by two-stage folding in $T$-jump experiments \cite{kubelka2006sub}.
Using Alg. \ref{alg:MDG} with Alg. \ref{alg:IPRG} for information projections, we estimate conditional-maximum-likelihood TCL-GME-DT propagators with a timestep of 1 ns that fully satisfy Eqs. \ref{eq:convex_orthant}-\ref{eq:convex_stochastic} and \ref{eq:convex_db}-\ref{eq:convex_commutation} to parameterize a valid inhomogeneous Markov chain.
The propagators plateau by a cutoff of 32 ns [Fig. \ref{fig:HP35}(b)].
Initializing $10^4$ trajectories in both unfolded and near-native macrostates, we monitor first-passage times (FPTs) to the near-native and native macrostates, respectively, during evolution by the TCL-GME-DT.
Unfolded to near-native FPTs in the MD trajectory have mean $546.4\pm 129.4$ ns and are not converged to a smooth distribution, but the TCL-GME-DT nevertheless recovers a smooth FPT distribution with mean 859 ns (0.95 CI: [724 ns, 1058 ns]) closely matching the experimental mean of $730\pm 98$ ns [Fig. \ref{fig:HP35}(c)].
The constraints and conditional likelihood maximization used in TCL-GME-DT parameterization evidently regularize kinetic estimates.
Similarly, near-native to native FPTs in the MD trajectory have mean $18.0\pm 3.0$ ns and are not converged to a smooth distribution;
the TCL-GME-DT still recovers a smooth FPT distribution overlapping the experimentally inferred exponential distribution [Fig. \ref{fig:HP35}(d)],
though the TCL-GME-DT distribution has a heavy right tail and mean 151 ns (0.95 CI: [123 ns, 183 ns]) greater than the approximate 70 ns experimental mean.
Using the same 1 ns timestep and three-macrostate coarse-graining, we also estimated a conditional-maximum-likelihood MSM satisfying reversibility and stationarity constraints using Alg. \ref{alg:MDU} (or, equivalently, the estimator of Refs. \cite{prinz2011markov,trendelkampschroer2015estimation}).
The MSM underestimates unfolded to near-native and near-native to native mean FPTs as 281 ns (0.95 CI: [237 ns, 329 ns]) and 24 ns (0.95 CI: [22 ns, 25 ns]), respectively.
These results demonstrate the advantages of our approach relative to MSMs in modeling equilibrium kinetics, especially in the context of experimentally motivated coarse-graining.

\section{OUTLOOK}
We present a principled and general approach to maximum-likelihood parameterization of discrete-time generalized master equations (GME-DTs) conditioned on physical and numerical constraints in and out of equilibrium.
As extensive protein examples illustrate, our approach recapitulates experimentally validated kinetic parameters inaccessible to Markov state models (MSMs) with the same timestep and coarse-graining.
Our results exactly generalize Nakajima-Zwanzig (NZ) GMEs \cite{nakajima1958quantum,zwanzig1961memory} to discrete time by direct application of the Zwanzig-Mori formalism, improving on previous quadrature-based methods \cite{makri2025discrete,cerrillo2014nonmarkovian}.
Through a diagrammatic expansion, we recursively express memory kernels of NZ-GME-DTs in terms of readily estimated macrostate transition matrices.
Introducing two projected-mirror-descent algorithms, we efficiently parameterize both NZ and time-convolutionless (TCL) \cite{tokuyama1975statistical,tokuyama1976statistical} GME-DTs.
Unlike previous methods, our algorithms impose reversibility and stationarity constraints while conditionally maximizing likelihoods of observed transition counts:
these constraints ensure that, unlike na\"ive estimation, our TCL-GME-DT parameterization always forms valid evolution equations.
Compared to na\"ive estimation under realistic sampling, our constrained NZ-GME-DT parameterization empirically improves subasymptotic statistical efficiency while provably preserving or improving the asymptotic-convergence exponent (Appendix \ref{app:benchmarks}).
Our GME-DTs recover relaxation rates, irreversible activity, dwell times, and first-passage times from protein systems with experimentally prescribed coarse-grainings,
extending coarse-grained kinetic modeling to nonequilibrium and experimental contexts that resist MSMs.

Interestingly, the recursion for discrete-time memory kernels in Eq. \ref{eq:recursion} resembles previously discovered expressions of the form
\begin{equation*}
    \mathscr{G}^{(n)}(t) = \frac{\partial \mathscr{P}^{(n)}(t)}{\partial t}-\sum_{m=1}^{n-1}\mathscr{G}^{(m)}(t)\mathscr{P}^{(n-m)}(t)
\end{equation*}
where $\mathscr{G}^{(n)}(t)$ are order-$n$ perturbations of TCL generator $\mathscr{G}(t)$ at time $t$ \cite{tokuyama1975statistical,tokuyama1976statistical}.
Here $\mathscr{G}^{(0)}(t) = 0$ and $\mathscr{P}^{(m)}(t)$ are either integrated bath momenta \cite{gasbarri2018recursive} or time-dependent perturbations of continuous-time transfer operators \cite{gu2020diagrammatic,gu2024diagrammatic};
either interpretation permits a diagrammatic expression similar to ours.
The relationship between these diagrammatic expressions remains to be elucidated,
but we believe that estimating transition matrices is generally less involved than estimating integrated bath momenta, transfer-operator perturbations, or time derivatives thereof.

Algs. \ref{alg:MDU} and \ref{alg:MDG} use the Sinkhorn-Knopp algorithm \cite{knopp1967concerning} to project transition-matrix iterates onto the feasible set,
suggesting connections between entropic optimal transport---a celebrated use of the Sinkhorn-Knopp algorithm \cite{cuturi2013sinkhorn}---and coarse-grained kinetic modeling.
Approaches that interpret flux matrices as transport plans between macrostates may merit further investigation.
Previous works minimize ``kinetic distances'' \cite{mcgibbon2013learning,noe2015kinetic} between microstates sharing a macrostate so as to satisfy Markovianity in models of equilibrium MD simulations;
these approaches may benefit from closed-form iterative schemes similar to ours.
Though we make no attempt to generalize our results to quantum-mechanical systems,
we note that projected mirror descent finds natural applications in optimizing trace-unity density matrices \cite{li2019convergence}.
Further work should establish the complementarity of conditional likelihood maximization schemes like ours to tomographically reconstructing quantum GMEs \cite{pollock2018tomograph}.
Indeed, our results suggest extensions of coarse-grained kinetic modeling beyond nonequilibrium and experimental biomolecular systems,
for which we have demonstrated the broad applicability and utility of conditional-maximum-likelihood discrete-time GMEs.

\section*{ACKNOWLEDGEMENTS}
This study was supported by a National Science Foundation Graduate Research Fellowship (to C.-W.J.L.),
the National Institute of General Medical Sciences Molecular Biophysics Training Program at Stanford University (T32GM136568),
and the U.S. Department of Energy, Office of Science, Office of Basic Energy Sciences, under Award No. DE-SC0022917.
We thank Jesper Levring for CFTR data and referees for insightful comments.

\section*{DATA AVAILABILITY}
Publicly available scripts and data \cite{liu2026gmex} reproduce all presented numerical results.
Protein visualizations use UCSF ChimeraX \cite{meng2023chimerax}.

\appendix
\section{PRELIMINARY IDENTITIES}
\label{app:identities}
Recall that we have defined microstates $\mu$, macrostates $M$, microscopic Markov-chain propagator $L$, microscopic stationary vector $\pi =L\pi$, the coarse-graining $\mathcal{M}:\mu\to M$ such that $A_{ij}:=\mathbbm{1}[\mathcal{M}(\mu_i)=M_j]$, macroscopic stationary vector $\Pi:=A^\mathsf{T}\pi$, and $\Omega:= \mathsf{D}_\pi A\mathsf{D}_\Pi^{-1}$ where $\mathsf{D}_\cdot=\text{diag}(\cdot)$.
Consider the $\pi$- and $\Pi$-weighted inner products respectively denoted $\braket{\cdot,\cdot}_\pi := \cdot^{\mathsf{T}} \mathsf{D}_\pi\cdot$ and $\braket{\cdot,\cdot}_\Pi = \cdot^{\mathsf{T}} \mathsf{D}_\Pi\cdot$.
We first show that the Mori projector $\mathcal{P}:=A\Omega^\mathsf{T}$ onto the relevant subspace is self-adjoint with respect to the $\pi$-weighted inner product:
\begin{align}
    \mathcal{P}^\dagger \mathsf{D}_\pi &:= \mathsf{D}_\pi\mathcal{P}\notag\\
    &= \mathsf{D}_\pi A \Omega^{\mathsf{T}}\notag\\
    &:= \mathsf{D}_\pi A \mathsf{D}_\Pi^{-1} A^{\mathsf{T}} \mathsf{D}_\pi\notag\\
    \implies \mathcal{P}^\dagger  &= \mathsf{D}_\pi A \mathsf{D}_\Pi^{-1} A^{\mathsf{T}}
    =:  \Omega A^{\mathsf{T}} = \mathcal{P}^{\mathsf{T}}
    \label{eq:p_self_adjoint}
\end{align}
Thus $\mathcal{P}$ and its complement $\mathcal{Q}$ are self-adjoint with respect to the $\pi$-weighted inner product.
Moreover, as
\begin{align}
    \mathcal{P}^\dagger \mathsf{D}_\pi\mathcal{Q} &=  \Omega A^{\mathsf{T}} \mathsf{D}_\pi(I- A \Omega^{\mathsf{T}})\notag\\
    &=  \Omega( A^{\mathsf{T}} \mathsf{D}_\pi -  A^{\mathsf{T}} \mathsf{D}_\pi  A \mathsf{D}_\Pi^{-1} A^{\mathsf{T}} \mathsf{D}_\pi)\notag\\
    &=  \Omega( A^{\mathsf{T}} \mathsf{D}_\pi -  A^{\mathsf{T}}  \Omega A^{\mathsf{T}} \mathsf{D}_\pi)\notag\\
    &=  \Omega( A^{\mathsf{T}} \mathsf{D}_\pi - I A^{\mathsf{T}} \mathsf{D}_\pi) = 0
\end{align}
$\mathcal{P}$ and $\mathcal{Q}$ are orthogonal with respect to the $\pi$-weighted inner product.

The probability vector $\pi$ representing the stationary density over microstates lies in the image of $\mathcal{P}^\dagger$:
\begin{align}
    \mathcal{P}^\dagger\pi &= \Omega A^{\mathsf{T}}\pi\notag\\
    &= \mathsf{D}_\pi A\mathsf{D}_\Pi^{-1}\Pi\notag\\
    &= \mathsf{D}_\pi A\mathbf{1}_{|M|}\notag\\
    &= \mathsf{D}_\pi\mathbf{1}_{|\mu|} = \pi
\end{align}
Thus $\pi$ lies in the kernel of $\mathcal{Q}^\dagger$ and the assumption in Eq. \ref{eq:initialization_assumption} holds for $\rho^{(0)}=\pi$.
The probability vector $\Pi := A^{\mathsf{T}}\pi$ representing the stationary density over macrostates is the stationary vector of $U^{(n)}$ at all lags $n\in\N$:
\begin{align}
    U^{(n)}\Pi &:= A^{\mathsf{T}} L^n\Omega\Pi\notag\\
    &= A^{\mathsf{T}} L^n \mathsf{D}_\pi A\mathsf{D}_\Pi^{-1}\Pi\notag\\
    &= A^{\mathsf{T}} L^n \mathsf{D}_\pi A\mathbf{1}_{|M|}\notag\\
    &= A^{\mathsf{T}} L^n \mathsf{D}_\pi\mathbf{1}_{|\mu|}\notag\\
    &= A^{\mathsf{T}} L^n\pi\notag\\
    &= A^{\mathsf{T}}\pi = \Pi
\end{align}
$U^{(n)}$ is detailed-balance with respect to $\Pi$ whenever $L$ is detailed-balance with respect to $\pi$, as
\begin{align}
    U^{(n)}\mathsf{D}_{\Pi} &:= A^{\mathsf{T}} L^n\Omega \mathsf{D}_\Pi\notag\\
    &= A^{\mathsf{T}} L^n \mathsf{D}_\pi A \mathsf{D}_\Pi^{-1}\mathsf{D}_\Pi\notag\\
    &= A^{\mathsf{T}} \mathsf{D}_\pi (L^{\mathsf{T}})^n A\notag\\
    &= \mathsf{D}_\Pi \mathsf{D}_\Pi^{-1}A^{\mathsf{T}} \mathsf{D}_\pi (L^{\mathsf{T}})^n A\notag\\
    &= \mathsf{D}_\Pi U^{(n)\mathsf{T}}
\end{align}
where we have used the self-adjointness of $L^\mathsf{T}$ with respect to the $\pi$-weighted inner product in the third equality.
$U^{(n)}$ converges to $\Pi\mathbf{1}_{|M|}^{\mathsf{T}}$ in the large-$n$ limit as
\begin{align}
    \lim_{n\to\infty}U^{(n)} &= \lim_{n\to\infty}A^{\mathsf{T}} e^{n\tau\mathcal{L}}\Omega\notag\\
    &= A^{\mathsf{T}}\pi\mathbf{1}_{|\mu|}^{\mathsf{T}} \mathsf{D}_\pi A \mathsf{D}_\Pi^{-1}\notag\\
    &= A^{\mathsf{T}}\pi\pi^{\mathsf{T}} A \mathsf{D}_\Pi^{-1}\notag\\
    &= \Pi\Pi^{\mathsf{T}} \mathsf{D}_\Pi^{-1} = \Pi\mathbf{1}_{|M|}^{\mathsf{T}}
\end{align}
where convergence of $e^{n\tau\mathcal{L}}$ to $\pi\mathbf{1}_{|\mu|}^{\mathsf{T}}$ in the second equality is a consequence of the ergodicity of the MJP.
Note that $\Pi\mathbf{1}_{|M|}^{\mathsf{T}}$ is singular by virtue of its one-dimensional span for coarse-grainings with $|M|>1$.

\section{ALTERNATIVE DEFINITION OF TIME-CONVOLUTIONLESS PROPAGATOR}
\label{app:alternative-tcl-gme-dt-def}
We may attempt to derive Eq. \ref{eq:tcl-gme-dt} in generality by extending that of a continuous-time TCL GME \cite{chaturvedi1979timeconvolutionless}.
Recall Dyson's identity \cite{dyson1949S}
\begin{align}
    u^{(n+1)} &= (\mathcal{A}+\mathcal{B})u^{(n)}\notag\\
    \implies u^{(n+1)} &= \mathcal{A}^{n+1}u^{(0)}+\sum_{m=0}^n \mathcal{A}^{n-m}\mathcal{B}u^{(m)}
    \label{eq:dyson}
\end{align}
for $\{u^{(n)}\}_{n=0}^\infty$ a sequence and $\mathcal{A}^n$ the power of a matrix $\mathcal{A}$.
We expand the microscopic Markov chain as
\begin{align}
    L^{n}\rho^{(0)} = (L\mathcal{Q}^\dagger &+ L\mathcal{P}^\dagger)L^{n-1}\rho^{(0)}\notag\\
    = (L\mathcal{Q}^\dagger)^{n}\rho^{(0)} &+ \sum_{m=0}^{n-1} (L\mathcal{Q}^\dagger)^{n-1-m}(L\mathcal{P}^\dagger)L^m\rho^{(0)}
    \label{eq:alternative-inhomogeneous-tcl-projected-micro-chain}
\end{align}
using Eq. \ref{eq:dyson}.
Eliminating inhomogeneity $(L\mathcal{Q}^\dagger)^{n}\rho^{(0)}$ of Eq. \ref{eq:alternative-inhomogeneous-tcl-projected-micro-chain} by the assumption in Eq. \ref{eq:initialization_assumption},
\begin{align}
    \mathcal{Q}^\dagger L^{n}\rho^{(0)} &= \sum_{m=0}^{n-1}(\mathcal{Q}^\dagger L)^{n-m}\mathcal{P}^\dagger L^m\rho^{(0)}\notag\\
    &= \Sigma^{(n)}\mathcal{P}^\dagger L^{n}\rho^{(0)}+\Sigma^{(n)}\mathcal{Q}^\dagger L^{n}\rho^{(0)}
    \label{eq:irrelevant_tcl_chain}
\end{align}
where we have assumed $L$ is invertible to define
\begin{align}
    \Sigma^{(n)} &= \sum_{m=0}^{n-1}(\mathcal{Q}^\dagger L)^{n-m}\mathcal{P}^\dagger L^{m-n}\notag\\
    &= \sum_{m=1}^{n}(\mathcal{Q}^\dagger L)^{m}\mathcal{P}^\dagger L^{-m}\notag\\
    &= \sum_{m=1}^{n}\left[(\mathcal{Q}^\dagger L)^m L^{-m}-(\mathcal{Q}^\dagger L)^m\mathcal{Q}^\dagger L^{-m}\right]\notag\\
    &= \sum_{m=1}^{n}\left[(\mathcal{Q}^\dagger L)^{m-1}\mathcal{Q}^\dagger L^{1-m}-(\mathcal{Q}^\dagger L)^m\mathcal{Q}^\dagger L^{-m}\right]\notag\\
    &= \mathcal{Q}^\dagger - (\mathcal{Q}^\dagger L)^{n}\mathcal{Q}^\dagger L^{-n}
\end{align}
and eliminate the time convolution for all $n\in\N_+$.

Rearranging Eq. \ref{eq:irrelevant_tcl_chain},
\begin{equation*}
    \mathcal{Q}^\dagger L^n\rho^{(0)}
    = (I-\Sigma^{(n)})^{-1}\Sigma^{(n)}\mathcal{P}^\dagger L^{n}\rho^{(0)}
\end{equation*}
for all $n\in\N_+$.
We substitute $\mathcal{Q}^\dagger L^n\rho^{(0)}$ into the Markov chain projected by $\mathcal{P}^\dagger$:
\begin{multline}
    \mathcal{P}^\dagger L^{n+1}\rho^{(0)} = \mathcal{P}^\dagger L\mathcal{P}^\dagger L^n\rho^{(0)} + \mathcal{P}^\dagger L\mathcal{Q}^\dagger L^n\rho^{(0)}\\
    = \mathcal{P}^\dagger L\mathcal{P}^\dagger L^n\rho^{(0)} + \mathcal{P}^\dagger L(I-\Sigma^{(n)})^{-1}\Sigma^{(n)}\mathcal{P}^\dagger L^{n}\rho^{(0)}\\
    = \mathcal{P}^\dagger L(I-\Sigma^{(n)})^{-1}\mathcal{P}^\dagger L^{n}\rho^{(0)}
    \label{eq:relevant_tcl_chain}
\end{multline}
In Eq. \ref{eq:relevant_tcl_chain}, we have used the identity
\begin{align}
    &I + (I-\mathcal{A})^{-1}\mathcal{A}\notag\\
    = &(I-\mathcal{A})^{-1}(I-\mathcal{A}) + (I-\mathcal{A})^{-1}\mathcal{A}\notag\\
    = &(I-\mathcal{A})^{-1}\notag
\end{align}
in the third equality.
We thus obtain
{
\allowdisplaybreaks
\begin{align}
    A^{\mathsf{T}}\mathcal{P}^\dagger L^{n+1}\rho^{(0)} &= A^{\mathsf{T}}\mathcal{P}^\dagger L(I-\Sigma^{(n)})^{-1}\mathcal{P}^\dagger L^{n}\rho^{(0)}\notag\\
    = A^{\mathsf{T}} L^{n+1}\rho^{(0)} &=
    A^{\mathsf{T}} L(I-\Sigma^{(n)})^{-1}\Omega A^{\mathsf{T}} L^{n}\rho^{(0)}\notag\\
    = P^{(n+1)} &= G^{(n+1)} P^{(n)}\notag
\end{align}
}
where
\begin{align}
    G^{(n+1)} &:= A^{\mathsf{T}} L(I-\Sigma^{(n)})^{-1}\Omega\notag\\
    &= A^{\mathsf{T}} L[\mathcal{P}^\dagger+\mathcal{Q}^\dagger (L\mathcal{Q}^\dagger)^{n} L^{-n}]^{-1}\Omega
    \label{eq:tcl-gme-dt-alternative-propagator}
\end{align}
with $G^{(0)}:=U^{(0)}=I$ and $G^{(1)}:=U^{(1)}=A^{\mathsf{T}} L\Omega$.

As defined in Eq. \ref{eq:tcl-gme-dt-alternative-propagator},
$G^{(n+1)}$ clearly exists for $n\geq 1$ only if $I-\Sigma^{(n)}$ is invertible.
Unfortunately, $I-\Sigma^{(n)}$ is not always invertible.
Observe that, as $A^{\mathsf{T}}\Omega=I$, $\Omega$ defines an injective map and its kernel is the zero vector.
Suppose that $I-\Sigma^{(n)}$ is invertible for $n$ such that $U^{(n)}$ is singular.
Then for some nonzero $P\in\R^{|M|}$ such that $U^{(n)}P=0$, $L^n\Omega P\neq 0$ by the invertibility of $L$ and yet
\begin{align}
    (I-&\Sigma^{(n)})L^n\Omega P = [\mathcal{P}^\dagger+\mathcal{Q}^\dagger (L\mathcal{Q}^\dagger)^{n} L^{-n}]L^n\Omega P\notag\\
    &= \mathcal{P}^\dagger L^n\Omega P + (\mathcal{Q}^\dagger L)^{n}\mathcal{Q}^\dagger\Omega P\notag\\
    &= \Omega A^{\mathsf{T}} L^n\Omega P  + (\mathcal{Q}^\dagger L)^{n}(I-\Omega A^{\mathsf{T}})\Omega P\notag\\
    &= \Omega U^{(n)} P + (\mathcal{Q}^\dagger L)^{n}(\Omega-\Omega) P = 0\notag
\end{align}
contradicting the invertibility of $I-\Sigma^{(n)}$.
Thus singularity of $U^{(n)}$ implies singularity of $I-\Sigma^{(n)}$.
$G^{(n)}$ as given in Eq. \ref{eq:tcl-gme-dt-alternative-propagator} is undefined at $n$ with singular $U^{(n-1)}$;
$G^{(n)}$ as given in Eq. \ref{eq:tcl-gme-dt-propagator} is more general.

\section{TRIPLET IRREVERSIBILITY}
\label{app:triplet-irreversibility}
As shown in Eqs. \ref{eq:G_interpretation_start}-\ref{eq:G_interpretation_end},
the TCL-GME-DT in Eq. \ref{eq:tcl-gme-dt} defines statistics of the triplet $\{X^{(n)},X^{(n-1)},X^{(0)}\}$.
Reversibility implies $\{X^{(n)},X^{(n-1)},X^{(0)}\}$ has the same statistics as its time-reverse $\{X^{(0)},X^{(1)},X^{(n)}\}$,
and, in particular,
\begin{widetext}
\begin{align}
    \prob_\Pi (X^{(n)}=M_i,X^{(n-1)}=M_j,X^{(0)}=M_k) &= \prob_\Pi (X^{(n)}=M_k,X^{(1)}=M_j,X^{(0)}=M_i)\\
    \implies \prob_\Pi (X^{(n)}=M_i|X^{(n-1)}=M_j,X^{(0)}=M_k)&\prob_\Pi (X^{(n-1)}=M_j|X^{(0)}=M_k)\prob_\Pi (X^{(0)}=M_k)\notag\\
    = \prob_\Pi (X^{(n)}=M_k|X^{(1)}=M_j,X^{(0)}=M_i)&\prob_\Pi (X^{(1)}=M_j|X^{(0)}=M_i)\prob_\Pi (X^{(0)}=M_i)\\
    \implies G_{ij}^{(n)}U^{(n-1)}_{jk}\Pi_k &= \prob_\Pi (X^{(n)}=M_k|X^{(1)}=M_j,X^{(0)}=M_i)U^{(1)}_{ji}\Pi_i
\end{align}
gives Eq. \ref{eq:triplet-reversibility}.
We may be tempted to identify
\begin{equation}
    \prob_\Pi (X^{(n)}=M_k|X^{(1)}=M_j,X^{(0)}=M_i)
    \overset{!}{=}\(\prod_{m=0}^{n-2}G^{(n-m)}\)_{kj}
\end{equation}
for all integer lags $n > 1$ and further constrain estimates $\{\hat{G}^{(n)}\}_{n=0}^p$.
Note, however, the simple counterexample
{
\allowdisplaybreaks
\begin{align}
    &\prob_\Pi (X^{(3)}=M_k|X^{(1)}=M_j,X^{(0)}=M_i)\notag\\
    &= \sum_{\ell=1}^{|M|}\prob_\Pi (X^{(3)}=M_k|X^{(2)}=M_\ell,X^{(1)}=M_j,X^{(0)}=M_i)\prob_\Pi (X^{(2)}=M_\ell|X^{(1)}=M_j,X^{(0)}=M_i)\notag\\
    &= \sum_{\ell=1}^{|M|}\prob_\Pi (X^{(3)}=M_k|X^{(2)}=M_\ell,X^{(1)}=M_j,X^{(0)}=M_i)G_{\ell j}^{(2)}\notag\\
    &\neq \sum_{\ell=1}^{|M|}\prob_\Pi (X^{(3)}=M_k|X^{(2)}=M_\ell,X^{(0)}=M_i)G_{\ell j}^{(2)}
    = \sum_{\ell=1}^{|M|}G_{k\ell}^{(3)}G_{\ell j}^{(2)}
\end{align}
}
\begin{equation}
    \implies\prob_\Pi (X^{(3)}=M_k|X^{(1)}=M_j,X^{(0)}=M_i)\neq\(\prod_{m=0}^{3-2}G^{(3-m)}\)_{kj}
\end{equation}
\end{widetext}
that reflects our ignorance of triplet-conditioned probabilities such as $\prob_\Pi (X^{(3)}=M_k|X^{(2)}=M_\ell,X^{(1)}=M_j,X^{(0)}=M_i)$.

The triplet $\{X^{(2)}, X^{(1)}, X^{(0)}\}$ is an exception.
Here
\begin{multline}
    \prob_\Pi (X^{(2)}=M_k|X^{(1)}=M_j,X^{(0)}=M_i)\\
    = \(\prod_{m=0}^{2-2}G^{(2-m)}\)_{kj} = G^{(2)}_{kj}
\end{multline}
holds and reversibility implies
\begin{multline}
    \prob_\Pi (X^{(2)}=M_i,X^{(2-1)}=M_j,X^{(0)}=M_k)\\
    = \prob_\Pi (X^{(2)}=M_k,X^{(1)}=M_j,X^{(0)}=M_i)\notag
\end{multline}
\begin{align}
    \iff G_{ij}^{(2)}U_{jk}^{(1)}\Pi_k &= G_{kj}^{(2)}U_{ji}^{(1)}\Pi_i\notag\\
    \iff G_{ij}^{(2)}U_{kj}^{(1)}\Pi_j &= G_{kj}^{(2)}U_{ij}^{(1)}\Pi_j\notag\\
    \iff \frac{G_{ij}^{(2)}}{G^{(2)}_{kj}} &=  \frac{U_{ij}^{(1)}}{U^{(1)}_{kj}} =: \frac{G_{ij}^{(1)}}{G^{(1)}_{kj}}
\end{align}
for all $i,j,k\in\{1,\cdots,|M|\}$,
where we have repeatedly used the detailed balance of both $G^{(2)}$ and $U^{(1)}=G^{(1)}$ with respect to $\Pi$.
Then as both $G^{(1)}$ and $G^{(2)}$ are column-stochastic,
reversible statistics of $\{X^{(2)}, X^{(1)}, X^{(0)}\}$ imply $G_{ij}^{(1)} = G_{ij}^{(2)}$ for all $i,j\in\{1,\cdots,|M|\}$.

\section{DERIVATION OF NAKAJIMA-ZWANZIG GENERALIZED MASTER EQUATION IN DISCRETE TIME}
\label{app:nzgmedt}
We first apply adjoint projectors to the Markov chain:
\begin{align}
    \rho^{(n+1)} &=L\rho^{(n)} =L\mathcal{P}^\dagger \rho^{(n)} + L\mathcal{Q}^\dagger \rho^{(n)}\implies
    \label{eq:mc-nzgmedt-appendix}\\
	\mathcal{P}^\dagger \rho^{(n+1)} &=\mathcal{P}^\dagger L\rho^{(n)} = \mathcal{P}^\dagger  L\mathcal{P}^\dagger \rho^{(n)} + \mathcal{P}^\dagger  L\mathcal{Q}^\dagger \rho^{(n)}
    \label{eq:pmc-nzgmedt-appendix}
\end{align}
Applying Dyson's identity (Eq. \ref{eq:dyson}) \cite{dyson1949S},
Eq. \ref{eq:mc-nzgmedt-appendix} has formal solution
\begin{multline}
    \rho^{(n+1)} = (L\mathcal{Q}^\dagger )^{n+1}\rho^{(0)} + \sum_{m=0}^n (L\mathcal{Q}^\dagger )^{n-m}L\mathcal{P}^\dagger \rho^{(m)}
    \label{eq:nzgmedt-dyson-soln}
\end{multline}
for all $n\in\N$.
Projecting Eq. \ref{eq:nzgmedt-dyson-soln} with $\mathcal{Q}^\dagger$,
\begin{multline}
    \mathcal{Q}^\dagger \rho^{(n+1)} =\\ (\mathcal{Q}^\dagger  L)^{n+1}\mathcal{Q}^\dagger \rho^{(0)} + \sum_{m=0}^{n}(\mathcal{Q}^\dagger  L)^{n+1-m}\mathcal{P}^\dagger \rho^{(m)}\notag
\end{multline}
for all $n\in\N$,
which we substitute into Eq. \ref{eq:pmc-nzgmedt-appendix}:
\begin{multline}
    \mathcal{P}^\dagger \rho^{(n+1)} =\\ \mathcal{P}^\dagger  L(\mathcal{Q}^\dagger  L)^n\mathcal{Q}^\dagger \rho^{(0)} + \mathcal{P}^\dagger  L\sum_{m=0}^{n}(\mathcal{Q}^\dagger  L)^{n-m}\mathcal{P}^\dagger \rho^{(m)}
    \label{eq:P-projected-nzgmedt-chain}
\end{multline}
Decomposing $\mathcal{P}^\dagger=\Omega A^{\mathsf{T}}$ and applying coarse-graining $A^{\mathsf{T}}$ to both sides of Eq. \ref{eq:P-projected-nzgmedt-chain},
we obtain
\begin{multline}
    A^{\mathsf{T}} \Omega A^{\mathsf{T}}\rho^{(n+1)} =
    A^{\mathsf{T}}\Omega A^{\mathsf{T}} L(\mathcal{Q}^\dagger L)^n\mathcal{Q}^\dagger \rho^{(0)}
    \\+ A^{\mathsf{T}} \Omega A^{\mathsf{T}} L\sum_{m=0}^{n}(\mathcal{Q}^\dagger  L)^{n-m} \Omega A^{\mathsf{T}}\rho^{(m)}\notag
\end{multline}
simplifying to
\begin{multline}
    A^{\mathsf{T}}\rho^{(n+1)} =
    A^{\mathsf{T}} L(\mathcal{Q}^\dagger L)^n\mathcal{Q}^\dagger \rho^{(0)}\\
    + \sum_{m=0}^{n} A^{\mathsf{T}} L(\mathcal{Q}^\dagger L)^{n-m} \Omega A^{\mathsf{T}}\rho^{(m)}
    \label{eq:nzgmedt-expanded}
\end{multline}
as $A^{\mathsf{T}}\Omega =I$.
At last identifying $P^{(m)} := A^{\mathsf{T}}\rho^{(m)}$ for all $m\in\N$ on both sides of Eq. \ref{eq:nzgmedt-expanded},
we define discrete-time memory kernels
\begin{equation}
    K^{(n)} :=  A^{\mathsf{T}} L(\mathcal{Q}^\dagger  L)^{n} \Omega =  A^{\mathsf{T}} L(L-\mathcal{P}^\dagger L)^{n} \Omega
    \label{eq:nzgmedt-kernel-def-appendix}
\end{equation}
to obtain an exact Nakajima-Zwanzig generalized master equation in discrete time (NZ-GME-DT)
\begin{equation}
    P^{(n+1)} = A^{\mathsf{T}} L(\mathcal{Q}^\dagger L)^n\mathcal{Q}^\dagger \rho^{(0)} + \sum_{m=0}^{n}K^{(n-m)} P^{(m)}
    \label{eq:inhomogeneous-nzgmedt}
\end{equation}
for all $n\in\N$,
where the inhomogeneity $A^{\mathsf{T}} L(\mathcal{Q}^\dagger L)^n\mathcal{Q}^\dagger \rho^{(0)}$ vanishes upon assuming Eq. \ref{eq:initialization_assumption}.
Applying Dyson's identity (Eq. \ref{eq:dyson}) \cite{dyson1949S} again,
\begin{align}
    K^{(n)} &= A^{\mathsf{T}}L(L-\mathcal{P}^\dagger L)^n\Omega\notag\\
    &= A^{\mathsf{T}}L\left[L^n + \sum_{m=0}^{n-1}L^{n-1-m}(-\mathcal{P}^\dagger L)(L-\mathcal{P}^\dagger L)^m\right]\Omega\notag\\
    &= A^{\mathsf{T}}L^{n+1}\Omega - \sum_{m=0}^{n-1}A^{\mathsf{T}}L^{n-m}\Omega A^{\mathsf{T}}L(\mathcal{Q}^\dagger L)^m\Omega\notag\\
    &= U^{(n+1)} - \sum_{m=1}^{n}U^{(m)}K^{(n-m)}
\end{align}
we may nondiagrammatically recover Eq. \ref{eq:recursion}.

\section{DISCRETE-TIME FLUCTUATION-DISSIPATION THEOREM}
\label{app:fdt}
Inhomogeneities due to $\mathcal{Q}^\dagger\rho^{(0)}\neq 0$---violations of the assumption in Eq. \ref{eq:initialization_assumption}---are in principle observable.
This is transparent in the special case of reversible MJPs,
which have $L^\mathsf{T}$ self-adjoint with respect to the $\pi$-weighted inner product.
Proceeding from the definition in Eq. \ref{eq:nzgmedt-kernel-def-appendix},
\begin{align}
    K^{(n)} &:= A^{\mathsf{T}} (L\mathcal{Q}^\dagger)^{n}L\Omega\notag\\
    &= \Xi^{(n)}L \mathsf{D}_\pi A\mathsf{D}_\Pi^{-1}\notag\\
    &= \Xi^{(n)} \mathsf{D}_\pi L^{\mathsf{T}} A\mathsf{D}_\Pi^{-1}\notag\\
    &= \Xi^{(n)} \mathsf{D}_\pi\mathcal{P} L^{\mathsf{T}} A\mathsf{D}_\Pi^{-1} +  \Xi^{(n)} \mathsf{D}_\pi\mathcal{Q} L^{\mathsf{T}} A\mathsf{D}_\Pi^{-1}\notag\\
    &= \Xi^{(n)} \mathsf{D}_\pi\mathcal{Q} L^{\mathsf{T}} A\mathsf{D}_\Pi^{-1}\notag\\
    &= \Xi^{(n)}\mathsf{D}_\pi\Xi^{(1)\mathsf{T}} \mathsf{D}_\Pi^{-1}
    \label{eq:dtfdt}
\end{align}
for all $n\in\N_+$,
where we have defined the time-dependent fluctuating-force \cite{langevin1908} operator $\Xi^{(n)}:= A^{\mathsf{T}} (L\mathcal{Q}^\dagger)^{n}$ at times $n\tau$,
the third equality follows from the self-adjointness of $L^\mathsf{T}$ with respect to the $\pi$-weighted inner product,
and
\begin{equation}
    \Xi^{(n)} \mathsf{D}_\pi\mathcal{P} L^{\mathsf{T}} A\mathsf{D}_\Pi^{-1} = \Xi^{(n-1)} L(\mathcal{Q}^\dagger \mathsf{D}_\pi\mathcal{P}) L^{\mathsf{T}} A\mathsf{D}_\Pi^{-1}
\end{equation}
vanishes in the fifth equality as $\mathcal{Q}^\dagger \mathsf{D}_\pi\mathcal{P}=0$.
Eq. \ref{eq:dtfdt} is an exact fluctuation-dissipation theorem (FDT) of the second kind \cite{callen1951irreversibility,kubo1956fdt} in discrete time.
However, note from Eq. \ref{eq:inhomogeneous-nzgmedt} that only
\begin{equation}
    \Xi^{(n+1)}\rho^{(0)} = P^{(n+1)}-\sum_{m=0}^{n}K^{(n-m)}P^{(m)}\notag
\end{equation}
and thus $\Xi^{(n)}\rho^{(0)}\rho^{(0)\mathsf{T}}\Xi^{(1)\mathsf{T}}\mathsf{D}_\Pi^{-1}$ where $\rho^{(0)}\rho^{(0)\mathsf{T}}\neq \mathsf{D}_\pi$ are readily accessible without recourse to microstates,
impeding use of Eq. \ref{eq:dtfdt} in parameterization of discrete-time memory kernels.

\section{INFORMATION-PROJECTION ALGORITHMS}
\label{app:projectors}
Here $\|\cdot\|_{\rm TV} = \|\cdot\|_{1}/2$ denotes total-variation distance from the origin,
\texttt{ClampBelow}$(\cdot,\epsilon)$ clamps to a minimum of $\epsilon$,
and $\text{sqrt}(\cdot)$ denotes the elementwise square root.

\begin{algorithm}[H] \label{alg:SK}
\caption{\protect\mbox{\texttt{SinkhornKnopp}($\hat{\Phi}$, $\hat{\Pi}$, $\epsilon$)} \cite{knopp1967concerning}}
\textbf{input} infeasible flux estimate $\hat{\Phi}$, stationary-vector estimate $\hat{\Pi}$, tolerance $\epsilon$\;
$\delta^{(c)}\gets\infty$, $\delta^{(r)}\gets\infty$\;
$c\gets\mathbf{1}_{|M|}$, $r\gets\mathbf{1}_{|M|}$\;
$\hat{\Phi}\gets\texttt{ClampBelow}(\hat{\Phi},\epsilon)$\;
\While{$\max\{\|\delta^{(c)}\|_{\rm TV},\|\delta^{(r)}\|_{\rm TV}\} >\epsilon$}{
    $\delta^{(c)}\gets\hat{\Pi}\oslash\hat{\Phi}^{\mathsf{T}} r-c$, $c\gets c+\delta^{(c)}$\;
    $\delta^{(r)}\gets\hat{\Pi}\oslash\hat{\Phi}c-r$, $r\gets r+\delta^{(r)}$\;
}
\textbf{return} feasible flux estimate $\mathsf{D}_r\hat{\Phi}\mathsf{D}_c$
\end{algorithm}

\begin{algorithm}[H] \label{alg:KRU}
\caption{\protect\mbox{\texttt{KnightRuizU\c{c}ar}($\hat{\Theta}$, $\hat{\Pi}$, $\epsilon$)} \cite{knight2014symmetry}}
\textbf{input} infeasible flux estimate $\hat{\Theta}$, stationary-vector estimate $\hat{\Pi}$, tolerance $\epsilon$\;
$\delta\gets\infty$, $r\gets\mathbf{1}_{|M|}$\;
$\hat{\Theta}\gets\texttt{ClampBelow}(\hat{\Theta},\epsilon)$\;
$\hat{\Theta}\gets\text{sqrt}(\hat{\Theta}\odot\hat{\Theta}^{\mathsf{T}})$\;
\While{$\|\delta\|_{\rm TV} >\epsilon$}{
    $\delta\gets r\odot\text{sqrt}[\hat{\Pi}\oslash (r\odot\hat{\Theta} r)]-r$\;
    $r\gets r+\delta$\;
}
\textbf{return} feasible flux estimate $\mathsf{D}_r\hat{\Theta} \mathsf{D}_r$
\end{algorithm}

The Sinkhorn-Knopp algorithm (Alg. \ref{alg:SK}) \cite{knopp1967concerning} projects flux matrices $\hat{\Phi}$ onto the feasible set defined by Eqs. \ref{eq:flux_nn}-\ref{eq:flux_cols}.
The usual Knight-Ruiz-U\c{c}ar algorithm \cite{knight2014symmetry} differs from Alg. \ref{alg:KRU} in lacking the initial geometric symmetrization $\hat{\Theta}\gets\text{sqrt}(\hat{\Theta}\odot\hat{\Theta}^{\mathsf{T}})$;
it only preserves the symmetry of already symmetric input fluxes $\hat{\Theta}$ during projection onto the feasible set defined by Eqs. \ref{eq:convex_orthant}-\ref{eq:convex_stochastic}.
Alg. \ref{alg:KRU} nevertheless minimizes $D_\mathrm{KL}(\Theta'\|\hat{\Theta})$ subject to Eqs. \ref{eq:convex_orthant}-\ref{eq:convex_stochastic} and \ref{eq:convex_db}
with initial geometric symmetrization of the input.
To see this, observe that
\begin{align}
    &\argmin_{\Theta'} D_\mathrm{KL}(\Theta'\|\hat{\Theta})\notag\\
    :=&\argmin_{\Theta'}\(\sum_{ij}\Theta'_{ij}\ln\frac{\Theta'_{ij}}{\hat{\Theta}_{ij}}-\Theta'_{ij}+\hat{\Theta}_{ij}\)\notag\\
    =&\argmin_{\Theta'}\(\sum_{ij}\Theta'_{ij}\ln\frac{\Theta'_{ij}}{\hat{\Theta}_{ij}}\)\label{eq:KRU_KL_sum1}\\
    =&\argmin_{\Theta'}\(\sum_{ij}\Theta'_{ij}\ln\frac{\Theta'_{ij}}{\sqrt{\hat{\Theta}_{ij}\hat{\Theta}_{ji}}}\)\label{eq:KRU_KL_sym}\\
    =&\argmin_{\Theta'} D_\mathrm{KL}\bigl(\Theta'\|\text{sqrt}(\hat{\Theta}\odot\hat{\Theta}^{\mathsf{T}})\bigr)\label{eq:KRU_KL_sum2}
\end{align}
for any feasible $\Theta'$.
Eqs. \ref{eq:KRU_KL_sum1} and \ref{eq:KRU_KL_sum2} follow as $\Theta'_{ij}$ always sum to unity by Eqs. \ref{eq:convex_stationary}-\ref{eq:convex_stochastic} while $\hat{\Theta}_{ij}$ and $\sqrt{\hat{\Theta}_{ij}\hat{\Theta}_{ji}}$ summands are invariant to $\Theta'$.
Eq. \ref{eq:KRU_KL_sym} follows from Eq. \ref{eq:convex_db}.
Thus symmetry-preserving projection of $\text{sqrt}(\hat{\Theta}\odot\hat{\Theta}^{\mathsf{T}})$ onto the set satisfying Eqs. \ref{eq:convex_orthant}-\ref{eq:convex_stochastic}
equals the projection of $\hat{\Theta}$ onto the set satisfying Eqs. \ref{eq:convex_orthant}-\ref{eq:convex_stochastic} and \ref{eq:convex_db}.

We now describe a Levenberg-Marquardt algorithm \cite{levenberg1944method,marquardt1963algorithm} for projecting local-flux estimates $\hat{\Theta}$ onto the subset satisfying Eq. \ref{eq:affine_commutation}.
We note that maximizing Eq. \ref{eq:dual_objective} is equivalent to minimizing the loss $\mathbf{1}_{|M|^2}^{\mathsf{T}}\theta^*(\lambda)$,
which has gradient $-\mathscr{C}\theta^*(\lambda)$ and Hessian $\mathscr{C}\mathsf{D}_{\theta^*(\lambda)}\mathscr{C}^{\mathsf{T}}$ with respect to $\lambda$.
Then
\begin{multline}
    \argmin_{\theta'}\{D_\mathrm{KL}\bigl(\theta'\|\text{vec}(\hat{\Theta})\bigr):\hat{U}\text{mat}(\theta')-\text{mat}(\theta')\hat{U}^{\mathsf{T}}=0\}\\
    = \theta^*\(\argmin_{\lambda}\{\mathbf{1}_{|M|^2}^{\mathsf{T}}\theta^*(\lambda)\}\)
    \label{eq:lm_problem}
\end{multline}
for $\theta^*(\cdot)$ defined in Eq. \ref{eq:theta_optimum}.
For numerical stability, we also orthonormalize $\mathscr{C}\theta' = 0$ to the equivalent system $Q\theta' =0$ by discarding singular vectors with singular values smaller than tolerance $\epsilon$;
we redefine $\theta^*(\cdot)$ by replacing $\mathscr{C}$ in Eq. \ref{eq:theta_optimum} with $Q$.
The damping $\varrho$ chosen is the smallest such that the effective Hessian
\begin{equation}
    Q\mathsf{D}_{\theta^*(\lambda)}Q^{\mathsf{T}} + \varrho I
\end{equation}
is numerically positive definite, that is, it has no singular values smaller than tolerance $\epsilon$.
Note that $Q\mathsf{D}_{\theta^*(\lambda)}Q^{\mathsf{T}}$ is already positive semidefinite as all elements of $\theta^*(\lambda)$ are positive.
We use Armijo line search \cite{armijo1966minimization} to select learning-rate $\eta$ and aid the guaranteed convergence \cite{yamashita2001rate,zhang2003convergence} to a global solution of Eq. \ref{eq:lm_problem}:

\begin{algorithm}[H] \label{alg:LM}
\caption{\protect\mbox{\texttt{LevenbergMarquardt}($\hat{\Theta},\hat{U},\epsilon,\alpha$)} \cite{levenberg1944method,marquardt1963algorithm}}
\textbf{input} infeasible local-flux estimate $\hat{\Theta}$, previous-lag transition-matrix estimate $\hat{U}$, tolerance $\epsilon$, Armijo sufficient-improvement coefficient $\alpha$\;
$Q\gets\texttt{Orthonormalize}(I\otimes \hat{U}-\hat{U}\otimes I)$\;
$\theta\gets\mathrm{vec}(\hat{\Theta})$, $\lambda\gets 0$\;
\While{$\|Q\theta^*(\lambda)\|_\infty >\epsilon$}{
    $\{\varsigma\}\gets\texttt{SingularValues}(Q\mathsf{D}_{\theta^*(\lambda)}Q^{\mathsf{T}})$\;
    $\varrho\gets\max\{0,\epsilon-\min\{\varsigma\}\}$\;
    $v\gets (Q\mathsf{D}_{\theta^*(\lambda)}Q^{\mathsf{T}}+\varrho I)^{-1}Q\theta^*(\lambda)$\;
    $\eta\gets 1$\;
    $\lambda^+\gets \lambda + \eta v$\;
    \While{$\mathbf{1}_{|M|^2}^{\mathsf{T}}\theta^*(\lambda^+) - \mathbf{1}_{|M|^2}^{\mathsf{T}}\theta^*(\lambda) > -\alpha\eta \theta^*(\lambda)^{\mathsf{T}} Q^{\mathsf{T}} v$}{
        $\eta\gets \eta / 2$\;
        $\lambda^+\gets \lambda + \eta v$\;
    }
    $\lambda\gets\lambda^+$\;
}
\textbf{return} $\hat{\Theta}\gets\mathrm{mat}[\theta^*(\lambda)]$
\end{algorithm}

\section{METADATA, PREPROCESSING, ESTIMATION, AND ERROR}
\label{app:estimation}

\begin{table}[!h]
\centering
\begin{tabular}{ll}
Hyperparameter & Value \\
\hline
$\eta$ & $0.1$ (toy models), $1$ (proteins) \\
$\epsilon$ & $10^{-24}$ (\texttt{ClampBelow}), $10^{-12}$ (elsewhere) \\
$\alpha$ & $10^{-4}$ (HP35), not applicable (elsewhere) \\
$\nabla_{\max}$ & $1000$ \\
\hline
\end{tabular}
\caption{Projection and mirror-descent hyperparameters.}
\label{tab:hyperparameters}
\end{table}

Table \ref{tab:hyperparameters} lists hyperparameters used to estimate $U^{(n)}$ and $G^{(n)}$.
At each exponentiated-gradient step, we scale (``clip'') gradient matrix $\nabla f$ ($\nabla f_U$ or $\nabla f_G$) by constant
\begin{equation}
    \min\left\{1.0, \frac{\nabla_{\max}}{\max_{ij}\{|(\nabla f)_{ij}|\}}\right\}
\end{equation}
for numerical stability \cite{goodfellow2016deep}.

\begin{table}[!h]
\centering
\begin{tabular}{llll}
Dataset & Rate & $N$ & $B$ \\
\hline
Toy models & 1 a.u. & $10^2$ to $10^6$ & 1 to 32 \\
CFTR WT apo & 10 Hz & 1458859 & 4042 \\
CFTR WT +GLPG1837 & 10 Hz & 593025 & 1688 \\
CFTR G551D apo & 10 Hz & 2658309 & 5134 \\
CFTR G551D +GLPG1837 & 10 Hz & 1308487 & 2781 \\
CFTR L927P apo & 10 Hz & 1726716 & 4061 \\
CFTR L927P +GLPG1837 & 10 Hz & 648953 & 1927 \\
BMF1 +ATP$\gamma$S & $10^3$ Hz & 6000 & 1 \\
BMF1 +ATP & $4.5\times 10^4$ Hz & 39011 & 1 \\
HP35 & $5\times 10^9$ Hz & 1526040 & 314 \\
\hline
\end{tabular}
\caption{Metadata. ``Rate'' denotes sampling rate. $N$ denotes total samples over $B$ records.}
\label{tab:metadata}
\end{table}

Table \ref{tab:metadata} lists metadata.
In practice, count matrices are summed across $B$ records and computed up to the length of the longest record rather than $N$.
All count matrices are computed using sliding windows \cite{trendelkampschroer2015estimation}.
We also scale each count matrix by a constant so that its entries sum to unity;
$\eta$ and magnitudes of $\nabla f$ are thus comparable across sample sizes (observation times).
Note that scaling count matrix $C^{(n)}$ by a constant is equivalent to scaling $f_U$ or $f_G$ by a constant;
optima over $\Phi'$ and $\Theta'$ (and thus optimal $\hat{U}^{(n)}$ and $\hat{G}^{(n)}$) are invariant to such scaling.
We symmetrize count matrices using
\begin{equation}
    C^{(n)}\gets\frac{C^{(n)}+C^{(n)\mathsf{T}}}{2}
    \label{eq:count_symmetrization}
\end{equation}
in reversible cases.
In such reversible cases,
\begin{align}
    \Theta' \mathsf{D}_{\hat{\Pi}}^{-1}\hat{U}^{(n-1)}\mathsf{D}_{\hat{\Pi}} &= \Theta'\hat{U}^{(n-1)\mathsf{T}}\notag\\
    &= (\hat{U}^{(n-1)}\Theta')^{\mathsf{T}}\notag\\
    &= (\Theta'\hat{U}^{(n-1)\mathsf{T}})^{\mathsf{T}}\notag\\
    &= (\Theta' \mathsf{D}_{\hat{\Pi}}^{-1}\hat{U}^{(n-1)}\mathsf{D}_{\hat{\Pi}})^{\mathsf{T}}
\end{align}
using detailed balance of $\hat{U}^{(n-1)}$ in the first and fourth equalities,
Eq. \ref{eq:convex_db} in the second,
and Eq. \ref{eq:affine_commutation} in the third.
Thus
\begin{multline}
    f_G(\Theta';C^{(n)}|\hat{\Pi},\hat{U}^{(n-1)})\\
    := -\sum_{ij}C^{(n)}_{ij}\ln (\Theta' \mathsf{D}_{\hat{\Pi}}^{-1}\hat{U}^{(n-1)}\mathsf{D}_{\hat{\Pi}})_{ij}\\
    = -\sum_{ij}\frac{C^{(n)}_{ij}+C^{(n)}_{ji}}{2}\ln (\Theta' \mathsf{D}_{\hat{\Pi}}^{-1}\hat{U}^{(n-1)}\mathsf{D}_{\hat{\Pi}})_{ij}
\end{multline}
and optimal $\Theta'$ satisfying Eqs. \ref{eq:convex_orthant}-\ref{eq:convex_commutation} are unchanged by Eq. \ref{eq:count_symmetrization} in reversible cases.

WT and mutant CFTR FRET traces were previously idealized \cite{levring2023CFTR} to at most three or four states, respectively, using SPARTAN \cite{juette2016single}.
As all traces without photobleaching were previously excluded \cite{levring2023CFTR},
the final idealized efficiency of each trace is annotated as photobleached.
For each trace, all idealized efficiencies lower than the final are also annotated as photobleached; as SPARTAN idealization models efficiencies as normally distributed around reference values with standard deviation 0.1, efficiencies less than 0.05 are annotated as significantly lower than the 0.25 reference of the low-FRET state and thus also photobleached.
Photobleached efficiencies observed between non-photobleached efficiencies are annotated as blinks \cite{vogelsang2008reducing} and backfilled.
For each trace, remaining idealized efficiencies are mapped bijectively to reference efficiencies of 0.25 (low), 0.31 (intermediate in L927P mutant), 0.37 (intermediate in G551D mutant), and 0.49 (high) by minimizing mean-squared differences between paired efficiencies;
if idealized efficiencies are normally distributed around reference values with the same standard deviation (assumed 0.1 in SPARTAN),
this is the maximum-likelihood bijective mapping.

Defining $\hat{g}^{(0)}_i:=1$ and $\hat{g}^{(m)}_i = \hat{G}_{ii}^{(m)}$,
we estimate mean BMF1 dwell times in macrostate $M_i$ as
\begin{equation}
    \sum_{n=1}^\infty n\tau\(\prod_{m=0}^{n-1} \hat{g}_i^{(m)}-\prod_{m=0}^{n} \hat{g}_i^{(m)}\)
\end{equation}
where $\hat{G}^{(n)}$ is replaced by its cutoff value past the cutoff lag.
Note that MSMs are TCL-GME-DT models with a cutoff lag of one timestep \cite{dominic2023building}.
In practice, we truncate the sum past a dwell of $n=1000$ timesteps.

Reported and plotted error margins of all statistics represent 0.95 confidence intervals.
In toy examples, confidence intervals are from standard errors over models trained on ten independent samples unless otherwise stated.
In experimental examples, confidence intervals of MSM- and GME-based statistics are estimated by training 1000 models on independent bootstraps of $B$ records and forming bootstrap estimates using each such model.
Confidence intervals of CFTR ATP hydrolysis rates shown in Fig. \ref{fig:CFTR}(c) are from standard errors of means in NADH-coupled assays \cite{levring2023CFTR}.
Confidence intervals on the experimental dimerization curve shown in Fig. \ref{fig:CFTR}(e) are from normal estimated sample-mean distributions with standard deviations given by standard errors of means over 42 imaging patches \cite{levring2023CFTR};
initial conditions of bootstrap MSM and GME relaxation curves in Fig. \ref{fig:CFTR}(e) are sampled from the estimated sample-mean distribution of the initial dimerized fraction.
Confidence intervals of mean dwell times from change-point detection \cite{kobayashi2020BMF1} shown in Fig. \ref{fig:BMF1}(d) use previously reported errors of exponential fits to distributions of detected dwell times.
Confidence intervals of folding mean FPTs in the HP35 MD trajectory \cite{piana2012HP35} are from standard errors over means;
confidence intervals of folding mean FPTs in $T$-jump experiments follow a previously reported sensitivity analysis \cite{kubelka2006sub}.

\section{GLOBALLY MAXIMUM-LIKELIHOOD ESTIMATION OF TIME-CONVOLUTIONLESS PROPAGATORS IS NONCONVEX}

\label{app:nonconvex}
Unlike Eq. \ref{eq:flux_obj}, Eq. \ref{eq:nonconvex_obj} composes a strictly convex function with transformations that are nonlinear in parameters $\{G'^{(n)}\}_{n=1}^p$:
we must examine whether the problem
\begin{align}
    \min_{\{G'^{(n)}\}_{n=1}^p} & f_\mathrm{NC}(\{G'^{(n)}\}_{n=1}^p;\{C^{(n)}\}_{n=1}^p)\label{eq:nonconvex_min}\\
    \text{s.t.}\quad&  G'^{(n)}_{ij} > 0\quad\forall i,j,n\\
    & G'^{(n)}\hat{\Pi} = \hat{\Pi}\quad\forall n\\
    & \mathbf{1}_{|M|}^{\mathsf{T}} G'^{(n)} = \mathbf{1}_{|M|}^{\mathsf{T}}\quad\forall n\label{eq:nonconvex_constraint}
\end{align}
is amenable to convex optimization.
We see by a counterexample that it is not.
Consider count matrices
\begin{equation*}
    C^{(1)}_{(\mathrm{NC})} = C^{(2)}_{(\mathrm{NC})} = \begin{pmatrix} 2 & 1 \\ 1 & 2 \end{pmatrix}
\end{equation*}
and the parameterizations
{
\allowdisplaybreaks
\begin{align*}
    U'^{(1)}_{(A)} = G'^{(1)}_{(A)} &= \begin{pmatrix} 0.4 & 0.6 \\ 0.6 & 0.4 \end{pmatrix}\\
    G'^{(2)}_{(A)} &= \begin{pmatrix} 0.1 & 0.9 \\ 0.9 & 0.1 \end{pmatrix}\\
    U'^{(1)}_{(B)} = G'^{(1)}_{(B)} &= \begin{pmatrix} 0.6 & 0.4 \\ 0.4 & 0.6 \end{pmatrix}\\
    G'^{(2)}_{(B)} &= \begin{pmatrix} 0.9 & 0.1 \\ 0.1 & 0.9 \end{pmatrix}
\end{align*}
}
where $U'^{(2)}_{(A)} = G'^{(2)}_{(A)}G'^{(1)}_{(A)}$, $U'^{(2)}_{(B)} = G'^{(2)}_{(B)}G'^{(1)}_{(B)}$, and $p=2$. As $G'^{(1)}_{(A)}$, $G'^{(2)}_{(A)}$, $G'^{(1)}_{(B)}$, and $G'^{(2)}_{(B)}$ are bistochastic,
they lie in the feasible set with uniform stationary vector $\hat{\Pi} = (0.5\;\; 0.5)^{\mathsf{T}}$.

\bottomnextwidefloat
\begin{figure*}[!t]
\centering
\includegraphics[width=\textwidth]{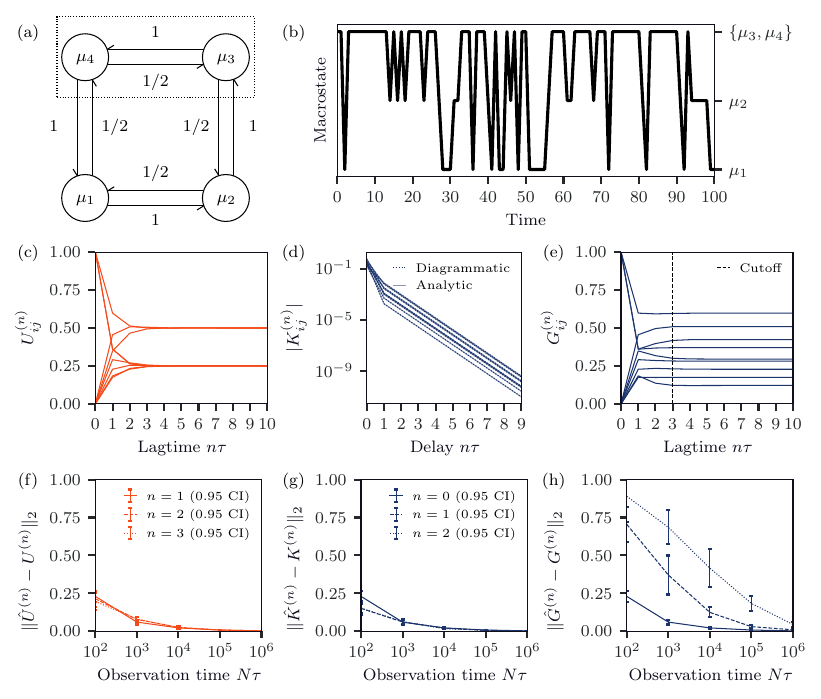}
\caption{
    Convergence in an adversarial toy example.
    The coarse-grained MJP in (a) generates example macrostate series (b) and lacks separation of timescales.
    Transition rates between microstates $\mu_3$ and $\mu_4$, which are coarse-grained into a single macrostate,
    equal transition rates among macrostates $\mu_1$, $\mu_2$, and $\{\mu_3,\mu_4\}$.
    (c) $U^{(n)}$ is nearly $\Pi\mathbf{1}_{|M|}^{\mathsf{T}}$ by lag $n=3$.
    (d) Analytic and diagrammatic memory-kernel elements $K^{(n)}_{ij}$ still coincide;
    (e) TCL propagator $G^{(n)}$ is nearly nonunique by plateau lag $n=3$ as $U^{(3)} \approx \Pi\mathbf{1}_{|M|}^{\mathsf{T}} \approx \mathcal{A} U^{(2)}$ for any propagator $\mathcal{A}$ with stationary vector $\Pi$.
    Nevertheless, projected-mirror-descent estimates of (f) $U^{(n)}$, (g) $K^{(n)}$, and (h) $G^{(n)}$ converge to ground truth in the large-sample limit for all lags up to cutoff $n=3$.
}
\label{fig:adversarial}
\end{figure*}

The midpoint
\begin{align*}
    U'^{(1)}_{(C)} &= G'^{(1)}_{(C)} = \frac{G'^{(1)}_{(A)}+G'^{(1)}_{(B)}}{2}\\
    &= G'^{(2)}_{(C)} = \frac{G'^{(2)}_{(A)}+G'^{(2)}_{(B)}}{2} = \begin{pmatrix} 0.5 & 0.5 \\ 0.5 & 0.5 \end{pmatrix}
\end{align*}
of these parameterizations is itself a feasible parameterization with $U'^{(2)}_{(C)} = G'^{(2)}_{(C)}G'^{(1)}_{(C)}$.
If $f_\mathrm{NC}(\{G'^{(n)}\}_{n=1}^{p=2};\{C_{(\mathrm{NC})}^{(n)}\}_{n=1}^{p=2})$ is convex with respect to $\{G'^{(n)}\}_{n=1}^{p=2}$, then
\begin{equation*}
    f_\mathrm{NC}(\{G'^{(n)}_{(C)}\}_{n=1}^{p=2};\{C_{(\mathrm{NC})}^{(n)}\}_{n=1}^{p=2}) \approx 8.318
\end{equation*}
should be no greater than the approximate mean $8.195$ of
\begin{equation*}
    f_\mathrm{NC}(\{G'^{(n)}_{(A)}\}_{n=1}^{p=2};\{C_{(\mathrm{NC})}^{(n)}\}_{n=1}^{p=2}) \approx 8.601
\end{equation*}
and
\begin{equation*}
    f_\mathrm{NC}(\{G'^{(n)}_{(B)}\}_{n=1}^{p=2};\{C_{(\mathrm{NC})}^{(n)}\}_{n=1}^{p=2}) \approx 7.790
\end{equation*}
by Jensen's inequality \cite{jensen1906inequality}, a contradiction.
Thus the problem given in Eqs. \ref{eq:nonconvex_min}-\ref{eq:nonconvex_constraint} is generally nonconvex.

\section{ADVERSARIAL EXAMPLE}
\label{app:adversarial}
Here we present an adversarial example in which Algs. \ref{alg:MDU} and \ref{alg:MDG} nevertheless converge.
We use a toy coarse-grained MJP with no separation of timescales [Fig. \ref{fig:adversarial}(a)] to generate macrostate series [Fig. \ref{fig:adversarial}(b)].
The transition matrices $U^{(n)}$ converge rapidly to $\Pi\mathbf{1}_{|M|}^{\mathsf{T}}$;
analytic (Eq. \ref{eq:nzgmedt-kernel-def}) and diagrammatic (Eq. \ref{eq:recursion}) memory kernels $K^{(n)}$ still coincide [Fig. \ref{fig:adversarial}].
We must use a cutoff lag of $3\tau$ for estimating TCL propagators $G^{(n)}$ [Fig. \ref{fig:adversarial}(e)], but note that $U^{(2)}$ is already approximately $\Pi\mathbf{1}_{|M|}^{\mathsf{T}}$ and nearly singular.
Thus the loss in Eq. \ref{eq:convex_obj} is nearly flat around its minimum, slowing optimization.
Alg. \ref{alg:MDU} still converges rapidly to ground truth in estimating $U^{(n)}$ [Fig. \ref{fig:adversarial}(f)] and $K^{(n)}$ [Fig. \ref{fig:adversarial}(g)];
Alg. \ref{alg:MDG} converges slowly, but eventually in estimating $G^{(n)}$ [Fig. \ref{fig:adversarial}(h)].

\section{COMPARISON TO NA\"IVE ESTIMATORS}
\label{app:benchmarks}
\begin{figure*}[!t]
    \centering
    \includegraphics[width=\linewidth]{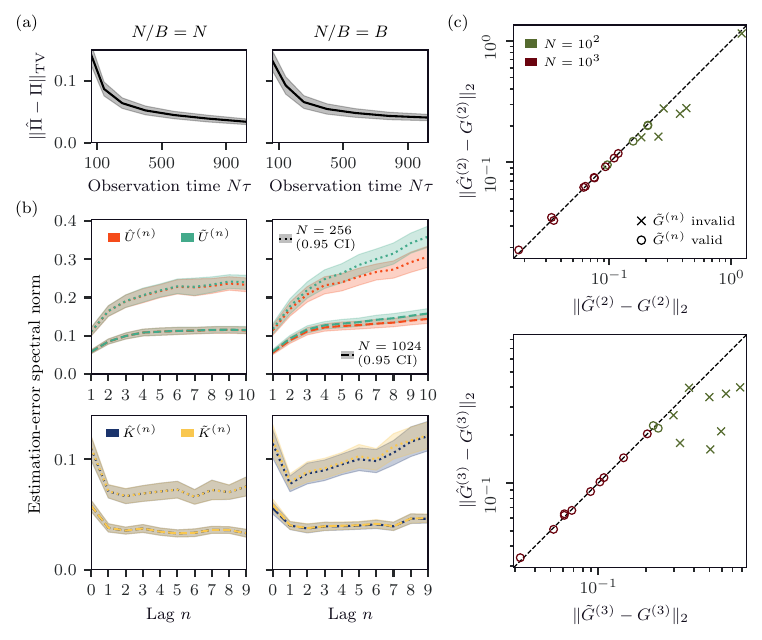}
    \caption{
    Constrained and na\"ive estimates benchmarked on the toy model from Fig. \ref{fig:schematic}.
    (a) Using single length-$N$ records or $B$ length-$B$ records, estimated stationary distribution $\hat{\Pi}$ converges rapidly to ground truth $\Pi$.
    (b) Using single length-$N$ records, constrained estimates $\hat{U}$ show slightly lower estimation error than na\"ively normalized $\tilde{U}$.
    Using $B$ length-$B$ records, constrained estimates $\hat{U}$ show clearly and consistently lower estimation error than na\"ively normalized $\tilde{U}$.
    Error comparisons hold for NZ-GME-DT kernels $\hat{K}$ and $\tilde{K}$ from applying Eq. \ref{eq:recursion} to $\hat{U}$ and $\tilde{U}$ respectively.
    (c) Constrained TCL-GME-DT parameters $\hat{G}$ show lower estimation error than na\"ive $\tilde{G}$ from Eq. \ref{eq:G-hat-from-inversion}.
    Eq. \ref{eq:G-hat-from-inversion} often fails to parameterize valid column-stochastic propagators with nonnegative entries.
    Applied to the adversarial toy model from Fig. \ref{fig:adversarial}, Eq. \ref{eq:G-hat-from-inversion} fails to parameterize any valid TCL propagators at either $N=10^2$ or $N=10^3$.
    }
    \label{fig:benchmark}
\end{figure*}

Though strict physical feasibility of coarse-grained models is desirable even without improved statistical efficiency \cite{trendelkampschroer2015estimation},
maximum-likelihood $\hat{U}$ conditioned on $\hat{\Pi}$ exhibits lower error relative to ground truth $U$ than na\"{i}vely normalized transition counts $\tilde{U}:=C\mathsf{D}_{C^\mathsf{T}\mathbf{1}_{|M|}}^{-1}$ in the finite-data regime.
To compare estimates to ground truth and empirically demonstrate the superior statistical efficiency of $\hat{U}$, we study the toy model from Fig. \ref{fig:schematic} at subasymptotic sample sizes $N<10^4$.
Curves in Fig. \ref{fig:benchmark} show means and confidence intervals from standard errors over 64 independent samples by each of two schemes.
The first sampling scheme supplies a $B=1$ record of $N$ observations;
the second sampling scheme supplies $B=\sqrt{N}$ independent records totaling $N$ observations.
Both schemes use both $N=256$ and $N=1024$.

Stationary-vector estimates $\hat{\Pi}$ converge rapidly under both schemes [Fig. \ref{fig:benchmark}(a)].
Statistical-efficiency comparisons differ between sampling schemes for $\hat{U}$ and $\tilde{U}$ as well as kernels $\hat{K}$ and $\tilde{K}$ respectively computed from those $U$ estimates using the recursion in Eq. \ref{eq:recursion}.
When using single $N$-observation trajectories, the estimation-error spectral norms of constrained $\hat{U}$ and $\hat{K}$ are only slightly lower than those of na\"ive $\tilde{U}$ and $\tilde{K}$ at larger lags [left panels of Fig. \ref{fig:benchmark}(b)].
However, when using $\sqrt{N}$ independent $\sqrt{N}$-observation trajectories, the estimation-error spectral norms of constrained $\hat{U}$ and $\hat{K}$ are clearly and consistently lower than those of na\"ive $\tilde{U}$ and $\tilde{K}$ [right panels of Fig. \ref{fig:benchmark}(b)], especially for $U$ estimates.
We also note that estimation-error spectral norms of constrained $\hat{G}^{(n)}$ are generally lower than those of matched-data na\"ive estimates for the independent replicates from Fig. \ref{fig:schematic}(j);
as expected, the na\"ive estimator $\tilde{G}^{(n)}:=\tilde{U}^{(n)}(\tilde{U}^{(n-1)})^{-1}$ given in Eq. \ref{eq:G-hat-from-inversion} often fails to parameterize valid column-stochastic propagators with nonnegative entries [Fig. \ref{fig:benchmark}(c)].
The discrepancy between Alg. \ref{alg:MDG} and Eq. \ref{eq:G-hat-from-inversion} is even greater for the adversarial toy example from Fig. \ref{fig:adversarial};
Eq. \ref{eq:G-hat-from-inversion} fails to parameterize any valid TCL propagators at lag $n=3$ for either sample size $N=10^2$ or $N=10^3$. 

Sampling schemes with multiple short records are representative of both molecular-dynamics simulations \cite{perez2016assessment} and single-molecule imaging experiments.
Indeed, allowing record length to grow with record number as $\sqrt{N}$ may understate the dispersion of observations over records characteristic of real experiments.
Single-molecule fluorescence assays \textit{in vivo}, for example, typically combine thousands of records no longer than 20 frames each \cite{elf2019single}.
Though the statistical-efficiency improvement of $\hat{K}$ relative to $\tilde{K}$ may seem slight even under multiple-record sampling,
observe that integrating Eq. \ref{eq:nz-gme-dt} to time $n\tau$ for any initial condition using $\tilde{K}$ recovers a linear combination of $\tilde{U}^{(n)}$ columns by self-consistency.
Errors in $K$ estimates may accumulate over lags as physically spurious fluctuations during NZ-GME-DT integration.
The statistical-efficiency improvement of $\hat{U}$ relative to $\tilde{U}$ better reflects the practical consequences of imposing $\hat{\Pi}$.
For a modest computational cost---all 1280 $\hat{U}$ parameterizations shown in Fig. \ref{fig:benchmark}(b) completed in five minutes total on one CPU core---\texttt{MirrorDescentU}
(Alg. \ref{alg:MDU}) avoids qualitative flaws of na\"ive normalization while noticeably decreasing error on finite data.

We also show that $\hat{\Phi}:=\hat{U}\mathsf{D}_{\hat{\Pi}}$ preserves or improves on the $\mathcal{O}_\mathbbm{P}(N^{-1/2})$ convergence of $\tilde{\Phi}:=\tilde{U}\mathsf{D}_{\tilde{\Phi}^\mathsf{T}\mathbf{1}_{|M|}}=(\mathbf{1}^\mathsf{T}C\mathbf{1})^{-1}C$ at each fixed lag.
This result is a consequence of the linearity of the constraints in Eqs. \ref{eq:flux_rows}-\ref{eq:flux_cols} and the strict convexity of the loss in Eq. \ref{eq:mirrordescentu_loss}.
Recall that for stationary processes with finite integrated autocorrelation time \cite{sokal1997mc}
\begin{equation}
    \|\hat{\Pi}-\Pi\|=\mathcal{O}_\mathbbm{P}(N^{-1/2})
\end{equation}
\begin{equation}
    \|\tilde{\Phi}-\Phi\| = \mathcal{O}_\mathbbm{P}(N^{-1/2})
\end{equation}
and estimation errors of $\hat{\Pi}$ and $\tilde{\Phi}$ are bounded in probability by $N^{-1/2}$ to constants \cite{fonollosa1995sample,wu2011asymptotic}.
For $\hat{\Phi}$ to preserve or improve on the $\mathcal{O}_\mathbbm{P}(N^{-1/2})$ asymptotic convergence of $\tilde{\Phi}$ at each fixed lag, it suffices that for some constant $c$
\begin{equation}
    \|\hat{\phi}-\phi\|\leq c(\|\tilde{\phi}-\phi\| + \|\hat{\Pi}-\Pi\|)
    \label{eq:convergence-exponent-sufficient-condition}
\end{equation}
where $\hat{\phi}:=\mathrm{vec}(\hat{\Phi})$ and $\tilde{\phi}:=\mathrm{vec}(\tilde{\Phi})$.

We proceed by proving that $\hat{\phi}$ (equivalently $\hat{\Phi}$) is a continuously differentiable function of $\hat{\Pi}$ and $\tilde{\phi}$ on an open neighborhood of ground-truth stationary vector $\Pi$ and $\phi:=\mathrm{vec}(\Phi)$.
For $\phi':=\mathrm{vec}(\Phi')$, the linear marginal constraints in Eqs. \ref{eq:flux_rows} and \ref{eq:flux_cols} may be written as
\begin{equation*}
    \begin{pmatrix}\mathbf{1}_{|M|}^\mathsf{T}\otimes I\\ I\otimes\mathbf{1}_{|M|}^\mathsf{T}\end{pmatrix}\phi' = \begin{pmatrix}\hat{\Pi}\\\hat{\Pi}\end{pmatrix}
\end{equation*}
or, upon a suitable orthogonalization, the reduced system $\mathsf{Q}\phi'=\mathsf{b}(\hat{\Pi})$ where $\mathsf{Q}$ has full row rank by construction.
Observe that the loss $f_U$ from Eq. \ref{eq:mirrordescentu_loss} equals
\begin{equation*}
    f_\phi (\phi'; \tilde{\phi}|\hat{\Pi}) := -\sum_{k}^{|M|^2} \tilde{\phi}_k\ln \phi_k'
\end{equation*}
to a constant.
We thus construct the Lagrangian
\begin{equation}
    \mathsf{L}(\phi',\Lambda;\hat{\Pi},\tilde{\phi}) := \Lambda^\mathsf{T}[\mathsf{Q}\phi'-\mathsf{b}(\hat{\Pi})]-\sum_{k}^{|M|^2} \tilde{\phi}_k\ln \phi_k'
\end{equation}
with gradient
\begin{equation}
    \mathscr{R}(\hat{\Pi},\tilde{\phi},\phi',\Lambda) := \nabla_{(\phi',\Lambda)}\mathsf{L} = \begin{pmatrix}\mathsf{Q}^\mathsf{T}\Lambda - \tilde{\phi}\oslash\phi'\\\mathsf{Q}\phi'-\mathsf{b}(\hat{\Pi})\end{pmatrix}
\end{equation}
itself continuously differentiable where $\phi'_k>0$ for all $k\in\{1,2,\dots,|M|^2\}$.
This constraint from Eq. \ref{eq:flux_nn} is physically justified:
no ground-truth $\Phi_{ij}$ can vanish at positive lags for ergodic finite-state microscopic Markov jump processes (MJPs), which we assume.

$\mathscr{R}(\hat{\Pi},\tilde{\phi},\phi',\Lambda)$ is suggestively a function of data $(\hat{\Pi},\tilde{\phi})$ and parameters $(\phi',\Lambda)$:
we may choose any point where $\mathscr{R}(\hat{\Pi},\tilde{\phi},\phi',\Lambda) = 0$ to apply the implicit function theorem.
Observe that $(\phi',\hat{\Pi})=(\phi,\Pi)$ satisfies the constraint $\mathsf{Q}\phi'-\mathsf{b}(\hat{\Pi})=0$.
Then as $\phi\oslash\phi=\mathbf{1}_{|M|^2}$ is in the row space of $\mathbf{1}_{|M|}^\mathsf{T}\otimes I$,
a subset of the $\mathsf{Q}$ row space by construction,
we may choose $(\phi',\Lambda)=(\phi,\Lambda^*)$ such that
\begin{equation}
    \mathscr{R}(\Pi,\phi,\phi,\Lambda^*) = \begin{pmatrix}\mathsf{Q}^\mathsf{T}\Lambda^* - \phi\oslash\phi\\\mathsf{Q}\phi-\mathsf{b}(\Pi)\end{pmatrix} = 0
\end{equation}
for some $\Lambda^*$.
Now let
\begin{equation}
    \mathrm{D}\mathscr{R}(\hat{\Pi},\tilde{\phi},\phi',\Lambda) = \begin{pmatrix}\mathrm{D}_{(\hat{\Pi},\tilde{\phi})}\mathscr{R}\vert\mathrm{D}_{(\phi',\Lambda)}\mathscr{R}\end{pmatrix}
\end{equation}
be the Jacobian matrix of $\mathscr{R}(\hat{\Pi},\tilde{\phi},\phi',\Lambda)$.
Its right block
\begin{equation}
    \mathrm{D}_{(\phi',\Lambda)}\mathscr{R} = \begin{pmatrix}\mathsf{D}_{\tilde{\phi}\oslash (\phi'\odot\phi')} & \mathsf{Q}^\mathsf{T} \\ \mathsf{Q} & 0\end{pmatrix}
\end{equation}
is invertible at the point $(\hat{\Pi},\tilde{\phi},\phi',\Lambda) = (\Pi,\phi,\phi,\Lambda^*)$ we have chosen.
To see this, note that $\mathsf{D}_{\phi\oslash (\phi\odot\phi)}=\mathsf{D}_{\mathbf{1}_{|M|^2}\oslash\phi}$ has strictly positive diagonal elements and is positive-definite.
Suppose that for some $v$ and $w$
\begin{equation*}
    (\mathrm{D}_{(\phi',\Lambda)}\mathscr{R})(\Pi,\phi,\phi,\Lambda^*)\begin{pmatrix}v\\w\end{pmatrix} = \begin{pmatrix}\mathsf{D}_{\mathbf{1}_{|M|^2}\oslash\phi}v+\mathsf{Q}^\mathsf{T}w\\\mathsf{Q}v\end{pmatrix}
\end{equation*}
vanishes. Then $\mathsf{Q}v = 0$ and
\begin{align}
    \mathsf{D}_{\mathbf{1}_{|M|^2}\oslash\phi}v+\mathsf{Q}^\mathsf{T}w &= 0 \notag\\
    \implies v^\mathsf{T}\mathsf{D}_{\mathbf{1}_{|M|^2}\oslash\phi}v + (\mathsf{Q}v)^\mathsf{T}w &= 0 \notag\\
    \implies v^\mathsf{T}\mathsf{D}_{\mathbf{1}_{|M|^2}\oslash\phi}v &= 0
\end{align}
so that $v=0$ by the positive-definiteness of $\mathsf{D}_{\mathbf{1}_{|M|^2}\oslash\phi}$.
As $\mathsf{Q}$ has full row rank by construction,
\begin{equation}
    \mathsf{D}_{\mathbf{1}_{|M|^2}\oslash\phi}v+\mathsf{Q}^\mathsf{T}w = \mathsf{Q}^\mathsf{T}w = 0
\end{equation}
then implies $w=0$.
The kernel of $\mathrm{D}_{(\phi',\Lambda)}\mathscr{R}$ therefore contains only the origin when evaluated at $(\hat{\Pi},\tilde{\phi},\phi',\Lambda) = (\Pi,\phi,\phi,\Lambda^*)$: it is invertible.
By the implicit function theorem,
there exists a continuously differentiable function $h(\hat{\Pi}, \tilde{\phi}) = (\phi^*,\Lambda)$ on an open neighborhood of $(\Pi,\phi)$ such that $\mathscr{R}(\hat{\Pi},\tilde{\phi},\phi^*,\Lambda) = 0$.
By continuity of $h$ and positivity of $\phi_k$ for all $k\in\{1,2,\dots,|M|^2\}$,
we may consider an open ball of positive radius centered on $(\Pi,\phi)$ that is a subset of the open neighborhood such that $\tilde{\phi}_k>0$ and $\phi^*_k>0$ for all $k\in\{1,2,\dots,|M|^2\}$.

We now identify $\phi^*$ with $\hat{\phi}$.
Observe that for $(\hat{\Pi},\tilde{\phi})$ in the open ball around $(\Pi,\phi)$
\begin{equation}
    \mathscr{R}(\hat{\Pi},\tilde{\phi},\phi^*,\Lambda) = 0
    \implies \mathsf{Q}^\mathsf{T}\Lambda=\tilde{\phi}\oslash\phi^*
\end{equation}
and, moreover, $\nabla_{\phi'}f_\phi=-\tilde{\phi}\oslash\phi'$ evaluates to $-\tilde{\phi}\oslash\phi^*=-\mathsf{Q}^\mathsf{T}\Lambda$ at $\phi'=\phi^*$.
Suppose some $\hat{\phi}$ minimizes $f_\phi$ over the feasible set satisfying marginal constraints $\mathsf{Q}\hat{\phi}=\mathsf{b}(\hat{\Pi})$ and orthant constraints $\hat{\phi}_k>0$ for all $k\in\{1,2,\dots,|M|^2\}$.
Then by convexity of $f_\phi$
\begin{align}
    f_\phi (\hat{\phi}) &\geq f_\phi (\phi^*)+[(\nabla_{\phi'}f_\phi)(\phi^*)]^\mathsf{T}(\hat{\phi}-\phi^*) \notag\\
    &=f_\phi (\phi^*)-\Lambda^\mathsf{T}\mathsf{Q}(\hat{\phi}-\phi^*) \notag\\
    &=f_\phi (\phi^*)-\Lambda^\mathsf{T}[\mathsf{b}(\hat{\Pi})-\mathsf{b}(\hat{\Pi})] \notag\\
    &= f_\phi(\phi^*)
\end{align}
where $\mathsf{Q}\phi^*=\mathsf{b}(\hat{\Pi})$ by $\mathscr{R}(\hat{\Pi},\tilde{\phi},\phi^*,\Lambda)=0$ in the penultimate equality.
As we also suppose $\phi^*_k>0$ for all $k\in\{1,2,\dots,|M|^2\}$ and $f_\phi(\hat{\phi})\leq f_\phi(\phi^*)$,
\begin{equation}
    f_\phi (\hat{\phi}) = f_\phi (\phi^*)\implies \hat{\phi} = \phi^*
\end{equation}
by strict convexity of $f_\phi$ implying a unique minimizer in the convex feasible set containing both $\hat{\phi}$ and $\phi^*$.

Finally, denote by $\|\cdot\|_2$ the spectral norm of $\cdot$ and let $c$ be a constant.
For some closed ball of positive radius concentric with and within the open ball where $\hat{\phi}=\phi^*$ is continuously differentiable with respect to $(\hat{\Pi},\tilde{\phi})$,
\begin{equation}
    \|\mathrm{D}_{(\hat{\Pi},\tilde{\phi})}\hat{\phi}\|_2\leq c
\end{equation}
by the boundedness of continuous derivatives.
As $\hat{\Pi}$ and $\tilde{\phi}$ are asymptotically consistent, both inputs belong to this ball with probability approaching unity as $N\to\infty$.
Upon this event,
\begin{align*}
    \|\hat{\phi}-\phi\| &= \|\hat{\phi}(\hat{\Pi},\tilde{\phi})-\hat{\phi}(\Pi,\phi)\|\\
    &\leq c\|(\hat{\Pi},\tilde{\phi})-(\Pi,\phi)\|\\
    &\leq c(\|\hat{\Pi}-\Pi\|+\|\tilde{\phi}-\phi\|)
\end{align*}
using the triangle inequality to recover Eq. \ref{eq:convergence-exponent-sufficient-condition}.
Thus
\begin{equation}
    \|\hat{\Phi}-\Phi\|=\mathcal{O}_\mathbbm{P}(N^{-1/2})
\end{equation}
and imposing stationary vector $\hat{\Pi}$ at worst preserves the $\mathcal{O}_\mathbbm{P}(N^{-1/2})$ asymptotic-convergence order of $\tilde{\Phi}$.

\bibliography{main.bib}

@article{mori1965transport,
    author = {Mori, Hajime},
    title = {Transport, collective motion, and {B}rownian motion},
    journal = {Prog. Theor. Exp. Phys.},
    year = {1965},
    volume = {33},
    pages = {423},
    doi = {10.1143/PTP.33.423}
}

@article{zwanzig1961memory,
    author = {Zwanzig, Robert},
    title = {Memory effects in irreversible thermodynamics},
    journal = {Phys. Rev.},
    year = {1961},
    volume = {124},
    pages = {983},
    doi = {10.1103/PhysRev.124.983}
}

@article{hummer2014optimal,
    author = {Hummer, Gerhard and Szabo, Attila},
    title = {Optimal dimensionality reduction of multistate kinetic and {M}arkov-state models},
    journal = {J. Phys. Chem. B},
    year = {2014},
    volume = {119},
    pages = {9029},
    doi = {10.1021/jp508375q}
}

@article{gb2024tutorial,
    author = {Gonzalez-{B}allestero, Carlos},
    title = {Tutorial: projector approach to master equations for open quantum systems},
    journal = {Quantum},
    year = {2024},
    volume = {8},
    pages = {1454},
    doi = {10.22331/q-2024-08-29-1454}
}

@article{dyson1949S,
    author = {Dyson, Freeman},
    title = {The {$S$} matrix in quantum electrodynamics},
    journal = {Phys. Rev.},
    year = {1949},
    volume = {75},
    pages = {1736},
    doi = {10.1103/PhysRev.75.1736}
}

@article{schilling2022coarse,
    author = {Schilling, Tanja},
    title = {Coarse-grained modeling out of equilibrium},
    journal = {Phys. Rep.},
    year = {2022},
    volume = {972},
    pages = {1},
    doi = {10.1016/j.physrep.2022.04.006}
}

@article{schilling2024evolution,
    author = {Schilling, Tanja},
    title = {Evolution equations for open systems and collective variables: Which equation would you like to solve by molecular dynamics simulation?},
    journal = {J. Chem. Phys.},
    volume = {161},
    number = {19},
    pages = {191501},
    year = {2024},
    month = {11},
    doi = {10.1063/5.0239400}
}

@article{cao2020advantages,
    author = {Cao, Siqin and Montoya-Castillo, Andr\'{e}s and Wang, Wei and Markland, Thomas E and Huang, Xuhui},
    title = {On the advantages of exploiting memory in {M}arkov state models for biomolecular dynamics},
    journal = {J. Chem. Phys.},
    year = {2020},
    volume = {153},
    pages = {014105},
    doi = {10.1063/5.0010787}
}

@article{nagel2023toward,
    author = {Nagel, Daniel and Sartore, Sofia and Stock, Gerhard},
    title = {Toward a benchmark for {M}arkov state models: the folding of {HP}35},
    journal = {J. Phys. Chem. Lett.},
    year = {2023},
    volume = {14},
    pages = {6956},
    doi = {10.1021/acs.jpclett.3c01561}
}

@article{vanvliet2006equilibration,
    title = {Equilibration of experimentally determined protein structures for molecular dynamics simulation},
    author = {Walton, Emily B. and VanVliet, Krystyn J.},
    journal = {Phys. Rev. E},
    volume = {74},
    pages = {061901},
    year = {2006},
    doi = {10.1103/PhysRevE.74.061901}
}

@article{stella1998equilibration,
    title = {Equilibration and sampling in molecular dynamics simulations of biomolecules},
    author = {Stella, Lorenzo and Melchionna, Simone},
    journal = {J. Chem. Phys.},
    volume = {109},
    pages = {10115},
    numpages = {8},
    year = {1998},
    doi = {10.1063/1.477703}
}

@article{gasbarri2018recursive,
    author = {Gasbarri, Giulio and Ferialdi, L},
    title = {Recursive approach for non-{M}arkovian time-convolutionless master equations},
    journal = {Phys. Rev. A},
    year = {2018},
    volume = {97},
    pages = {022114},
    doi = {10.1103/PhysRevA.97.022114}
}

@article{gu2020diagrammatic,
    author = {Gu, Bing},
    title = {Diagrammatic time-local master equation for open quantum systems},
    journal = {Phys. Rev. A},
    year = {2020},
    volume = {101},
    pages = {012121},
    doi = {10.1103/PhysRevA.101.012121}
}

@article{husic2018markov,
    author = {Husic, Brooke E and Pande, Vijay S},
    title = {Markov state models: from an art to a science},
    journal = {J. Am. Chem. Soc.},
    year = {2018},
    volume = {140},
    pages = {2386},
    doi = {10.1021/jacs.7b12191}
}

@article{pollock2018tomograph,
    author = {Pollock, Felix A and Modi, Kavan},
    title = {Tomographically reconstructed master equations for any open quantum dynamics},
    journal = {Quantum},
    year = {2018},
    volume = {2},
    pages = {76},
    doi = {10.22331/q-2018-07-11-76}
}

@article{gu2024diagrammatic,
    title = {Diagrammatic representation and nonperturbative approximations of the exact time-convolutionless master equation},
    author = {Gu, Bing},
    journal = {J. Chem. Phys.},
    volume = {160},
    pages = {204113},
    year = {2024},
    doi = {10.1063/5.0187191}
}

@article{buser2017initial,
    title = {Initial system-environment correlations via the transfer-tensor method},
    author = {Buser, Maximilian and Cerrillo, Javier and Schaller, Gernot and Cao, Jianshu},
    journal = {Phys. Rev. A},
    volume = {96},
    pages = {062122},
    year = {2017},
    doi = {10.1103/PhysRevA.96.062122}
}

@article{makri2025discrete,
    title = {Discrete generalized quantum master equations},
    author = {Makri, Nancy},
    journal = {J. Chem. Theory Comput.},
    volume = {21},
    pages = {5037},
    year = {2025},
    doi = {10.1021/acs.jctc.5c00396}
}

@article{nakajima1958quantum,
    title = {On quantum theory of transport phenomena: steady diffusion},
    author = {Nakajima, Sadao},
    journal = {Prog. Theor. Exp. Phys.},
    volume = {20},
    pages = {948},
    year = {1958},
    doi = {10.1143/PTP.20.948}
}

@article{zwanzig1960ensemble,
    title = {Ensemble method in the theory of irreversibility},
    author = {Zwanzig, Robert},
    journal = {J. Chem. Phys.},
    volume = {33},
    pages = {1338},
    year = {1960},
    doi = {10.1063/1.1731409}
}

@article{callen1951irreversibility,
    author = {Callen, Herbert B and Welton, Theodore A},
    title = {Irreversibility and generalized noise},
    journal = {Phys. Rev.},
    year = {1951},
    volume = {83},
    pages = {34},
    doi = {10.1103/PhysRev.83.34}
}

@article{kubo1956fdt,
    author = {Kubo, Ryugo},
    title = {The fluctuation-dissipation theorem},
    journal = {Rep. Prog. Phys.},
    year = {1966},
    volume = {29},
    pages = {255},
    doi = {10.1088/0034-4885/29/1/306}
}

@article{langevin1908,
    author = {Langevin, Paul},
    title = {Sur la th\'{e}orie du mouvement brownien},
    journal = {C. R. Acad. Sci.},
    year = {1908},
    volume = {146},
    pages = {530},
    url = {https://gallica.bnf.fr/ark:/12148/bpt6k3100t}
}

@article{dominic2023building,
    author = {Dominic {III}, Anthony J. and Sayer, Thomas and Cao, Siqin and Markland, Thomas E and Huang, Xuhui and Montoya-{C}astillo, Andr\'{e}s},
    title = {Building insightful, memory-enriched models to capture long-time biochemical processes from short-time simulations},
    journal = {Proc. Natl. Acad. Sci. U.S.A.},
    year = {2023},
    volume = {120},
    pages = {e2221048120},
    doi = {10.1073/pnas.2221048120}
}

@article{chodera2015markov,
    author = {Chodera, John D and No\'{e}, Frank},
    title = {Markov state models of biomolecular conformational dynamics},
    journal = {Curr. Opin. Struct. Biol.},
    year = {2015},
    volume = {25},
    pages = {135},
    doi = {10.1016/j.sbi.2014.04.002}
}

@article{trendelkampschroer2015estimation,
    author = {Trendelkamp-{S}chroer, Benjamin and Wu, Hao and Paul, Fabian and No\'{e}, Frank},
    title = {Estimation and uncertainty of reversible {M}arkov models},
    journal = {J. Chem. Phys.},
    year = {2015},
    volume = {143},
    pages = {174101},
    doi = {10.1063/1.4934536}
}

@book{nemirovsky1979problem,
    title = {Slozhnost' zadach i effektivnost' metodov optimizacii},
    publisher = {Nauka},
    address = {Moscow, Russian Soviet Federative Socialist Republic, Union of Soviet Socialist Republics},
    isbn = {9780471103455},
    year = {1979},
    author = {Nemirovski, Arkadi Semenovich and Yudin, David Berkovich}
}

@article{bregman1967relax,
    author = {Bregman, Lev M},
    title = {A relaxation method of finding a common point of convex sets and its application to the solution of problems in convex programming},
    journal = {U.S.S.R. Comput. Math. Math. Phys.},
    year = {1967},
    volume = {7},
    pages = {200},
    doi = {10.1016/0041-5553(67)90040-7}
}

@book{beck2017first,
    title = {First-Order Methods in Optimization},
    publisher = {Society of Industrial and Applied Mathematics},
    address = {Philadelphia, Pennsylvania, United States of America},
    doi = {10.1137/1.9781611974997},
    series={{MOS-SIAM Series on Optimization}},
    year={2017},
    author={Beck, Amir}
}

@book{vishnoi2021algorithms,
    title = {Algorithms for Convex Optimization},
    publisher = {Cambridge University Press},
    address = {Cambridge, United Kingdom of Great Britain and Northern Ireland},
    isbn = {9781108699211},
    year = {2021},
    doi = {10.1017/9781108699211},
    author = {Vishnoi, Nisheeth K}
}

@article{li2019convergence,
    author = {Li, Yen-Huan and Cevher, Volkan},
    title = {Convergence of the exponentiated gradient method with {A}rmijo line search},
    journal = {J. Optim. Theory Appl.},
    year = {2019},
    volume = {181},
    pages = {588},
    doi = {10.1007/s10957-018-1428-9}
}

@article{armijo1966minimization,
    author = {Armijo, Larry},
    title = {Minimization of functions having {L}ipschitz continuous first partial derivatives},
    journal = {Pacific J. Math.},
    year = {1966},
    volume = {16},
    pages = {1},
    doi = {10.2140/pjm.1966.16.1}
}

@article{kld,
    author = {Kullback, Solomon and Leibler, Richard A},
    title = {On information and sufficiency},
    journal = {Ann. Math. Stat.},
    year = {1951},
    volume = {22},
    pages = {79},
    doi = {10.1214/aoms/1177729694}
}

@article{knopp1967concerning,
    author = {Knopp, Paul and Sinkhorn, Richard},
    title = {Concerning nonnegative matrices and doubly stochastic matrices},
    journal = {Pac. J. Math.},
    year = {1967},
    volume = {21},
    pages = {343},
    doi = {10.2140/pjm.1967.21.343}
}

@article{iproject,
    author = {Nielsen, Frank},
    title = {What is... an information projection?},
    journal = {Not. Am. Math. Soc.},
    year = {2018},
    volume = {65},
    pages = {321},
    doi = {10.1090/noti1647}
}

@article{knight2014symmetry,
    author = {Knight, Philip A and Ruiz, Daniel and U\c{c}ar, Bora},
    title = {A symmetry preserving algorithm for matrix scaling},
    journal = {{SIAM} J. Matrix Anal. Appl.},
    year = {2014},
    volume = {35},
    pages = {25},
    doi = {10.1137/110825753}
}

@inproceedings{cuturi2013sinkhorn,
    author = {Cuturi, Marco},
    title = {Sinkhorn distances: lightspeed computation of optimal transport},
    year = {2013},
    publisher = {Curran Associates Inc.},
    address = {Red Hook, New York, United States of America},
    booktitle = {Proceedings of the 27th International Conference on Neural Information Processing Systems - Volume 2},
    pages = {2292--2300},
    numpages = {9},
    location = {Lake Tahoe, Nevada},
    series = {NIPS'13},
    doi = {10.48550/arXiv.1306.0895}
}

@article{sayer2023compact,
    author = {Sayer, Thomas and Montoya-Castillo, Andr\'{e}s},
    title = {Compact and complete description of non-{M}arkovian dynamics},
    journal = {J. Chem. Phys.},
    year = {2023},
    volume = {158},
    pages = {014105},
    doi = {10.1063/5.0132614}
}

@article{jensen1906inequality,
    author = {Jensen, Johan Ludwig William Valdemar},
    title = {Sur les fonctions convexes et les in\'{e}galit\'{e}s entre les valeurs moyennes},
    journal = {Acta Math.},
    year = {1906},
    volume = {30},
    pages = {175},
    doi = {10.1007/2FBF02418571}
}

@article{Frobenius1878,
    author  = {Frobenius, G.},
    title   = {{Ueber lineare Substitutionen und bilineare Formen}},
    journal = {J. Reine Angew. Math.},
    volume  = {84},
    pages   = {1--63},
    year    = {1878},
    doi     = {10.1515/crll.1878.84.1}
}

@article{prinz2011markov,
    author = {Prinz, Jan-{H}endrik and Wu, Hao and Sarich, Marco and Keller, Bettina and Senne, Martin and Held, Martin and Chodera, John D and Sch\"{u}tte, Christof and No\'{e}, Frank},
    title = {Markov models of molecular kinetics: generation and validation},
    journal = {J. Chem. Phys.},
    year = {2011},
    volume = {134},
    pages = {174105},
    doi = {10.1063/1.3565032}
}

@article{strasberg2019nonmarkovianity,
    author = {Strasberg, Philipp and Esposito, Massimiliano},
    title = {Non-{M}arkovianity and negative entropy production rates},
    journal = {Phys. Rev. E},
    year = {2019},
    volume = {99},
    pages = {012120},
    doi = {10.1103/PhysRevE.99.012120}
}

@article{maldonado2012investigating,
    title = {Investigating the generality of time-local master equations},
    author = {Maldonado-Mundo, Daniel and \"Ohberg, Patrik and Lovett, Brendon W. and Andersson, Erika},
    journal = {Phys. Rev. A},
    volume = {86},
    issue = {4},
    pages = {042107},
    numpages = {6},
    year = {2012},
    month = {Oct},
    publisher = {American Physical Society},
    doi = {10.1103/PhysRevA.86.042107}
}

@article{tokuyama1975statistical,
    title = {Statistical-mechanical approach to random frequency modulations and the {G}aussian memory function},
    author = {Tokuyama, Michio and Mori, Hazime},
    journal = {Prog. Theor. Exp. Phys.},
    volume = {54},
    issue = {3},
    pages = {918},
    numpages = {3},
    year = {1975},
    doi = {10.1143/PTP.54.918}
}

@article{tokuyama1976statistical,
    title = {Statistical-mechanical theory of random frequency modulations and generalized {B}rownian motions},
    author = {Tokuyama, Michio and Mori, Hazime},
    journal = {Prog. Theor. Exp. Phys.},
    volume = {55},
    issue = {2},
    pages = {411},
    numpages = {19},
    year = {1976},
    doi = {10.1143/PTP.55.411}
}

@article{chaturvedi1979timeconvolutionless,
    title = {Time-convolutionless projection operator formalism for elimination of fast variables. {A}pplications to {B}rownian motion},
    author = {Chaturvedi, S and Shibata, F},
    journal = {Z. Phys. B},
    volume = {35},
    pages = {297},
    numpages = {12},
    year = {1979},
    doi = {10.1007/BF01319852}
}

@article{brandner2025dynamics,
    title = {Dynamics of microscale and nanoscale systems in the weak-memory regime: A mathematical framework beyond the {M}arkov approximation},
    author = {Brandner, Kay},
    journal = {Phys. Rev. E},
    volume = {111},
    issue = {1},
    pages = {014137},
    numpages = {23},
    year = {2025},
    month = {Jan},
    publisher = {American Physical Society},
    doi = {10.1103/PhysRevE.111.014137}
}

@article{laine2012local,
    title = {Local-in-time master equations with memory effects: applicability and interpretation},
    author = {Laine, E M and Luoma, K and Piilo, J},
    journal = {J. Phys. B},
    volume = {45},
    pages = {154004},
    year = {2012},
    month = {Jul},
    doi = {10.1088/0953-4075/45/15/154004}
}

@article{chapman1928brownian,
    author  = {Chapman, Sydney},
    title   = {On the {B}rownian displacements and thermal diffusion of grains suspended in a non-uniform fluid},
    journal = {Proceedings of the Royal Society of London. Series A, Containing Papers of a Mathematical and Physical Character},
    year    = {1928},
    volume  = {119},
    number  = {781},
    pages   = {34--54},
    doi     = {10.1098/rspa.1928.0082}
}

@article{kolmogoroff1931ueber,
    author  = {Kolmogoroff, A.},
    title   = {{\"U}ber die analytischen {M}ethoden in der {W}ahrscheinlichkeitsrechnung},
    journal = {Mathematische Annalen},
    year    = {1931},
    volume  = {104},
    pages   = {415--458},
    doi     = {10.1007/BF01457949}
}

@article{shabani2005completely,
    title = {Completely positive post-{M}arkovian master equation via a measurement approach},
    author = {Shabani, A. and Lidar, D. A.},
    journal = {Phys. Rev. A},
    volume = {71},
    issue = {2},
    pages = {020101},
    numpages = {4},
    year = {2005},
    month = {Feb},
    publisher = {American Physical Society},
    doi = {10.1103/PhysRevA.71.020101}
}

@article{matthew2023full,
    title = {Full counting statistics and coherences: Fluctuation symmetry in heat transport with the unified quantum master equation},
    author = {Gerry, Matthew and Segal, Dvira},
    journal = {Phys. Rev. E},
    volume = {107},
    issue = {5},
    pages = {054115},
    numpages = {14},
    year = {2023},
    month = {May},
    publisher = {American Physical Society},
    doi = {10.1103/PhysRevE.107.054115},
}

@article{nestmann2021quantum,
    title = {How Quantum Evolution with Memory is Generated in a Time-Local Way},
    author = {Nestmann, K. and Bruch, V. and Wegewijs, M. R.},
    journal = {Phys. Rev. X},
    volume = {11},
    issue = {2},
    pages = {021041},
    numpages = {22},
    year = {2021},
    month = {May},
    publisher = {American Physical Society},
    doi = {10.1103/PhysRevX.11.021041}
}

@article{modi2021quantum,
    title = {Quantum Stochastic Processes and Quantum non-{M}arkovian Phenomena},
    author = {Milz, Simon and Modi, Kavan},
    journal = {PRX Quantum},
    volume = {2},
    issue = {3},
    pages = {030201},
    numpages = {81},
    year = {2021},
    month = {Jul},
    publisher = {American Physical Society},
    doi = {10.1103/PRXQuantum.2.030201}
}

@article{pomeau1982symmetry,
    title = {Sym\'{e}trie des fluctuations dans le renversement du temps},
    author = {Pomeau, Yves},
    journal = {J. Phys. (Paris)},
    volume = {43},
    pages = {859},
    numpages = {9},
    year = {1982},
    doi = {10.1051/jphys:01982004306085900}
}

@book{feller1991introduction,
    title = {An Introduction to Probability Theory and Its Applications},
    publisher = {Wiley},
    address = {New York, New York, United States of America},
    isbn = {9780471257080},
    edition = {3},
    year = {1991},
    author = {Feller, William}
}

@article{dieball2025time,
    author = {Dieball, Cai and Godec, Alja\v{z}},
    title = {Perspective: time irreversibility in systems observed at coarse resolution},
    journal = {J. Chem. Phys.},
    year = {2025},
    volume = {162},
    pages = {090901},
    doi = {10.1063/5.0251089}
}

@article{ro2022modelfree,
    title = {Model-Free Measurement of Local Entropy Production and Extractable Work in Active Matter},
    author = {Ro, Sunghan and Guo, Buming and Shih, Aaron and Phan, Trung V. and Austin, Robert H. and Levine, Dov and Chaikin, Paul M. and Martiniani, Stefano},
    journal = {Phys. Rev. Lett.},
    volume = {129},
    issue = {22},
    pages = {220601},
    numpages = {6},
    year = {2022},
    doi = {10.1103/PhysRevLett.129.220601}
}

@article{lynn2022emergence,
    title = {Emergence of local irreversibility in complex interacting systems},
    author = {Lynn, Christopher W. and Holmes, Caroline M. and Bialek, William and Schwab, David J.},
    journal = {Phys. Rev. E},
    volume = {106},
    issue = {3},
    pages = {034102},
    numpages = {14},
    year = {2022},
    doi = {10.1103/PhysRevE.106.034102}
}

@article{ge2006reversibility,
    author = {Ge, Hao and Jiang, Da Quan and Qian, Min},
    title = {Reversibility and entropy production of inhomogeneous {M}arkov chains},
    journal = {J. Appl. Prob.},
    volume = {43},
    issue = {4},
    pages = {1028},
    numpages = {16},
    year = {2006},
    doi = {10.1239/jap/1165505205}
}

@article{roldan2012entropy,
    title = {Entropy production and {K}ullback-{L}eibler divergence between stationary trajectories of discrete systems},
    author = {Rold\'{a}n, \'{E}dgar and Parrondo, Juan M. R.},
    journal = {Phys. Rev. E},
    volume = {85},
    pages = {031129},
    numpages = {12},
    year = {2012},
    doi = {10.1103/PhysRevE.85.031129}
}

@article{ehrich2021tightest,
    title = {Tightest bound on hidden entropy production from partially observed dynamics},
    author = {Ehrich, Jannik},
    journal = {J. Stat. Mech.},
    volume = {2021},
    pages = {083214},
    year = {2021},
    doi = {10.1088/1742-5468/ac150e}
}

@article{benamou2015iterative,
    title = {Iterative {B}regman projections for regularized transportation problems},
    author = {Benamou, Jean-{D}avid and Carlier, Guillaume and Cuturi, Marco and Nenna, Luca and Peyr\'{e}, Gabriel},
    journal = {{SIAM} J. Sci. Comput.},
    volume = {37},
    pages = {A1111--A1138},
    year = {2015},
    doi = {10.1137/141000439}
}

@article{levenberg1944method,
    title = {A method for the solution of certain non-linear problems in least squares},
    author = {Levenberg, Kenneth},
    journal = {Quart. Appl. Math.},
    volume = {2},
    pages = {164--168},
    year = {1944},
    doi = {10.1090/qam/10666}
}

@article{marquardt1963algorithm,
    title = {An algorithm for least-squares estimation of nonlinear parameters},
    author = {Marquardt, Donald W},
    journal = {SIAM J. Appl. Math.},
    volume = {11},
    pages = {431--441},
    year = {1963},
    doi = {10.1137/0111030}
}

@article{zhang2003convergence,
    title = {On the convergence properties of the {L}evenberg-{M}arquardt method},
    author = {Zhang, Ju-Liang},
    journal = {Optimization},
    volume = {52},
    pages = {739--756},
    year = {2003},
    doi = {10.1080/0233193031000163993}
}

@inproceedings{yamashita2001rate,
    author = {Yamashita, Nobuo and Fukushima, Masao},
    title = {On the rate of convergence of the {L}evenberg-{M}arquardt method},
    year = {2001},
    publisher = {Springer},
    address = {Vienna, Republic of Austria},
    booktitle = {Topics in Numerical Analysis},
    series = {Computing Supplementa},
    pages = {239--249},
    doi = {10.1007/978-3-7091-6217-0_18}
}

@article{hartich2024comment,
    title = {Comment on ``{I}nferring broken detailed balance in the absence of observable currents''},
    author = {Hartich, David and Godec, Alja\v{z}},
    journal = {Nat. Commun.},
    volume = {15},
    pages = {8678},
    year = {2024},
    doi = {10.1038/s41467-024-52602-0}
}

@article{bisker2024reply,
    title = {Reply to: {C}omment on ``{I}nferring broken detailed balance in the absence of observable currents''},
    author = {Bisker, Gili and Mart\'{i}nez, Ignacio A and Horowitz, Jordan M and Parrondo, Juan M R},
    journal = {Nat. Commun.},
    volume = {15},
    pages = {8679},
    year = {2024},
    doi = {10.1038/s41467-024-52603-z}
}

@article{chruscinski2010nonmarkovian,
    title = {Non-{M}arkovian Quantum Dynamics: Local versus Nonlocal},
    author = {Chru\ifmmode \acute{s}\else \'{s}\fi{}ci\ifmmode \acute{n}\else \'{n}\fi{}ski, Dariusz and Kossakowski, Andrzej},
    journal = {Phys. Rev. Lett.},
    volume = {104},
    issue = {7},
    pages = {070406},
    year = {2010},
    doi = {10.1103/PhysRevLett.104.070406}
}

@article{hou2012singularity,
    title = {Singularity of dynamical maps},
    author = {Hou, S. C. and Yi, X. X. and Yu, S. X. and Oh, C. H.},
    journal = {Phys. Rev. A},
    volume = {86},
    pages = {012101},
    year = {2012},
    doi = {10.1103/PhysRevA.86.012101}
}

@article{dominic2023memory,
    author = {Dominic {III}, Anthony J. and Cao, Siqin and Montoya-Castillo, Andr{\'e}s and Huang, Xuhui},
    title = {Memory Unlocks the Future of Biomolecular Dynamics: Transformative Tools to Uncover Physical Insights Accurately and Efficiently},
    journal = {Journal of the American Chemical Society},
    volume = {145},
    number = {18},
    pages = {9916--9927},
    year = {2023},
    doi = {10.1021/jacs.3c01095},
}

@article{sinclair2025influence,
    title = {Influence of solute induced memory on interface migration},
    author = {Sinclair, Chad W. and Rottler, Joerg},
    journal = {Phys. Rev. Mater.},
    volume = {9},
    pages = {123402},
    year = {2025},
    doi = {10.1103/c4jh-g8fr},
}

@article{bement2025models,
    author = {Bement, Phillip and Rottler, Jörg},
    title = {Models for polymer dynamics from dimensionality reduction techniques},
    journal = {J. Chem. Phys.},
    volume = {163},
    pages = {104902},
    year = {2025},
    doi = {10.1063/5.0289397}
}

@article{yue2024tutorials,
    author = {Wu, Yue and Cao, Siqin and Qiu, Yunrui and Huang, Xuhui},
    title = {Tutorial on how to build non-{M}arkovian dynamic models from molecular dynamics simulations for studying protein conformational changes},
    journal = {J. Chem. Phys.},
    volume = {160},
    pages = {121501},
    year = {2024},
    doi = {10.1063/5.0189429}
}

@misc{schwarz2025consistent,
      title={Consistent time reversal and reliable and accurate inference in the presence of memory}, 
      author={Tassilo Schwarz and Anatoly B. Kolomeisky and Aljaž Godec},
      year={2025},
      eprint={2410.11819},
      archivePrefix={arXiv},
      primaryClass={cond-mat.stat-mech}
}

@article{cerrillo2014nonmarkovian,
    title = {Non-{M}arkovian Dynamical Maps: Numerical Processing of Open Quantum Trajectories},
    author = {Cerrillo, Javier and Cao, Jianshu},
    journal = {Phys. Rev. Lett.},
    volume = {112},
    pages = {110401},
    year = {2014},
    doi = {10.1103/PhysRevLett.112.110401}
}

@article{timm20115tcl,
    title = {Time-convolutionless master equation for quantum dots: Perturbative expansion to arbitrary order},
    author = {Timm, Carsten},
    journal = {Phys. Rev. B},
    volume = {83},
    pages = {115416},
    year = {2011},
    doi = {10.1103/PhysRevB.83.115416}
}

@article{kidon2015exact,
    author = {Kidon, Lyran and Wilner, Eli Y. and Rabani, Eran},
    title = {Exact calculation of the time convolutionless master equation generator: Application to the nonequilibrium resonant level model},
    journal = {J. Chem. Phys.},
    volume = {143},
    pages = {234110},
    year = {2015},
    doi = {10.1063/1.4937396}
}

@article{espanol2002coarse,
    author = {Espa\~{n}ol, Pep and V\'{a}zquez, Federico},
    title = {Coarse-graining from coarse-grained descriptions},
    journal = {Phil. Trans. R. Soc. A},
    volume = {360},
    pages = {383},
    year = {2002},
    doi = {10.1098/rsta.2001.0935}
}

@article{cohen2011memory,
    title = {Memory effects in nonequilibrium quantum impurity models},
    author = {Cohen, Guy and Rabani, Eran},
    journal = {Phys. Rev. B},
    volume = {84},
    pages = {075150},
    year = {2011},
    doi = {10.1103/PhysRevB.84.075150}
}

@article{tian2022explicit,
    author = {Tian, Xiaofei and Xu, Xiaolei and Chen, Ye and Chen, Jizhong and Xu, Wen-Sheng},
    title = {Explicit analytical form for memory kernel in the generalized {L}angevin equation for end-to-end vector of {R}ouse chains},
    journal = {J. Chem. Phys.},
    volume = {157},
    pages = {224901},
    year = {2022},
    doi = {10.1063/5.0124925}
}

@article{netz2024derivation,
    title = {Derivation of the nonequilibrium generalized {L}angevin equation from a time-dependent many-body {H}amiltonian},
    author = {Netz, Roland R.},
    journal = {Phys. Rev. E},
    volume = {110},
    pages = {014123},
    year = {2024},
    doi = {10.1103/PhysRevE.110.014123}
}

@article{vroylandt2022likelihood,
    author = {Vroylandt, Hadrien and Goudenège, Ludovic and Monmarch\'{e}, Pierre and Pietrucci, Fabio and Rotenberg, Benjamin},
    title = {Likelihood-based non-{M}arkovian models from molecular dynamics},
    journal = {Proc. Natl. Acad. Sci. U.S.A.},
    volume = {119},
    pages = {e2117586119},
    year = {2022},
    doi = {10.1073/pnas.2117586119}
}

@article{qiang2003new,
    author = {Shi, Qiang and Geva, Eitan},
    title = {A new approach to calculating the memory kernel of the generalized quantum master equation for an arbitrary system–bath coupling},
    journal = {J. Chem. Phys.},
    volume = {119},
    pages = {12063},
    year = {2003},
    doi = {10.1063/1.1624830}
}

@article{mulvihill2021roadmap,
    author = {Mulvihill, Ellen and Geva, Eitan},
    title = {A Road Map to Various Pathways for Calculating the Memory Kernel of the Generalized Quantum Master Equation},
    journal = {J. Phys. Chem. B},
    volume = {125},
    pages = {9834},
    year = {2021},
    doi = {10.1021/acs.jpcb.1c05719}
}

@book{goodfellow2016deep,
    title={Deep Learning},
    author={Goodfellow, Ian and Bengio, Yoshua and Courville, Aaron},
    publisher={MIT Press},
    address={Cambridge, Massachusetts, United States of America},
    url={http://www.deeplearningbook.org},
    year={2016}
}

@article{levring2023CFTR,
    author = {Levring, Jesper and Terry, Daniel S and Kilic, Zeliha and Fitzgerald, Gabriel and Blanchard, Scott C and Chen, Jue},
    title = {{CFTR} function, pathology, and pharmacology at single-molecule resolution},
    journal = {Nature},
    volume = {616},
    pages = {606},
    year = {2023},
    doi = {10.1038/s41586-023-05854-7}
}

@article{noe2020machine,
    title = {Machine learning for protein folding and dynamics},
    journal = {Curr. Opin. Struct. Biol.},
    volume = {60},
    pages = {77--84},
    year = {2020},
    doi = {10.1016/j.sbi.2019.12.005},
    author = {No\'{e}, Frank and {De Fabritiis}, Gianni and Clementi, Cecilia}
}

@article{noe2015kinetic,
    author = {No{\'e}, Frank and Clementi, Cecilia},
    title = {Kinetic Distance and Kinetic Maps from Molecular Dynamics Simulation},
    journal = {J. Chem. Theory Comput.},
    volume = {11},
    number = {10},
    pages = {5002--5011},
    year = {2015},
    doi = {10.1021/acs.jctc.5b00553}
}

@article{beauchamp2012simple,
    author = {Beauchamp, Kyle A and McGibbon, Robert and Lin, Yu-Shan and Pande, Vijay S},
    title = {Simple few-state models reveal hidden complexity in protein folding},
    journal = {Proc. Natl. Acad. Sci. U.S.A.},
    volume = {109},
    number = {44},
    pages = {17807--17813},
    year = {2012},
    doi = {10.1073/pnas.1201810109}
}

@article{sartore2025markov,
    author = {Sartore, Sofia and Teichmann, Franziska and Stock, Gerhard},
    title = {Markov-Type State Models to Describe Non-{M}arkovian Dynamics},
    journal = {J. Chem. Theory Comput.},
    volume = {21},
    number = {5},
    pages = {2757--2765},
    year = {2025},
    doi = {10.1021/acs.jctc.4c01630}
}

@article{bowman2012improved,
    author = {Bowman, Gregory R},
    title = {Improved coarse-graining of {M}arkov state models via explicit consideration of statistical uncertainty},
    journal = {J. Chem. Phys.},
    volume = {137},
    number = {13},
    pages = {134111},
    year = {2012},
    doi = {10.1063/1.4755751}
}

@article{ph2013identification,
    author = {P\'{e}rez-Hern\'{a}ndez, Guillermo and Paul, Fabian and Giorgino, Toni and {De Fabritiis}, Gianni and No\'{e}, Frank},
    title = {Identification of slow molecular order parameters for {M}arkov model construction},
    journal = {J. Chem. Phys.},
    volume = {139},
    number = {1},
    pages = {015102},
    year = {2013},
    doi = {10.1063/1.4811489}
}

@article{schwantes2013improvements,
    author = {Schwantes, Christian R and Pande, Vijay S},
    title = {Improvements in {M}arkov State Model Construction Reveal Many Non-Native Interactions in the Folding of {NTL9}},
    journal = {J. Chem. Theory Comput.},
    volume = {9},
    number = {4},
    pages = {2000--2009},
    year = {2013},
    doi = {10.1021/ct300878a}
}

@article{diez2022correlation,
    author = {Diez, Georg and Nagel, Daniel and Stock, Gerhard},
    title = {Correlation-Based Feature Selection to Identify Functional Dynamics in Proteins},
    journal = {J. Chem. Theory Comput.},
    volume = {18},
    number = {8},
    pages = {5079--5088},
    year = {2022},
    doi = {10.1021/acs.jctc.2c00337}
}

@article{prinz2014spectral,
    title = {Spectral Rate Theory for Two-State Kinetics},
    author = {Prinz, Jan-Hendrik and Chodera, John D. and No\'{e}, Frank},
    journal = {Phys. Rev. X},
    volume = {4},
    issue = {1},
    pages = {011020},
    numpages = {19},
    year = {2014},
    doi = {10.1103/PhysRevX.4.011020}
}

@article{swope2004describing,
    author = {Swope, William C and Pitera, Jed W and Suits, Frank},
    title = {Describing Protein Folding Kinetics by Molecular Dynamics Simulations. 1. {T}heory},
    journal = {The Journal of Physical Chemistry B},
    volume = {108},
    number = {21},
    pages = {6571-6581},
    year = {2004},
    doi = {10.1021/jp037421y}
}

@article{dill1991denatured,
    author = {Dill, Ken A and Shortle, David},
    title = {Denatured states of proteins}, 
    journal= {Annu. Rev. Biochem.},
    year = {1991},
    volume = {60},
    pages = {795--825},
    doi = {10.1146/annurev.bi.60.070191.004051}
}

@article{husic2018minimum,
    author = {Husic, Brooke E and McKiernan, Keri A and Wayment-Steele, Hannah K and Sultan, Mohammad M and Pande, Vijay S},
    title = {A Minimum Variance Clustering Approach Produces Robust and Interpretable Coarse-Grained Models},
    journal = {J. Chem. Theory Comput.},
    volume = {14},
    number = {2},
    pages = {1071--1082},
    year = {2018},
    doi = {10.1021/acs.jctc.7b01004}
}

@article{cao2023igme,
    author = {Cao, Siqin and Qiu, Yunrui and Kalin, Michael L and Huang, Xuhui},
    title = {Integrative generalized master equation: A method to study long-timescale biomolecular dynamics via the integrals of memory kernels},
    journal = {The Journal of Chemical Physics},
    volume = {159},
    number = {13},
    pages = {134106},
    year = {2023},
    doi = {10.1063/5.0167287}
}

@article{goonetilleke2025targeting,
    author = {Goonetilleke, Eshani C and Huang, Xuhui},
    title = {Targeting Bacterial {RNA} Polymerase: Harnessing Simulations and Machine Learning to Design Inhibitors for Drug-Resistant Pathogens},
    journal = {Biochemistry},
    volume = {64},
    number = {6},
    pages = {1169--1179},
    year = {2025},
    doi = {10.1021/acs.biochem.4c00751}
}

@article{lorpaiboon2024accurate,
    author = {Lorpaiboon, Chatipat and Guo, Spencer C. and Strahan, John and Weare, Jonathan and Dinner, Aaron R},
    title = {Accurate estimates of dynamical statistics using memory},
    journal = {The Journal of Chemical Physics},
    volume = {160},
    number = {8},
    pages = {084108},
    year = {2024},
    doi = {10.1063/5.0187145}
}

@article{unarta2024submillisecond,
    author = {Unarta, Ilona C and Cao, Siqin and Goonetilleke, Eshani C and Niu, Jiani and Gellman, Samuel H and Huang, Xuhui},
    title = {Submillisecond Atomistic Molecular Dynamics Simulations Reveal Hydrogen Bond-Driven Diffusion of a Guest Peptide in Protein–{RNA} Condensate},
    journal = {The Journal of Physical Chemistry B},
    volume = {128},
    number = {10},
    pages = {2347--2359},
    year = {2024},
    doi = {10.1021/acs.jpcb.3c08126}
}

@book{oppenheim2009discrete,
    author = {Oppenheim, Alan V and Schafer, Ronald W},
    title = {Discrete-Time Signal Processing},
    year = {2009},
    isbn = {0131988425},
    publisher = {Prentice Hall Press},
    address = {Upper Saddle River, New Jersey, United States of America},
    edition = {3rd}
}

@book{larsson2003partial,
    author = {Larsson, Stig and Thom\'{e}e, Vidar},
    title = {Partial Differential Equations with Numerical Methods},
    year = {2003},
    isbn = {9783540017721},
    publisher = {Springer Berlin},
    address = {Heidelberg, Baden-W\"{u}rttemberg, Federal Republic of Germany}
}

@article{meng1993mle,
    author = {Meng, Xiao-Li and Rubin, Donald B},
    title = {Maximum likelihood estimation via the {ECM} algorithm: A general framework},
    journal = {Biometrika},
    volume = {80},
    number = {2},
    pages = {267--278},
    year = {1993},
    doi = {10.1093/biomet/80.2.267}
}

@article{piana2012HP35,
    author = {Piana, Stefano and Lindorff-Larsen, Kresten and Shaw, David E},
    title = {Protein folding kinetics and thermodynamics from atomistic simulation},
    journal = {Proc. Natl. Acad. Sci. U.S.A.},
    volume = {109},
    number = {44},
    pages = {17845--17850},
    year = {2012},
    doi = {10.1073/pnas.1201811109}
}

@article{kobayashi2020BMF1,
    author = {Kobayashi, Ryohei and Ueno, Hiroshi and Li, Chun-Biu and Noji, Hiroyuki},
    title = {Rotary catalysis of bovine mitochondrial {F}$_1$-{ATP}ase studied by single-molecule experiments},
    journal = {Proc. Natl. Acad. Sci. U.S.A.},
    volume = {117},
    number = {3},
    pages = {1447--1456},
    year = {2020},
    doi = {10.1073/pnas.1909407117}
}

@article{peyre2019computational,
    author = {Peyr\'{e}, Gabriel and Cuturi, Marco},
    title = {Computational Optimal Transport},
    year = {2019},
    issue_date = {Feb 2019},
    publisher = {Now Publishers Inc.},
    address = {Hanover, Massachusetts, United States of America},
    volume = {11},
    doi = {10.1561/2200000073},
    journal = {Found. Trends Mach. Learn.},
    pages = {355--607},
    numpages = {257}
}

@article{foerster1948zwischenmolekulare,
    author = {F\"{o}rster, Theodor},
    title = {Zwischenmolekulare {E}nergiewanderung und {F}luoreszenz},
    journal = {Ann. Phys. (Berl.)},
    volume = {437},
    number = {1--2},
    pages = {55--75},
    doi = {10.1002/andp.19484370105},
    year = {1948}
}

@article{knoch2019nonequilibrium,
    author = {Knoch, Fabian and Speck, Thomas},
    title = {Non-equilibrium {M}arkov state modeling of periodically driven biomolecules},
    journal = {J. Chem. Phys.},
    volume = {150},
    number = {5},
    pages = {054103},
    year = {2019},
    month = {2},
    doi = {10.1063/1.5055818}
}

@article{me,
    title = {Measuring Irreversibility from Learned Representations of Biological Patterns},
    author = {Li, Junang and Liu, Chih-Wei Joshua and Szurek, Michal and Fakhri, Nikta},
    journal = {PRX Life},
    volume = {2},
    issue = {3},
    pages = {033013},
    numpages = {14},
    year = {2024},
    month = {Sep},
    publisher = {American Physical Society},
    doi = {10.1103/PRXLife.2.033013},
    url = {https://link.aps.org/doi/10.1103/PRXLife.2.033013}
}

@article{li2025convergence,
    author = {Li, Mengmou and Laib, Khaled and Hatanaka, Takeshi and Lestas, Ioannis},
    title = {Convergence rate bounds for the mirror descent method: {IQC}s, {P}opov criterion and {B}regman divergence},
    journal = {Automatica},
    volume = {171},
    pages = {111973},
    year = {2025},
    doi = {10.1016/j.automatica.2024.111973}
}

@article{cutting2015cystic,
    author = {Cutting, Garry R.},
    title = {Cystic fibrosis genetics: from molecular understanding to clinical application},
    journal = {Nat. Rev. Genet.},
    volume = {16},
    pages = {45--56},
    year = {2015},
    doi = {10.1038/nrg3849}
}

@article{csanady2019structure,
    author = {Csan\'{a}dy, L\'{a}szl\'{o} and Vergani, Paola and Gadsby, David C.},
    title = {Structure, Gating, and Regulation of the {CFTR} Anion Channel},
    journal = {Physiol. Rev.},
    volume = {99},
    number = {1},
    pages = {707--738},
    year = {2019},
    doi = {10.1152/physrev.00007.2018}
}

@article{vanderplas2018drug,
    author = {Van der Plas, Steven E. and Kelgtermans, Hans and De Munck, Tom and Martina, S{\'e}bastien L. X. and Dropsit, S{\'e}bastien and Quinton, Evelyne and De Blieck, Ann and Joannesse, Caroline and Tomaskovic, Linda and Jans, Mia and Christophe, Thierry and van der Aar, Ellen and Borgonovi, Monica and Nelles, Luc and Gees, Maarten and Stouten, Pieter and Van Der Schueren, Jan and Mammoliti, Oscar and Conrath, Katja and Andrews, Martin},
    title = {Discovery of {N}-(3-Carbamoyl-5,5,7,7-tetramethyl-5,7-dihydro-4{H}-thieno[2,3-c]pyran-2-yl)-l{H}-pyrazole-5-carboxamide ({GLPG1837}), a Novel Potentiator Which Can Open Class {III} Mutant Cystic Fibrosis Transmembrane Conductance Regulator ({CFTR}) Channels to a High Extent},
    journal = {J. Med. Chem.},
    volume = {61},
    number = {4},
    pages = {1425--1435},
    year = {2018},
    doi = {10.1021/acs.jmedchem.7b01288}
}

@article{juette2016single,
    author = {Juette, Manuel F. and Terry, Daniel S and Wasserman, Michael R. and Altman, Roger B. and Zhou, Zhou and Zhao, Hong and Blanchard, Scott C.},
    title = {Single-molecule imaging of non-equilibrium molecular ensembles on the millisecond timescale},
    journal = {Nat. Methods},
    volume = {13},
    pages = {341--344},
    year = {2016},
    doi = {10.1038/nmeth.3769}
}

@article{jih2013vx,
    author = {Jih, Kang-{Y}ang and Hwang, Tzyh-{C}hang},
    title = {Vx-770 potentiates {CFTR} function by promoting decoupling between the gating cycle and {ATP} hydrolysis cycle},
    journal = {Proc. Natl. Acad. Sci. U.S.A.},
    volume = {110},
    number={11},
    pages = {4404--4409},
    year = {2013},
    doi = {10.1073/pnas.1215982110}
}

@article{kuehlbrandt2019structure,
    author = {K\"{u}hlbrandt, Werner},
    title = {Structure and mechanisms of {F}-type {ATP} synthases},
    journal = {Annu. Rev. Biochem.},
    volume = {88},
    pages = {514--549},
    year = {2019},
    doi = {10.1146/annurev-biochem-013118-110903}
}

@article{karplus2004biomolecular,
    author = {Karplus, Martin and Gao, Yi Qin},
    title = {Biomolecular motors: the {F}$_1$-{ATP}ase paradigm},
    journal = {Curr. Opin. Struct. Biol.},
    volume = {14},
    number={2},
    pages = {250--259},
    year = {2004},
    doi = {10.1016/j.sbi.2004.03.012}
}

@article{suzuki2014chemomechanical,
    author = {Suzuki, Toshiharu and Tanaka, Kazumi and Wakabayashi, Chiaki and Saita, Ei-{i}chiro and Yoshida, Masasuke},
    title = {Chemomechanical coupling of human mitochondrial {F}$_1$-{ATP}ase motor},
    journal = {Nat. Chem. Biol.},
    volume = {10},
    pages = {930--936},
    year = {2014},
    doi = {10.1038/nchembio.1635}
}

@article{zhang2018molecular,
    author = {Zhang, Zhe and Liu, Fangyu and Chen, Jue},
    title = {Molecular structure of the {ATP}-bound, phosphorylated human {CFTR}},
    journal = {Proc. Natl. Acad. Sci. U.S.A.},
    volume = {115},
    number = {50},
    pages = {12757--12762},
    year = {2018},
    doi = {10.1073/pnas.1815287115}
}

@article{bason2015release,
    author = {Bason, John V. and Montgomery, Martin G. and Leslie, Andrew G. W. and Walker, John E.},
    title = {How release of phosphate from mammalian {F}$_1$-{ATP}ase generates a rotary substep},
    journal = {Proc. Natl. Acad. Sci. U.S.A.},
    volume = {112},
    number = {19},
    pages = {6009--6014},
    year = {2015},
    doi = {10.1073/pnas.1506465112}
}

@article{nagel2023selecting,
    author = {Nagel, Daniel and Sartore, Sofia and Stock, Gerhard},
    title = {Selecting Features for {M}arkov Modeling: A Case Study on {HP35}},
    journal = {J. Chem. Theory Comput.},
    volume = {19},
    number = {11},
    pages = {3391-3405},
    year = {2023},
    doi = {10.1021/acs.jctc.3c00240}
}

@article{kubelka2006sub,
    author = {Kubelka, Jan and Chiu, Thang K. and Davies, David R. and Eaton, William A. and Hofrichter, James},
    title = {Sub-microsecond protein folding},
    journal = {J. Mol. Biol.},
    volume = {359},
    pages = {546--553},
    year = {2006},
    doi = {10.1016/j.jmb.2006.03.034}
}

@article{eaton2021modern,
    author = {Eaton, William A.},
    title = {Modern Kinetics and Mechanism of Protein Folding: A Retrospective},
    journal = {J. Phys. Chem. B},
    volume = {125},
    number = {14},
    pages = {3452-3467},
    year = {2021},
    doi = {10.1021/acs.jpcb.1c00206}
}

@article{bretscher1979villin,
    author = {Bretscher, Anthony and Weber, Klaus},
    title = {Villin: the major microfilament-associated protein of the intestinal microvillus},
    journal = {Proc. Natl. Acad. Sci. U.S.A.},
    volume = {76},
    number = {5},
    pages = {2321--2325},
    year = {1979},
    doi = {10.1073/pnas.76.5.2321}
}

@article{mcknight1997nmr,
    author = {Mc{K}night, C. James and Matsudaira, Paul T. and Kim, Peter S.},
    title = {{NMR} structure of the 35-residue villin headpiece subdomain},
    journal = {Nat. Struct. Mol. Biol.},
    volume = {4},
    pages = {180--184},
    year = {1997},
    doi = {10.1038/nsb0397-180}
}

@misc{liu2026gmex,
    author = {Liu, Chih-Wei Joshua and Klinger, J\'{e}r\'{e}mie and Rotskoff, Grant M.},
    year = {2026},
    title = {{GME}x},
    url = {https://github.com/rotskoff-group/GMEx}
}

@article{vogelsang2008reducing,
    author = {Vogelsang, Jan and Kasper, Robert and Steinhauer, Christian and Person, Britta and Heilemann, Mike and Sauer, Markus and Tinnefeld, Philip},
    title = {A Reducing and Oxidizing System Minimizes Photobleaching and Blinking of Fluorescent Dyes},
    journal = {Angew. Chem.},
    volume = {47},
    number = {29},
    pages = {5465--5469},
    doi = {https://doi.org/10.1002/anie.200801518},
    year = {2008}
}

@article{mcgibbon2013learning,
    author = {Mc{G}ibbon, Robert T. and Pande, Vijay S.},
    title = {Learning kinetic distance metrics for {M}arkov state models of protein conformational dynamics},
    journal = {J. Chem. Theory Comput.},
    volume = {9},
    number = {7},
    pages = {2900--2906},
    year = {2013},
    doi = {10.1021/ct400132h}
}

@book{zwanzig2001nonequilibrium,
    author = {Zwanzig, Robert},
    title = {Nonequilibrium Statistical Mechanics},
    publisher = {Oxford University Press},
    address = {New York, New York, United States of America},
    year = {2001},
    month = {04},
    isbn = {9780195140187},
    doi = {10.1093/oso/9780195140187.001.0001}
}

@book{elber2020molecular,
    author = {Elber, Ron and Makarov, Dmitrii E. and Orland, Henri},
    title = {Molecular Kinetics in Condensed Phases: Theory, Simulation, and Analysis},
    publisher = {John Wiley \& Sons Ltd},
    address = {Chichester, United Kingdom of Great Britain and Northern Ireland},
    year = {2020},
    isbn = {9781119176770},
    doi = {10.1002/9781119176800}
}

@article{widder2025generalized,
    author = {Widder, Christoph and Zimmer, Johannes and Schilling, Tanja},
    title = {On the generalized {L}angevin equation and the {M}ori projection operator technique},
    journal = {J. Phys. A: Math. Theor.},
    volume = {58},
    pages = {405001},
    year = {2025},
    doi = {10.1088/1751-8121/ae02cc}
}

@article{zhu2018estimation,
    author = {Zhu, Yuanran and Dominy, Jason M. and Venturi, Daniele},
    title = {On the estimation of the {M}ori-{Z}wanzig memory integral},
    journal = {J. Math. Phys.},
    volume = {59},
    pages = {103501},
    year = {2018},
    doi = {10.1063/1.5003467}
}

@article{meyer2020numerical,
    author = {Meyer, Hugues and Wolf, Steffen and Stock, Gerhard and Schilling, Tanja},
    title = {A numerical procedure to evaluate memory effects in non-equilibrium coarse-grained models},
    journal = {Adv. Theory Simul.},
    volume = {4},
    pages = {2000197},
    year = {2020},
    doi = {10.1002/adts.202000197}
}

@article{walker1931periodicity,
    author = {Walker, Gilbert Thomas},
    title = {On periodicity in series of related terms},
    journal = {Proc. R. Soc. A},
    volume = {131},
    pages = {518--532},
    year = {1931},
    doi = {10.1098/rspa.1931.0069}
}

@article{yule1927method,
    author = {Yule, George Udny},
    title = {On a method of investigating periodicities in disturbed series, with special reference to {W}olfer's sunspot numbers},
    journal = {Phil. Trans. R. Soc. A},
    volume = {226},
    pages = {267--298},
    year = {1927},
    doi = {10.1098/rsta.1927.0007}
}

@inproceedings{anton,
    author = {Shaw, David E. and Dror, Ron O. and Salmon, John K. and Grossman, J. P. and Mackenzie, Kenneth M. and Bank, Joseph A. and Young, Cliff and Deneroff, Martin M. and Batson, Brannon and Bowers, Kevin J. and Chow, Edmond and Eastwood, Michael P. and Ierardi, Douglas J. and Klepeis, John L. and Kuskin, Jeffrey S. and Larson, Richard H. and Lindorff-Larsen, Kresten and Maragakis, Paul and Moraes, Mark A. and Piana, Stefano and Shan, Yibing and Towles, Brian},
    title = {Millisecond-scale molecular dynamics simulations on Anton},
    year = {2009},
    isbn = {9781605587448},
    publisher = {Association for Computing Machinery},
    address = {New York, New York, United States of America},
    doi = {10.1145/1654059.1654126},
    booktitle = {Proceedings of the Conference on High Performance Computing Networking, Storage and Analysis},
    articleno = {65},
    numpages = {11},
    location = {Portland, Oregon},
    series = {SC '09}
}

@article{henry2013comparing,
    author = {Henry, Eric R. and Best, Robert B. and Eaton, William A.},
    title = {Comparing a simple theoretical model for protein folding with all-atom molecular dynamics simulations},
    journal = {Proc. Natl. Acad. Sci. U.S.A.},
    volume = {110},
    pages = {17880--17885},
    year = {2013},
    doi = {10.1073/pnas.1317105110}
}

@article{banushkina2013hires,
    author = {Banushkina, Polina V. and Krivov, Sergei V.},
    title = {High-resolution free-energy landscape analysis of $\alpha$-helical protein folding: {HP}35 and its double mutant},
    journal = {J. Chem. Theory Comput.},
    volume = {9},
    pages = {5257--5266},
    year = {2013},
    doi = {10.1021/ct400651z}
}

@article{lindorfflarsen2011how,
    author = {Lindorff-Larsen, Kresten and Piana, Stefano and Dror, Ron O. and Shaw, David E.},
    title = {How fast-folding proteins fold},
    journal = {Science},
    volume = {334},
    pages = {517--520},
    year = {2011},
    doi = {10.1126/science.1208351}
}

@article{wu2011asymptotic,
    author = {Wu, Wei Biao},
    title = {Asymptotic theory for stationary processes},
    journal = {Stat. Its Interface},
    volume = {4},
    pages = {207},
    year = {2011},
    doi = {10.4310/SII.2011.v4.n2.a15}
}

@article{fonollosa1995sample,
    author = {Fonollosa, Jos\'{e} A. R.},
    title = {Sample cumulants of stationary processes: asymptotic results},
    journal = {IEEE Trans. Signal Process.},
    volume = {43},
    pages = {967},
    year = {1995},
    doi = {10.1109/78.376848}
}

@book{sokal1997mc,
    author = {Sokal, Alan},
    editor={De{W}itt-{M}orette, Cecile and Cartier, Pierre and Folacci, Antoine},
    title = {Functional Integration},
    chapter = {6},
    series = {{NATO} Science Series B},
    publisher={Springer},
    address={New York, New York, United States of America},
    doi={10.1007/978-1-4899-0319-8},
    isbn={978-1-4899-0319-8},
    year = {1997},
    volume = {361},
    pages = {131--192},
}

@article{perez2016assessment,
    author = {Perez, Juan J. and Tomas, M. Santos and Rubio-Martinez, Jaime},
    title = {Assessment of the sampling performance of multiple-copy dynamics versus a single trajectory},
    journal = {J. Chem. Inf. Model.},
    year = {2016},
    volume = {56},
    pages = {1950--1962},
    doi = {10.1021/acs.jcim.6b00347}
}

@article{elf2019single,
    author = {Elf, Johan and Barkefors, Irmeli},
    title = {Single-molecule kinetics in living cells},
    journal = {Annu. Rev. Biochem.},
    year = {2019},
    volume = {88},
    pages = {635--659},
    doi = {10.1146/annurev-biochem-013118-110801}
}

@article{meng2023chimerax,
    author = {Meng, Elaine C. and Goddard, Thomas D. and Pettersen, Eric F. and Couch, Greg S. and Pearson, Zach J. and Morris, John H. and Ferrin, Thomas E.},
    title = {{UCSF ChimeraX}: tools for structure building and analysis},
    journal = {Protein Sci.},
    volume = {32},
    pages = {e4792},
    doi = {10.1002/pro.4792},
    year = {2023}
}

@article{gill1990survey,
    author = {Gill, Richard D. and Johansen, Soren},
    title = {A survey of product-integration with a view toward application in survival analysis},
    journal = {Ann. Stat.},
    volume = {18},
    pages = {1501},
    doi = {10.1214/aos/1176347865},
    year = {1990}
}
\end{document}